\documentclass[aps, rmp, twocolumn, 10pt,floatfix, longbibliography, superscriptaddress]{revtex4-1}

\usepackage{amsmath} % needed for \tfrac, \bmatrix, etc.
\usepackage{amsfonts} % needed for bold Greek, Fraktur, and blackboard bold
\usepackage{graphicx} % needed for figures
\usepackage[english]{babel}
\usepackage{blindtext}
\usepackage{microtype}
\usepackage[table]{xcolor}
\usepackage{hyphenat}
\definecolor{xbrown}{HTML}{744f18}
\definecolor{xblue}{HTML}{273e90}
\definecolor{xgreen}{HTML}{056901}
\definecolor{xaqua}{HTML}{0ba5b2}

\usepackage{ragged2e}
\usepackage{caption}
\DeclareCaptionJustification{justify}{\justify}
\usepackage{subcaption}
\usepackage{tabularx}
\usepackage{booktabs} % For better tables

\usepackage{enumerate}
\usepackage{multirow}

\usepackage{url}

\usepackage[inline]{enumitem}

\usepackage{tikz} % for figures
\usetikzlibrary{calc, arrows.meta,shadows}
\usetikzlibrary{decorations.pathmorphing, decorations.markings}

\usepackage{siunitx}

\newcommand{\n}{\textendash}
\newcommand{\m}{\textemdash}

\newenvironment{changemargin}[2]{%
\begin{list}{}{%
\setlength{\topsep}{0pt}%
\setlength{\leftmargin}{#1}%
\setlength{\rightmargin}{#2}%
\setlength{\listparindent}{\parindent}%
\setlength{\itemindent}{\parindent}%
\setlength{\parsep}{\parskip}%
}%
\item[]}{\end{list}}

\makeatletter
\let\cat@comma@active\@empty
\makeatother

\definecolor{darkblue}{rgb}{0,0,0.7}
\definecolor{darkred}{rgb}{0.9,0,0}
\usepackage[colorlinks]{hyperref}
\hypersetup{linkcolor = darkblue, citecolor = darkred, urlcolor = darkred}
\usepackage{hypcap}
\begin{document}

\title{Understanding quantum physics through simple experiments: from wave\hyp particle duality to Bell's theorem}

\author{Ish Dhand}
\affiliation{Institut f{\"u}r Theoretische Physik and Center for Integrated Quantum Science and Technology (IQST),\\
Albert-Einstein-Allee 11, Universit{\"a}t Ulm, 89069 Ulm, Germany}
\author{Adam D'Souza}
\affiliation{
Department of Community Health Sciences, Cumming School of Medicine,\\ University of Calgary, Calgary T2N4N1, Alberta, Canada }
\author{Varun Narasimhachar}
\affiliation{School of Physical \& Mathematical Sciences,\\ Nanyang Technological University, 21 Nanyang Link Singapore 637371}
\author{Neil Sinclair}
\affiliation{Department of Physics and Astronomy and Institute for Quantum Science and Technology (IQST),\\ University of Calgary, Calgary T2N1N4, Alberta, Canada}
\affiliation{
Department of Physics, Mathematics, and Astronomy and Alliance for Quantum Technologies (AQT),\\  California Institute of Technology,  1200 East California Blvd., Pasadena, California 91125, USA}	
\author{Stephen Wein}
\affiliation{Department of Physics and Astronomy and Institute for Quantum Science and Technology (IQST),\\ University of Calgary, Calgary T2N1N4, Alberta, Canada}
\author{Parisa Zarkeshian}
\affiliation{Department of Physics and Astronomy and Institute for Quantum Science and Technology (IQST),\\ University of Calgary, Calgary T2N1N4, Alberta, Canada}
\author{Alireza~Poostindouz}
\affiliation{Department of Physics and Astronomy and Institute for Quantum Science and Technology (IQST),\\ University of Calgary, Calgary T2N1N4, Alberta, Canada}
\affiliation{Department of Computer Science,\\ University of Calgary, Calgary T2N1N4, Alberta, Canada}
\author{Christoph Simon}
\email{christoph.simon@gmail.com}
\affiliation{Department of Physics and Astronomy and Institute for Quantum Science and Technology (IQST),\\ University of Calgary, Calgary T2N1N4, Alberta, Canada}
\date{\today}

\begin{abstract}
\begin{changemargin}{0cm}{-3cm}
Quantum physics, which describes the strange behavior of light and matter at the smallest scales, is one of the most successful descriptions of reality, yet it is notoriously inaccessible. Here we provide an approachable explanation of quantum physics using simple thought experiments. We derive all relevant quantum predictions using minimal mathematics, without introducing the advanced calculations that are typically used to describe quantum physics. We focus on the two key surprises of quantum physics, namely {\it wave{\n}particle duality}, a term that was introduced to capture the fact that single quantum particles in some respects behave like waves and in other respects like particles, and {\it entanglement}, which applies to two or more quantum particles and brings out the inherent contradiction between quantum physics and seemingly obvious assumptions regarding the nature of reality. Following arguments originally made by John Bell and Lucien Hardy, we show that so-called {\it local hidden variables} are inadequate at explaining the behavior of entangled quantum particles. This means that one either has to give up on hidden variables, i.e., the idea that the outcomes of measurements on quantum particles are determined before an experiment is actually carried out, or one has to relinquish the principle of locality, which requires that no causal influences should be faster than the speed of light and is a cornerstone of Einstein's theory of relativity. Finally, we describe how these remarkable predictions of quantum physics have been confirmed in experiments. We have successfully used the present approach in a course that is open to all undergraduate students at the University of Calgary, without any prerequisites in mathematics or physics.
\end{changemargin}
\end{abstract}

\maketitle
\tableofcontents 

\section{Introduction}
Quantum physics, our most successful description for the physics of the tiny, is notoriously hard to understand.
The difficulty stems from the conflict between the predictions of quantum physics and common-sense human intuition, which is relatively at ease with the laws of \textit{classical} physics. Experts of quantum physics, such as Einstein, Bohr, Heisenberg, de Broglie, and Feynman, have admitted to being baffled by it.
Thus, it might seem utterly hopeless to explain the theory to non{\hyp}experts.
Nevertheless, in this manuscript, we attempt to convey to non{\hyp}experts the essence of the \emph{strangeness} of quantum mechanics.
We convey this strangeness without invoking daunting technical details, but at the same time, without resorting to inaccurate analogies or compromising any essential rigor in treatment.

Now, to be sure, popular culture abounds with accounts of the strangeness of quantum mechanics{\m}jokes, motifs in science fiction, and so on.
But the prevalent notion of the strangeness of quantum mechanics is no different from that which one may associate with \emph{any} idea from a field in which one lacks expertise.
In addition, because of similarity in language, jargon from quantum mechanics has been misapplied to new{\hyp}age thought and this can propagate a false sense of mystique and obscurity.
Such a feeling is clearly different from the bafflement that a physicist may experience by studying quantum mechanics.
The latter results not from a lack of comprehension of the technical ideas, but on the contrary, from a precise understanding both of these technical ideas and of their conflict with commonsense notions.
It is this latter sense of strangeness which we attempt to communicate here.

Readers that have encountered (secondary school-level) mathematical concepts of probability, square roots, and imaginary numbers are qualified to grasp our arguments. 
Nonetheless, for clarity, we include sufficient introductions to these concepts to supplement our physical discussions.

We focus on two key ways in which quantum mechanics is at odds with our intuition about the macroscopic world. 
The first of these conflicts is the so\hyp called \emph{wave\n particle duality}.
Through a discussion of simple experiments involving light, we describe how tiny particles seem to possess both wave- and particle-like behavior. 
That is, some experiments can be explained by thinking of light as composed of particles and others can be explained by treating it as a wave.
A more satisfactory explanation of single-particle experiments is provided by quantum physics, which requires us to consider all the different possible paths leading to any given detector.
Under certain conditions, these paths can {\it interfere} in a manner that is analogous to classical waves like water or sound waves. After introducing the classical concepts of waves and interference in some detail, we show exactly how quantum physics explains the wave-like and the particle-like behavior of single-particle experiments.

We then enter even stranger territory: quantum interference between two or more particles. 
We approach such two\hyp particle phenomena by first describing \emph{quantum entanglement}, a property that collections of quantum particles can possess. 
Entangled particles exhibit {\it randomness} in their individual measurement outcomes, but these outcomes are random in a {\it correlated} manner, like synchronized dancers with unpredictable but coordinated moves. 
This interplay between correlation and randomness is such that entangled particles cannot be described as individual entities and must be described as one composite entity.
Entanglement is at the heart of the essential incompatibility of quantum physics with seemingly obvious assumptions about the nature of reality.
Using our single\hyp particle (example) experiments as building blocks, we piece together an accessible demonstration of this incompatibility, which is the second key revelation of quantum physics.

Despite the strangeness of single\hyp particle interference, its outcomes can also be explained completely by `mechanistic' theories in line with our everyday, classical, experience. 
In these so\hyp called \emph{hidden\hyp variable theories}, each particle carries information about the eventual outcomes of all measurements that can be performed on the particle, a bit like a hidden ``cheat sheet'' that the particle can refer to. Einstein, Podolsky and Rosen (EPR) suggested in 1935 that this approach might allow one to explain all of the apparent strangeness of quantum physics in an intuitive way~\cite{Einstein1935}.
 
However, in 1964 John Bell famously showed~\cite{Bell1964} that when one tries to apply the hidden variable approach to entangled particles, one encounters a major difficulty. Bell showed that so-called {\it local hidden variables} can not explain the predictions of quantum physics for entangled particles. One either has to give up on the idea of hidden variables altogether, or one has to accept that these hidden variables have to be {\it non-local}, i.e., the hidden variables associated with one particle sometimes need to be updated instantaneously based on the behavior of another, possibly very distant, particle. This latter possibility is at odds with the principles of Einstein's famous theory of relativity, which imposes a universal speed limit on the rate at which a cause can have an effect, a concept referred to as {\it locality} in the present context. 

It is worth pointing out immediately that despite this dramatic conclusion, entangled particles cannot be used to transmit information faster than at the speed of light. We will see that this important principle is protected by the randomness of the individual measurement outcomes for each particle.

After describing the EPR argument in detail, we piece together concepts from the described one- and two-particle experiments to present Lucien Hardy's version of Bell's theorem~\cite{Hardy1992}.  
We use this version because it allows us to derive both the predictions made by local hidden variables and those made by quantum physics with minimal mathematics. Moreover, because Hardy's approach shines a bright spotlight on the sharp contrast between quantum predictions and everyday thinking, it provides an intuitive understanding of the second key surprise of quantum physics.
 
We conclude with a brief overview of real\hyp life experiments that have demonstrated the strangeness of quantum physics.
Beginning with single\hyp particle experiments that show experimentalists' technical prowess in generating and controlling fragile systems of quantum particles, we describe experimental progress up to recent efforts that decisively ruled out local hidden variables by closing all the loopholes in previous experiments that alternate explanations had relied upon.
In this discussion, we hope to convey a sense of the extraordinary effort and thoroughness that scientists have put into testing quantum physics.

Our approach, which uses simple experiments to explain quantum principles, was inspired by the popular book of~\citet{Scarani2006}.
We would also like to mention the excellent second book by \citet{Scarani2010} and the equally excellent books by~\citet{Rudolph2017} and~\citet{Raymer2017}. In contrast to these authors, we do not introduce advanced mathematical concepts such as state vectors.
Remarkably, it is still possible to derive all the relevant quantum predictions just using the various relevant single-particle and two-particle histories.
The expert reader might anticipate correctly that these histories are basically generalized Feynman paths~\cite{feynman2010quantum,derbes1996feynman}. 
We also recommend articles by~\citet{Kwiat2000} as well as~\citet{christensen}, which pedagogically describe Hardy's version of Bell's theorem.

The spur for developing our approach was an annual undergraduate course, titled ``Quantum Mysteries and Paradoxes'', that we have taught for several years at the University of Calgary. This course is open to students from all Faculties without any prerequisites. It has been taken by close to a thousand students and always receives very positive reviews. Our success in communicating the principles of quantum physics to students with a broad range of backgrounds (majoring in the arts and business as well as in the various sciences) leads us to believe that our approach may also be useful in other educational contexts. Moreover, we have repeatedly and successfully taught a continuing education version of the course, which gives us confidence that this material can also be interesting for motivated individual readers with little background in science or mathematics.

This manuscript is organized as follows. Sec.~\ref{Sec:OneParticle} introduces single-particle interference and wave\hyp particle duality. Sec.~\ref{Sec:2Particle} discusses two-particle interference and entanglement. Sec.~\ref{Sec:EPR} introduces the Einstein-Podolsky-Rosen argument and the concept of local hidden variables. Sec.~\ref{Sec:Hardy} develops Hardy's version of Bell's theorem, demonstrating that local hidden variables cannot reproduce the predictions of quantum physics. Sec.~\ref{Sec:Experiments} summarizes the large body of experimental work that has verified these quantum predictions. Section~\ref{Sec:Discussion} summarizes these results and looks towards the future.

\section{Single-Particle Interference and Wave\hyp Particle Duality}
\label{Sec:OneParticle}

Here we introduce the first key surprise of quantum physics: that tiny particles display both wave-like and particle-like behavior.
To explain this so-called wave\hyp particle duality, we discuss simple experiments (which we name~\hyperref[exp1]{S1} through~\hyperref[exp6]{S6}) with single particles of light, i.e.\ photons.
These experiments can be and have been performed with other tiny particles, such as electrons or neutrons but we chose photons because it is more intuitive to think in terms of light sources, mirrors and detectors. 

Experiments~\hyperref[exp1]{S1} and~\hyperref[exp2]{S2} illustrate the particle aspect of single photons, whereas Experiments~\hyperref[exp3]{S3} and~\hyperref[exp4]{S4} reveal their wave aspect.
Experiments~\hyperref[exp5]{S5} and~\hyperref[exp6]{S6} help clarify the conditions under which one observes particle-like or wave-like characteristics of single photons. 

After a description of the experiments, we briefly recap the classical theory of waves and describe how this can explain the outcomes of Experiments~\hyperref[exp3]{S3} and~\hyperref[exp4]{S4}. 
Next, we present the unified explanation of all these experiments as provided by quantum physics.
This is done through a set of rules which quantum physics uses for predicting the experimental outcomes and which encapsulate two important principles of quantum physics, namely the superposition principle and the principle of indistinguishability.
We conclude this section with a surprising quantum phenomenon, referred to as ``interaction-free measurement'', that allows one to infer the presence of an object without directly interacting with it, which will play an important role in Hardy's version of Bell's theorem (Sec.~\ref{Sec:Hardy}).

% BEGIN \input{Figure_Single_particle_interference.tex}
\begin{figure}
\centering
\captionsetup[subfigure]{labelformat=empty}

\newcommand*{\figsScale}{1} % beamsplitter and reflectors scale up or down by this ratio

\begin{subfigure}[t]{0.45\columnwidth}
\includegraphics[width = 0.6\textwidth]{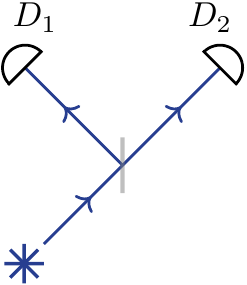}
\caption{(\hyperref[exp1]{S1})}
\label{Fig:Exp1}
\end{subfigure}
\begin{subfigure}[t]{0.5\columnwidth}
\includegraphics[width = \textwidth]{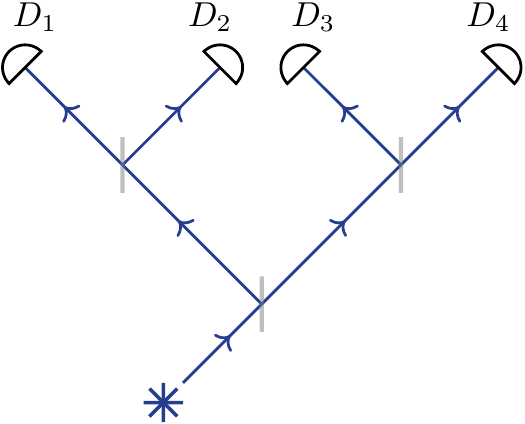}
\caption{(\hyperref[exp2]{S2})}\label{Fig:Exp2}
\end{subfigure}
\\
\begin{subfigure}[t]{0.45\columnwidth}
\includegraphics[width = 0.6\textwidth]{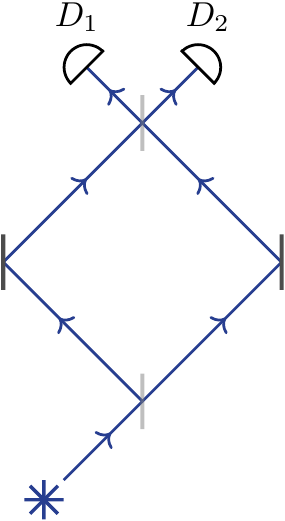}
\caption{(\hyperref[exp3]{S3})}\label{Fig:Exp3}
\end{subfigure}
\begin{subfigure}[t]{0.45\columnwidth}
\includegraphics[width = 0.6\textwidth]{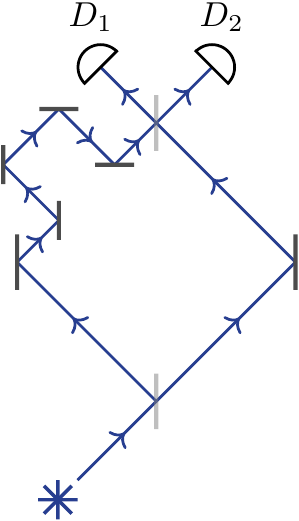}
\caption{(\hyperref[exp4]{S4})}\label{Fig:Exp4}
\end{subfigure}
%\captionsetup{width=1.0\columnwidth}
\caption{\textbf{Single-particle interference experiments.} (\hyperref[exp1]{S1}) Photons are emitted one by one by a single-photon source and directed towards a beam splitter. Photons that are reflected are detected by detector $D_1$, and photons that are transmitted are detected by detector $D_2$. 
 (\hyperref[exp2]{S2}) Extension of the previous experiment. Every photon now encounters a second beam splitter before being detected. (\hyperref[exp3]{S3}) Mach-Zehnder interferometer. Mirrors are used to make the two paths emerging from the first beam splitter cross again on a second beam splitter. There are now two possible ways for each photon to arrive at each detector. For example, to arrive at $D_2$, the photon could have been transmitted by the first beam splitter and reflected by the second beam splitter or vice versa. (\hyperref[exp4]{S4}) The left path in the Mach-Zehnder interferometer is made longer by inserting four mirrors.
}\label{Fig:SingleParticle}
\end{figure}
% END \input{Figure_Single_particle_interference.tex}

\subsection{Experiments with Photons}
\label{subsec:singlephotonexperiments}
Each of the following experiments involve a single\hyp photon source, some beam\hyp splitters, and some single\hyp photon detectors. A single\hyp photon source fires one photon at a time, ruling out the possibility of two photons being in the setup of the experiment at the same time. A beam\hyp splitter is a device that splits a continuous stream (i.e.\ a beam) of particles into two equal parts. In an experiment with a beam of light, one can think of a beam\hyp splitter as simply a ``semi\hyp transparent mirror''; it reflects half of the beam and transmits the rest of it (see Fig.~\hyperref[Fig:Exp1]{1:S1}). To observe photons, we use single\hyp photon detectors, i.e.,\ devices that can indicate the presence of a single photon. 
We will sometimes refer to this indication as a ``click'' (imagining an audible signal), but in practice the signal is typically electrical.
These simple building blocks are put together into six experiments, \hyperref[exp1]{S1}--\hyperref[exp6]{S6}, that enable one to appreciate the particle-like and wave-like properties of photons.

\subsubsection{Experiment~S1}\label{exp1}

The setup of the first experiment comprises a single-photon source, a beam\hyp splitter, and two detectors, as shown in Fig.~\hyperref[Fig:Exp1]{1:S1}.
We consider a situation in which the single photon source directs photons, one at a time, towards the  beam\hyp splitter. 
We know that a macroscopic beam of photons is split into two equal halves under these conditions, but what will happen to individual photons? 

To answer the question, the experiment is conducted and the detection outcomes are recorded.
The first important observation is that for a single photon there is always only a single detection. 
There is either a click in detector $D_1$ or a click in detector $D_2$, but never in both.
It is this observation because of which we can talk about the photon as if it is a particle of light; it cannot be split into two.

The second observation is that the clicks seem to occur randomly. 
On an average, half of the photons are detected in $D_1$ and half in $D_2$, but there is no discernible sequence to which detector will actually click. 
This suggests that when a photon reaches a beam\hyp splitter, it seems to make a random choice about which path to take: with probability $1/2$ it decides to take the path towards $D_1$, and with probability $1/2$ it takes the path towards $D_2$.

\subsubsection{Experiment~S2}\label{exp2}

The setup for Experiment~\hyperref[exp2]{S2} is depicted in Fig.~\hyperref[Fig:Exp2]{1:S2}. 
To better understand the effect of a beam\hyp splitter on single photons, we consider an experiment with two more beam\hyp splitters in each outgoing path of the first beam\hyp splitter. Where will the photons be detected now? 

Based on the previous experiment, we would expect that there will be only one detection for each photon sent. 
This is indeed what one observes in this experiment as well. 
Moreover we would expect that the photons will be detected in all four detectors with equal probability of $1/4$, since for each beam\hyp splitter the photon should have an equal probability ($1/2$) of being transmitted or reflected. 
So, for each detector the probability that the photon reaches it is $1/2 \times 1/2=1/4$.
This expectation is also confirmed if one performed this experiment: similar to the case of Experiment~\hyperref[exp1]{S1}, there is no way to predict which detector will actually click during a given trial, but on average the detectors click with equal probability.
This experiment thus reinforces the idea that the photon is a particle that seems to make a random choice at each  beam\hyp splitter.

\subsubsection{Experiment~S3}\label{exp3}
The setup of the third experiment, called a Mach--Zehnder interferometer, is depicted in Fig.~\hyperref[Fig:Exp3]{1:S3}. The first half of the setup is similar to that of Experiment~\hyperref[exp1]{S1}. Then, with the use of two perfect mirrors, we redirect the two outgoing paths of the first beam\hyp splitter so that they meet again. A second beam\hyp splitter is placed at the meeting point. Two single{\hyp}photon detectors are located at each outgoing path of this second beam\hyp splitter. 
Again sending in photons one by one, where will they be detected now? 

First of all, we again expect only a single detection for each photon sent into the setup, and this is indeed what is observed. 

Secondly, based on the experience with the first two experiments, one might expect the photons to be detected with an equal probability of $1/2$ in each detector. 
On reaching the first beam\hyp splitter, each photon should have a probability of $1/2$ to reflect or to transmit, i.e., to take the left path or the right path. 
Then at the second  beam\hyp splitter, each photon should have an equal probability to be transmitted or reflected, no matter which path it took previously. 
This should lead to equal probabilities for the two detectors. 

In more detail, one could argue as follows. There are two possibilities for the photon to arrive at detector $D_1$. It can take the left path (with probability 1/2) and then be reflected (with probability 1/2), giving a total probability of $1/2 \times 1/2=1/4$ for this possibility. Or it can take the right path (with probability 1/2) and then be transmitted (with probability 1/2), giving a total probability of $1/2 \times 1/2=1/4$ for this second possibility. Adding up the two possibilities should give a total probability of $1/4+1/4=1/2$ for the photon to be detected in $D_1$. An analogous argument can be made for detector $D_2$. 

However, this is not at all what is observed. In fact, \emph{all} the photons are detected by detector $D_2$ and none by $D_1$! This clearly contradicts the simple model of the photon making random choices at each beam\hyp splitter. Something different must be happening. 
As we will explain in detail in Sec.~\ref{sec:Single_particle_Theory}, the photons are showing a wave-like character here\m even though they are still arriving one by one at the detector $D_2$ in a particle-like manner! Now let us consider a slightly modified version of the Mach\n Zehnder interferometer.

\subsubsection{Experiment~S4}\label{exp4}
The setup of this experiment is identical to that of Experiment~\hyperref[exp3]{S3}, except that the left path has been extended with the help of some (perfect) mirrors (see Fig.~\hyperref[Fig:Exp4]{1:S4}).
What will happen to the percentage of photons detected in $D_1$ and $D_2$ as we vary the length of the path extension?

It is not immediately clear why the length of the path would matter. 
However, the path length has a dramatic effect on the outcome of the experiment: as we increase the path length on the left, detector $D_1$ starts clicking some of the time, with a probability initially increasing as we lengthen the path.
Of course, the probability of detection by $D_2$ decreases accordingly. 
At some point, with an extension by a certain length (call it $L$), we detect no photons at $D_2$: all the photons are now detected in $D_1$. 
The detection pattern has reversed from its original form! If we extend the length further, $D_2$ starts to register clicks again, with greater probability as we increase the length (with a simultaneous decrease in the click probability at $D_1$).
When the extension is $2 L$, the original pattern is recovered, and the photons are all detected in $D_2$ again. 
For an extension by $3 L$, the photons all go to $D_1$ again, and so on, the pattern reversing for each additional $L$ added to the path.

This behavior is difficult to understand if one tries to imagine the photon as a particle taking just one of the two paths within the interferometer. 
Considering the beam\hyp splitter that is placed right after the source, it is natural to suppose that half of the photons take the left path and the other half take the right path (based on our observations from Experiment~\hyperref[exp1]{S1}). 
Hence, any change made to one of the paths should change the outcomes of no more than half the photons.
Instead what is actually observed is that changing just one of the paths (the left one in our example) affects the behavior of all of the photons, and not (as one might incorrectly imagine) just the half that are supposed to have taken the left path. Thus, somehow all photons seem to be ``aware'' of this change in the left path! There is nothing special about the left path, by the way\m the same behavior can be observed by changing only the right path. Again all the photons are affected by changing only one of the two possible paths.

% BEGIN \input{Figure_NDD.tex}
\begin{figure}
\captionsetup[subfigure]{labelformat=empty}

\newcommand*{\figsScale}{1} % beamsplitter and reflectors scale up or down by this ratio

\begin{subfigure}[t]{0.45\columnwidth}
\includegraphics[width = 0.7\textwidth]{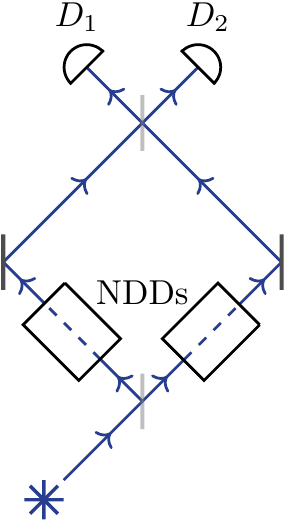}
\caption{(\hyperref[exp5]{S5})}\label{Fig:NDD1}
\end{subfigure}
\begin{subfigure}[t]{0.45\columnwidth}
\includegraphics[width = 0.73\textwidth]{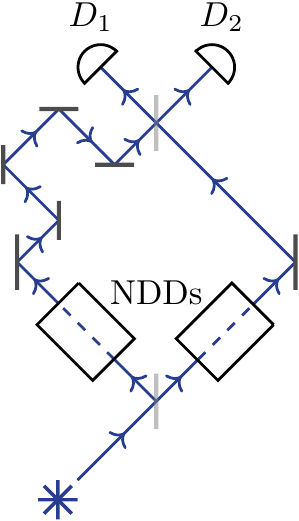}
\caption{(\hyperref[exp6]{S6})}\label{Fig:NDD2}
\end{subfigure}
\captionsetup{width=1.0\columnwidth}
\caption{\textbf{Single-particle experiments with non-destructive detectors (NDDs).} (\hyperref[exp5]{S5}) non-destructive detector (NDD) is placed in each path. Depending on which NDD clicks, one knows which path the photon took. (\hyperref[exp6]{S6}) The length of one path is varied while the NDDs are still present in each path.
}\label{Fig:NDD}
\end{figure}
% END \input{Figure_NDD.tex}

\subsubsection{Experiment~S5}
\label{exp5}

Experiments~\hyperref[exp3]{S3} and~\hyperref[exp4]{S4} may leave us wondering: how did all the photons sense a change in one of the paths if only half the photons actually seem to take that path? 
Did the photons take just one path or both, and if only one path, then which path did they take?

To answer this question we consider yet another modified version of Experiment~\hyperref[exp3]{S3}, in which we try to observe the path that the photon takes while keeping the setup otherwise unchanged. 
Let us assume that we have a device that can measure the presence of a photon in a given path without absorbing the photon or changing its intrinsic properties (for example, its speed or color). 
We call this device a ``non-destructive detector'' (NDD). What happens if we redo the experiment with such devices in the paths as depicted in Fig.~\hyperref[Fig:NDD]{2:S5}?

First of all, and as expected, for every photon sent into the setup, exactly one NDD will click each time.
That is, the photon is detected in either path with equal probability so half the photons are observed in the left path and half in the right one.

The mystery of single-particle behavior thickens when we focus on the probabilities of detectors $D_{1}$ and $D_{2}$ clicking.
In contrast to Experiment~\hyperref[exp3]{S3}, we no longer observe all the photons in detector $D_2$.
Instead, now detectors $D_1$ and $D_2$ click with equal probability! 
This experiment actually displays the seemingly obvious outcome that one might have naively expected for Experiment~\hyperref[exp3]{S3} based on Experiments~\hyperref[exp1]{S1} and~\hyperref[exp2]{S2}.
Now, when their paths are observed, the photons seem to behave like particles making random choices at the beam\hyp splitters again. 
Merely observing the photons' taken paths somehow prevents the photons from acting like they did in Experiment~\hyperref[exp3]{S3}.

One might wonder what would happen if, instead of placing NDDs in both paths, these are placed only in one of the two paths, say the left path.
In such a situation, if this NDD clicks, then we know for sure that the photon took the left path between the two beamsplitters.
Otherwise, we infer that the photon took the right path.
This means that a single NDD suffices to determine which path the photon took, and as a result the detection pattern will be identical to that seen if NDDs are placed in both paths. 
Going forward, we describe experiments with NDDs in each of the paths, but with the implicit understanding that sometimes we could get the same effect with fewer NDDs.

\subsubsection{Experiment~S6}
\label{exp6}

Another pertinent question for the NDD experiment is `what will happen if we change the length of one of the paths?'
That is, what would be the detection pattern for a modified version of Experiment~\hyperref[exp4]{S4} with an NDD in each path as depicted in Fig.~\hyperref[Fig:NDD2]{2:S6}.
Recall that without the NDD there to observe the photons' paths, the detection probabilities were dramatically effected by changing path length. 
In contrast, when this experiment is performed with the NDD included, then the changing path length turns out to have no effect whatsoever! 
The photons continue to arrive at detectors $D_1$ and $D_2$ with equal probability. 
This is again what one would have expected based on the straightforward particle model; the presence of the NDD changes the detection outcomes back to the particle-like behavior.

\subsubsection{Summary: Experiments with Photons and Wave-Particle Duality}

Looking back at the first four experiments, there is a clear contrast between Experiments~\hyperref[exp1]{S1} and \hyperref[exp2]{S2} on the one hand, and Experiments~\hyperref[exp3]{S3} and \hyperref[exp4]{S4} on the other. 
In the first two experiments, light seems to show a distinct particle-like behavior, with each photon making a random decision at each beam\hyp splitter on its way towards the detectors.
Note that in these two experiments, whenever a detector clicks, it is possible to unambiguously assign a specific path and imagine that the photon took this path to arrive at this detector.
In contrast, in Experiments~\hyperref[exp3]{S3} and~\hyperref[exp4]{S4}, each detector click admits two possible paths that the photon could have taken to arrive at that detector. In this case, each photon acts as if it traversed both paths at once, or at least somehow ``knew'' about both paths. 
it turns out that these two experiments admit a complete explanation in terms of classical waves, and we go through this explanation in detail in the following section. 
If we try to measure which of the two possible paths each photon took, like we did in Experiments~\hyperref[exp5]{S5} and~\hyperref[exp5]{S6}, then the photons behave like particles once again. That is, we can again predict the outcome by assuming that each photon makes a random choice each time it encounters a beam\hyp splitter.

These six experiments together capture the first key surprise of quantum physics: wave\hyp particle duality. Indeed, not just particles of light, but any quantum particle can display this seemingly-contradictory behavior, which has been tested and probed through decades of experiments (detailed in Sec.~\ref{Sec:Experiments}).
In Section~\ref{subsec:QuantumTheory}, we will take a look at how we can model this behavior and we work our way towards describing how quantum physics can accurately predict these surprising outcomes.
Before that, we describe how one can analyze Experiments~\hyperref[exp3]{S3} and~\hyperref[exp4]{S4} using the classical theory of waves.

\subsection{Classical Wave Explanation of Experiments~\hyperref[exp3]{S3} and~\hyperref[exp4]{S4}}\label{sec:Single_particle_Theory}

Here we highlight how the classical theory of waves can explain Experiments~\hyperref[exp3]{S3} and~\hyperref[exp4]{S4}.
We go through the explanation in detail as some of the features of wave theory, such as the superposition principle will be directly useful in the quantum physics of the following section.

\subsubsection{Classical Wave Theory}
\label{subsubsec:classicalwavetheory}

Recall that Experiments~\hyperref[exp1]{S1} and~\hyperref[exp2]{S2} can be understood by modeling the photon as a particle that makes random choices at each beam\hyp splitter. 
However, this model predicts incorrect outcomes for Experiments~\hyperref[exp3]{S3} and~\hyperref[exp4]{S4}. 
We now show that, on the other hand, the outcomes of Experiments \hyperref[exp3]{S3} and \hyperref[exp4]{S4} can be explained if light is described as a wave of electric and magnetic fields.
This description of light was suggested as early as 1865 by James Clerk Maxwell~\cite{Maxwell1865} and has been very successful in explaining a wide variety of phenomena related to light.
To distinguish it from the quantum description of light, we refer to this theory as the classical wave theory.

Classical wave theory can explain these experiments by introducing a phenomenon known as interference of waves. Before discussing interference, let us first understand what a wave is. 
Most waves that we are familiar with exhibit repetitive patterns, such as water ripples in an otherwise still pond caused by a stone tossed gently from the shore. The oscillations in the water surface have peaks and troughs. These peaks and troughs propagate along the surface of the water away from the point where the stone entered.

When we throw a large stone into the pond, we will notice that the water surface is disturbed more than when we throw a small stone. In general, a larger stone will produce taller peaks and deeper troughs than a smaller stone. 
From this observation we can define the first important property of a wave: the wave amplitude. 
Quite simply, the amplitude describes how much a wave disturbs the medium in which it propagates and it is quantified by a positive real number. 
The amplitude of a water wave is just the height by which the water surface changes when a wave travels on the surface.
For classical electromagnetic waves, the amplitude of a wave propagating in a medium is the maximum electric field magnitude that is caused by the wave in that medium.
The amplitude of an electromagnetic wave is an important property because the energy of a wave is proportional to the square of its amplitude.

If we now focus on the water surface at one position, we observe that the surface rises when a wave peak arrives. The surface then drops back to its original height, and then drops even more as a trough arrives. Finally, it returns to its original height before the process starts all over again. The complete process of starting from the original height, rising, falling and finally returning to the original height is called a cycle of the wave. The time that it takes for a wave to complete one cycle is called the period. However, if we instead take a picture of the wave at a fixed time, the cycle can also be represented by the distance between subsequent peaks, or equivalently the distance between subsequent troughs. This distance is called the wavelength. The period and wavelength of a wave are related to each other by the speed of the wave. For example, if two waves have the same speed but one has half the wavelength of the other, the shorter wave will have a period half that of the longer wave. The period, wavelength and speed of a wave are three important properties of a wave. For the experiments detailed in this manuscript, we only deal with waves that all have the same period, wavelength and speed\footnote{The speed of light in a vacuum is constant and equal to about \num{299792458} meters per second. Light cannot travel faster than this value, but it can travel slower if it travels through matter. For example, the speed of light in glass is roughly two-thirds of its speed in a vacuum.}.

We can now define another important property: the phase of a wave, which describes how much of a cycle the wave has completed at a given position. Each cycle of the wave is divided into 360 degrees. If a wave has just begun its cycle at a given position then the phase of the wave at that position is zero degrees. After completing half a cycle its phase is 180 degrees, and the phase reaches 360 degrees once the cycle is complete.

We can also define a relative phase between two separate waves that have the same period and speed (and hence, the same wavelength). The two waves are said to be in phase if their peaks and troughs arrive at the same time at a given position. The waves are said to be out of phase if their effect at a given position differs by half of a complete cycle (or 180 degrees). If two waves are out of phase, whenever a peak of the first wave arrives, it is accompanied by a trough of the second wave.

\begin{figure}
\newcommand*{\RedOpacity}{0.8}
\begin{subfigure}[t]{0.5\textwidth}
% \begin{tikzpicture}[domain=0:3.14*3/2,scale=1]
% 	\hspace{1.3mm}
%     \draw[->, thick] (0,0) -- (5,0) node[right] {$t$};
%     \draw[->, thick] (0,-1.5) -- (0,1.5) node[rotate=90, right] {$s$};
%     \draw[dashed,thick] (0,1) node[left] {$1$} -- (3*3.14/2,1);
%     \draw[ultra thick, xblue, line cap=round,opacity=\RedOpacity] plot[id=sin, samples=100] function{cos(2*x)}
%         node[black] at (4*3.14/4,1.5) {$a=1$};
%     \draw[](4*3.14/4,1.3)--(4*3.14/4,1);
%     \draw[ultra thick, loosely dotted, xaqua, line cap=round,opacity=\RedOpacity] plot[id=sin, samples=100] function{cos(2*x-3.14/2)}
%         node[black] at (3.14/4+0.1,1.9) {$a=-i$};
%     \draw[](3.14/4,1.7)--(3.14/4,1);
%     \draw[ultra thick, dashed, xgreen, line cap=round,opacity=\RedOpacity] plot[id=sin, samples=100] function{cos(2*x-3.14)}
%         node[black] at (2*3.14/4+0.1,1.5) {$a=-1$};
%     \draw[](2*3.14/4,1.3)--(2*3.14/4,1);
%     \draw[ultra thick, densely dashed, xbrown, line cap=round,opacity=\RedOpacity] plot[id=sin, samples=100] function{cos(2*x-3*3.14/2)}
%         node[black] at (3*3.14/4,1.9) {$a=i$};
%     \draw[](3*3.14/4,1.7)--(3*3.14/4,1);
%     \draw[](2*3.14/4,-1.3)--(2*3.14/4,-1);
%     \draw[](6*3.14/4,-1.3)--(6*3.14/4,-1);
%     \draw[](2*3.14/4,-1.2)--(6*3.14/4,-1.2);
%     \draw node at (4*3.14/4,-1.4) {$\text{Cycle}$};
% \end{tikzpicture}
\includegraphics[]{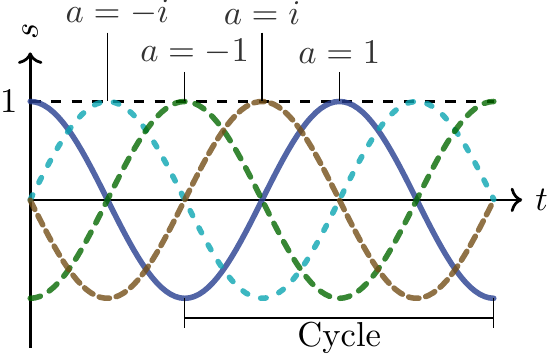}
\caption{}
\label{Fig:Phase}
\end{subfigure}

\begin{subfigure}[t]{0.5\textwidth}
% \begin{tikzpicture}[domain=0:3.14*3/2,scale=1]
% 	\hspace{-0.5mm}
%     \draw[->, thick] (0,0) -- (5,0) node[right] {$t$};
%     \draw[->, thick] (0,-1.5) -- (0,1.5) node[rotate=90, right] {$s$};
%     \draw[dashed,thick] (0,1/1.41) node[left] {$\dfrac{1}{\sqrt{2}}$} -- (3*3.14/4,1/1.41);
%     \draw[ultra thick, xblue, line cap=round,opacity=\RedOpacity] plot[id=sin, samples=100] function{cos(2*x)} 
%         node[right] {$ $};
%     \draw[ultra thick, densely dashed, xbrown, line cap=round,opacity=\RedOpacity] plot[id=sin, samples=100] function{cos(2*x+3.14/2)/1.41} 
%         node[right] {$ $};
%     \draw[] (4*3.14/4,1.3) node[above] {$a=1$} -- (4*3.14/4,1);
%     \draw[] (3-0.2,-1+0.2) node at (3,-1) {$a=\dfrac{i}{\sqrt{2}}$} --(3.3,-0.2);
% \end{tikzpicture}
\includegraphics[]{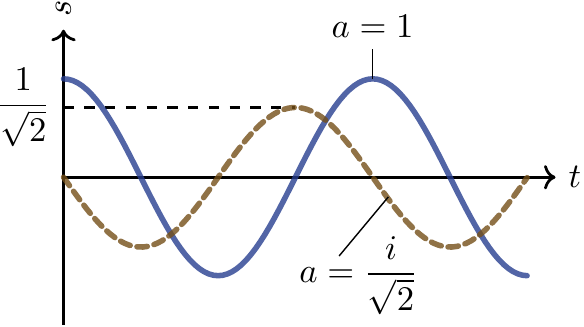}
\caption{}
\label{Fig:Amplitude}
\end{subfigure}
\caption{\textbf{A visual description of amplitude, cycle, and phase.} (a) The displacement $s$ plotted against time $t$ for four waves with the same amplitude ($|a|=1$) and period but different relative phase. The complex amplitude $a$ describes both the amplitude and phase of each wave. With respect to $a=1$, the waves labeled by $a=i$, $a=-1$, and $a=-i$ are shifted by one-quarter, one-half, and three-quarters of the cycle, respectively. In this case, since the horizontal axis is time, the labeled cycle is the wave period. (b) An example of a wave $(a=i/\sqrt{2})$ with both a different amplitude and phase relative to $a=1$. The wave $a=i/\sqrt{2}$ is reduced in height and delayed by one-quarter of a cycle compared to $a=1$. This is what happens for a reflected wave at a beam-splitter.}
\label{Fig:PhaseAmplitude}
\end{figure}

The amplitude and the relative phase can be combined into a single quantity.
This quantity is a complex number, denoted $a$, and is known as the complex amplitude. 
To explain the magnitude and phase of a wave in terms of the complex amplitude, we first define the imaginary unit as $i=\sqrt{-1}$ so that $i^2=-1$. 
The amplitude of the wave is the absolute value\footnote{ 
Recall that the absolute value of a complex number $a = x + yi$ is $|a|={\sqrt {x^{2}+y^{2}}}$.
Thus, the magnitude of $i=0 + 1i$ is $|i|=1$. 
Similarly, we have $|-1|=1$, from which it follows that $|i|=|-i|=|-1|=|1|=1$.} $|a|$ of its complex amplitude $a$. The phase of a wave is related to the argument of its complex amplitude, which is defined as the angle between the complex number and the real axis in the complex plane. For instance, consider a wave with an amplitude of $1$, then the absolute value of the complex amplitude is $|a| = 1$. We consider four illustrative complex amplitudes: $a=1$, $a=i$, $a=-1$, and $a=-i$.
In terms of wave cycles, $a=i$ implies that the wave is shifted by one-quarter of a cycle from $a=1$. That is, it has a relative phase of 90 degrees. Similarly, $a=-1$ and $a=-i$ are shifted by one-half (180 degrees) and three-quarters (270 degrees) of a cycle respectively. These four cases are illustrated in Fig~\ref{Fig:Phase}. See Fig~\ref{Fig:Amplitude} for an example of a wave with complex amplitude $a=i/\sqrt{2}$ and its comparison to a reference $a=1$ wave\footnote{For an electromagnetic wave, the energy is related to the squared amplitude. Therefore a wave with a complex amplitude of $a=i/\sqrt{2}$ ($|a|^2=1/2$) will have half the energy of a wave with $a=1$.}. For simplicity, hereafter we will refer to the complex amplitude as just the amplitude unless clarification is required.

% BEGIN \input{Figure_Phase_and_Amplitude.tex}
% END \input{Figure_Phase_and_Amplitude.tex}

\subsubsection{Interference of Waves and the Superposition Principle}
\label{subsubsec:interferenceofwaves}

Now that we have introduced the important properties that define a wave, we are in a position to describe their interference, which is a key concept in understanding not just classical wave theory but also quantum physics.
Let us first return to the analogy of the stone tossed into a still pond. 
This time, instead of a single stone, let us imagine tossing two stones into the pond simultaneously at separate but nearby locations. 
As the waves on the water surface propagate outwards from the two stones, they will meet each other to form an intricate pattern.
At some positions, the two waves are `in phase', such that peaks and troughs of the two waves will coincide, resulting in a wave with an even taller peak and deeper trough---the waves exhibit constructive interference. 
At other points, the waves are completely `out of phase' such that the peaks of one wave cancel the troughs of the other---the waves exhibit destructive interference and the water remains still.
This pattern of constructive and destructive interference can be explained using the superposition principle:

\vspace{6pt}
\textit{The superposition principle}.--- When two or more waves meet, they are superimposed and their amplitudes add together to create a single new wave.
\vspace{6pt}

\begin{figure}
\newcommand*{\RedOpacity}{0.8}
\begin{subfigure}[t]{0.5\textwidth}
% \begin{tikzpicture}[domain=0:3.14*3/2,scale=0.7]
%     \draw[->, thick] (0,-0.7) -- (5,-0.7) node[right] {$ $};
%     \draw[->, thick] (0,0.7) -- (5,0.7) node[right] {$ $};
%     \draw[->, thick] (7*3.14/4,0) node[left] {$=$} -- (10.5,0) node[right] {$t$};
%     \draw[->, thick] (0,-2) -- (0,2) node[above] {$ $};
%     \draw[ultra thick, xblue, line cap=round,opacity=\RedOpacity] plot[id=sin, samples=100] function{cos(6*x)/2+0.7}
%         node[black] at (4*3.14/4,1.5) {$ $};
%     \draw[ultra thick, xblue, line cap=round,opacity=\RedOpacity] plot[id=sin, samples=100] function{cos(6*x)/2-0.7}
%         node[black] at (4*3.14/4,1.5) {$ $};
%     \draw[ultra thick, xblue, line cap=round,opacity=\RedOpacity] plot[id=sin, samples=100,domain=7*3.14/4:7*3.14/4+3.14+3*3.14/6] function{sin(6*x)}
%         node[black] at (4*3.14/4,1.5) {$ $};
%     \draw node at (5,2.4) {$\text{Constructive Interference}$};
%     \draw[] node at (1.2,1.7) {$a_1=\frac{1}{2}$};
% 	\draw[] node at (1.2,-1.7) {$a_2=\frac{1}{2}$}
% node at (8,1.7) {$a_1+a_2=\frac{1}{2}+\frac{1}{2}=1$};
% \end{tikzpicture}
\includegraphics[]{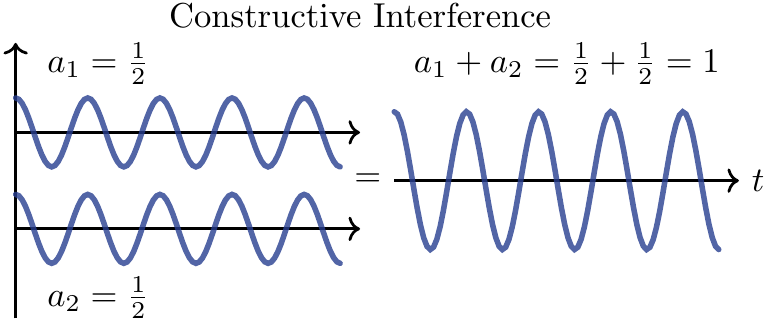}
\caption{}
\label{Fig:ConstructiveInterference}
\end{subfigure}

\vspace{4mm}
\begin{subfigure}[t]{0.5\textwidth}
% \begin{tikzpicture}[domain=0:3.14*3/2,scale=0.7]
%     \draw[->, thick] (0,-0.7) -- (5,-0.7) node[right] {$ $};
%     \draw[->, thick] (0,0.7) -- (5,0.7) node[right] {$ $};
%     \draw[->, thick] (7*3.14/4,0) node[left] {$=$} -- (10.5,0) node[right] {$t$};
%     \draw[->, thick] (0,-2) -- (0,2) node[above] {$ $};
%     \draw[ultra thick, xblue, line cap=round,opacity=\RedOpacity] plot[id=sin, samples=100] function{cos(6*x)/2+0.7}
%         node[black] at (4*3.14/4,1.5) {$ $};
%     \draw[ultra thick, xblue, line cap=round,opacity=\RedOpacity] plot[id=sin, samples=100] function{-cos(6*x)/2-0.7}
%         node[black] at (4*3.14/4,1.5) {$ $};
%     \draw[ultra thick, xblue, line cap=round,opacity=\RedOpacity] plot[id=sin, samples=100,domain=7*3.14/4:7*3.14/4+3.14+3*3.14/6] function{0}
%         node[black] at (4*3.14/4,1.5) {$ $};
%     \draw node at (5,2.4) {$\text{Destructive Interference}$} node at (1.2,1.7) {$a_1=\frac{1}{2}$}
% node at (1.35,-1.7) {$a_2=-\frac{1}{2}$}
% node at (8,0.6) {$a_1+a_2=\frac{1}{2}-\frac{1}{2}=0$};
% \end{tikzpicture}
\includegraphics[]{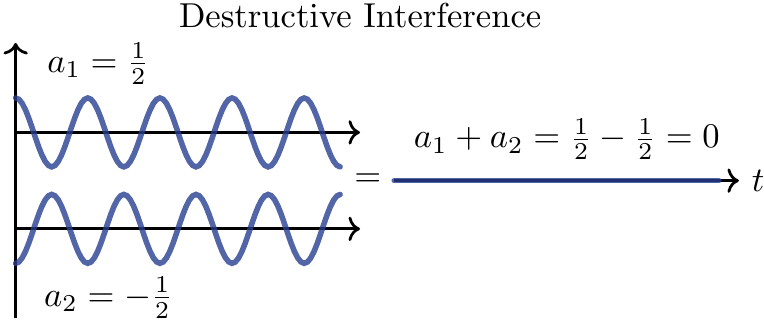}
\caption{}
\label{Fig:DestructiveInterference}
\end{subfigure}
\caption{\textbf{Illustrations of constructive and destructive interference of waves.} (a) Two waves of amplitude $a_1=a_2=1/2$ meet and constructively interfere to form a new wave $a_1+a_2=1$. (b) Two waves of amplitude $a_1=-a_2=1/2$ meet and destructively interfere to form a new wave $a_1+a_2=0$.}
\label{Fig:Interference}
\end{figure}

In other words, when two waves meet at a given position, the resulting motion of the surface is described by a third wave whose amplitude equals the sum of the amplitude of the original two waves.
This allows us to understand constructive and destructive interference as follows.
Let us consider a situation when a wave with an amplitude $a_1=1/2$ at a given position meets a wave described by the same amplitude $a_2=1/2$ at the same position, as depicted in Fig~\ref{Fig:ConstructiveInterference}.
This results in a new wave with an amplitude of $a_1+a_2 = 1$. 
Thus, the peaks are higher and the troughs are deeper.
We say that this new wave results from constructive interference of the original two waves.

The opposite situation occurs when a wave with $a_1=1/2$ meets with a wave with $a_2=-1/2$ at the same position. 
This results in $a_1+a_2 = 0$, so the resulting wave has zero amplitude!
More generally, when two waves of different amplitude meet such that the peaks of one wave coincide with the troughs of the other, the resultant wave has smaller amplitude (than the wave that originally had larger amplitude), and this is called destructive interference.
In the case that the amplitudes of two incoming waves are equal and if their peaks align perfectly with the troughs, then destructive interference could result in the two waves completely canceling each other out, as is depicted in Fig~\ref{Fig:DestructiveInterference}.

% BEGIN \input{Figure_Interference.tex}
% END \input{Figure_Interference.tex}

We are now equipped with the tools required to appreciate the wave-like behavior of light in Experiments \hyperref[exp3]{S3} and \hyperref[exp4]{S4}.
We start by constructing a model to describe the effect of a beam\hyp splitter on light, assuming that light is a electromagnetic wave.
Recall that the energy of an electromagnetic wave is equal to the square of its amplitude, $|a|^2$.
Let us denote the amplitude of the incoming light wave by $a$ and the amplitudes of the light that is outgoing along the two output paths by $a_\text{T}$ and $a_\text{R}$, where the subscripts refer either to transmitted ($\text{T}$) or reflected ($\text{R}$) light. 
Since the energy of the incoming light should be equal to the total energy of the outgoing light\footnote{This is in line with an important principle of physics, namely the conservation of energy. Strictly speaking, the incoming and outgoing energies being equal assumes that no light is lost on the way, for example through absorption at the beam\hyp splitter.}, we find that the square $|a|^2$ of the amplitude of the incoming light should be equal to the sum $|a_\text{T}|^2+|a_\text{R}|^2$ of squares of the outgoing amplitudes. 
Furthermore, because the beam\hyp splitter divides the energy of the incoming beam equally among the two outgoing paths, it follows that the energies of the two outgoing beams should be equal to each other and to half of the incoming energy.
Hence, the amplitudes of the outgoing beams are equal to each other and smaller than the incoming amplitude by a factor of $1/\sqrt{2}$.
Mathematically, we have $|a_\text{T}| = |a_\text{R}| = |a|/\sqrt{2}$.
This solution is valid because the incoming energy $|a|^2$ is the same as the total outgoing energy $|a/\sqrt{2}|^2 + |a/\sqrt{2}|^2 = |a|^2$.

In addition, if we measure the phase of the outgoing light, we can find that the wave reflected by a beam\hyp splitter is shifted by one-quarter of a cycle with respect to the transmitted wave. 
We can choose the transmitted wave $a_\text{T}$ as the reference wave: $a_\text{T}=a\times 1/\sqrt{2}$.
Then the reflected wave has a phase of $i$, so that $a_\text{R}=a\times i/\sqrt{2}$. 
From this, we can see that whenever a beam of light is incident on a beam\hyp splitter, the amplitude of the transmitted wave is determined by multiplying the original amplitude by $1/\sqrt{2}$. 
Similarly, the amplitude of the reflected wave is determined by multiplying the original amplitude by $i/\sqrt{2}$. This reflected wave is illustrated in Fig~\ref{Fig:Amplitude} as compared to the incoming wave with $a=1$.

Using this model for beam\hyp splitters and mirrors, the results from Experiments~\hyperref[exp3]{S3} and \hyperref[exp4]{S4} can be explained in terms of classical light waves. 
We will provide a detailed quantum explanation of these and the other experiments in the following section but for now we focus on the wave-like behavior exhibited by these experiments. 
In particular, the incoming light in Experiment~\hyperref[exp3]{S3} can be associated with a wave.
We can track its wave amplitude in order to predict the detection outcomes. 
Upon arrival at the first beamsplitter, the incident beam of light is split into two waves that travel through opposite arms of the Mach--Zehnder interferometer. 
The wave that is reflected at the beam\hyp splitter is shifted in phase by one-quarter of a cycle compared to the transmitted wave. When the two beams arrive at the second beam\hyp splitter, both the waves split yet again into a total of four waves, two of which take the path to $D_1$ and the other two towards $D_2$.
The constructive and destructive interference of the overlapping waves will lead to the observed detection rates.

Specifically, let us consider the two waves that contribute to light arriving at $D_1$. The paths that these two waves take are illustrated in Fig~\ref{Fig:SingleParticlePaths}. The first of these two waves is the one that was reflected at the first beam\hyp splitter and was once again reflected at the second beam\hyp splitter.
It acquires two phase shifts of one-quarter of a cycle each, so it is now shifted by one-half of a cycle compared to the reference.
The other contribution is from the wave that was transmitted through both beam\hyp splitters, and so it acquires no relative phase shift.
This means that the two waves arriving at $D_1$ have a relative phase of half a cycle (180 degrees), i.e., they are completely out of phase.
Hence, when the two waves recombine at detector $D_1$, they must superimpose destructively. Likewise, both the waves that recombine at detector $D_2$ have been once reflected and once transmitted, so they are in phase. 
Hence, they superimpose constructively.
This explains the outcome of Experiment~\hyperref[exp3]{S3}, that all the light is detected in one detector $D_{2}$ and no light is detected in $D_{1}$. 

The difference between reflection and transmission is not the only way to accumulate a relative phase. If one wave is physically delayed with respect to another, it will also acquire a relative phase.
In Experiment~\hyperref[exp4]{S4}, we observed that adding a path extension of a certain length $L$ in one path switched the outcomes for $D_1$ and $D_2$ so that all the light arrived at $D_1$. The interference pattern changes because one wave must take slightly longer to travel through the extension and so its arrival at the second beam\hyp splitter is delayed with respect to the wave in the other path.

When a path difference equal to half the wavelength of the incident light is introduced, the wave arriving at the second beam\hyp splitter is delayed by half of a period\m it acquired a relative phase of $180$ degrees by passing through the path extension.
For light arriving at detector $D_1$, this additional phase adds to the 180-degree phase acquired by the wave that experienced two reflections. Hence the two waves arriving at $D_1$ have a total relative phase of 360 degrees (or 0 degrees), and so they are once again in phase and interfere constructively.
On the other hand, the two waves arriving at $D_2$ also gain a relative phase of 180 degrees from the extension. Since they were in phase prior to adding the extension, they are now out of phase and will interfere destructively.
Thus, the interference pattern is reversed.
In fact, by continuously varying the path length, the model predicts that the interference effect alternates smoothly between constructive and destructive interference.
In other words, the light alternates smoothly between being detected entirely at $D_1$, to being detected partially at $D_{1}$ and partially at $D_{2}$, to being detected entirely at $D_2$ and then back again. 
This dependence of light intensity on path length is a signature of interference and the wave-like behavior of light.

So far, the explanation of wave interference in these experiments does not require any mysterious quantum concepts.
This interference can be explained purely as a result of classical physics.
Although the classical theory of light can explain the detection outcomes of Experiments~\hyperref[exp3]{S3} and~\hyperref[exp4]{S4}, it cannot explain the first two experiments. 

Recall that in Experiment~\hyperref[exp1]{S1}, we observed that detectors $D_1$ and $D_2$ will never click at the same time.
Indeed, it was observations such as this one that led physicists to realize that there exist quantities of light that cannot be divided further into smaller parts, and can therefore be considered as particles.
This particle-like behavior of light, as described first by Einstein the context of the photoelectric effect, was in-fact among the observations that led to the birth of quantum theory.

In contrast, if we model the incoming single photons as classical waves, then both the detectors will receive light of equal, albeit smaller, amplitude.
The smaller amplitude would explain that half the photons are observed in one detector and half in the other but it would also predict that both the detectors should click simultaneously in at least some cases.
However, this is not what is observed in actual experiments, in which photons are never split at the beam\hyp splitter but instead take either one path or the other. Furthermore, this wave model also cannot explain why adding NDDs in the arms of a Mach--Zehnder interferometer destroys the interference pattern, like we observed in Experiments \hyperref[exp5]{S5} and \hyperref[exp6]{S6}.

From analyzing the experiments using classical wave theory, we can now identify exactly why the light in Experiments \hyperref[exp3]{S3} and \hyperref[exp4]{S4} are said to exhibit wave-like behavior. However, this theory cannot explain the wave--particle duality of quantum particles. For that, we must turn to quantum physics.

\subsection{Quantum Physics Explanation of Experiments}
\label{subsec:QuantumTheory}

The duality between wave-like and particle-like properties of quantum particles that we observed in the previous section is quite different from the characteristic behavior of objects that we encounter on a day-to-day basis.
Thus, explaining this behavior requires a different way of looking at these experiments.
The framework that successfully describes this surprising behavior is given by quantum physics.
In this section, we develop a set of basic rules to predict the outcomes of experiments on quantum particles.
We conclude the section by applying these rules to explain the outcomes of the single-particle experiments.

\subsubsection{Histories and Their Associated Wave-Functions}

A good starting point is to understand exactly where classical wave theory breaks down in explaining phenomena involving photons or other tiny particles. 
Recall that a single photon incident on a beam\hyp splitter does not split into two photons that are then detected simultaneously at the two detectors.
Instead, the single photon remains a full single photon and travels either to one detector or the other, but not both.
However, classical wave theory makes the assumption that the energy of light is proportional to the squared absolute value of the wave amplitude.
This assumption would mean that the beam\hyp splitter, which splits light waves into waves of smaller amplitude, should split a single photon into two photons of smaller energy.
This splitting of photons is not observed in the experiment.
Hence, this assumption cannot be true at the quantum level. 

Quantum physics introduces a radical new way to predict the outcomes of experiments.
The quantum approach merges the idea of random choices made by particles with the interference behavior predicted by classical wave theory. This allows quantum physics to make accurate predictions for experiments with quantum particles\m capturing the effects of wave-particle duality.

In quantum physics, a wave is associated with the \emph{history} of each particle that results in a given outcome.
For single-particle experiments such as Experiments~\hyperref[exp1]{S1} through~\hyperref[exp6]{S6}, these single-particle histories are just the possible paths taken by the particle through the apparatus.
In other words, each path that the photon could in principle take to arrive at a given detector is a single-particle history that contributes to a detection at that detector.
For experiments involving more particles, a history refers to the set of paths taken by all the particles participating in the experiment and we will discuss multi-particle histories in more detail in the next section (Sec.~\ref{Sec:2Particle}).
Note that `history', here, is not intended to invoke the idea of time but rather the idea of possibilities; we could replace the word history with the word `possibility' and the quantum physics description would still be sensible.

What do these waves and their associated amplitudes represent? 
Recall that the classical wave theory explanation presented in Sec.~\ref{subsubsec:classicalwavetheory} envisaged a wave as a physical structure, something that could be directly observed, like a ripple on the surface of a pond.
In quantum physics on the other hand, the wave associated with each history represents something less tangible.
Quantum physics connects the amplitude of the wave with \textit{probability}.
More precisely, the squared absolute value $|a|^2$ of the amplitude of a wave does not represent energy as it did in classical wave theory; rather it represents the probability of observing a particular outcome for a particle or a set of particles. 
In quantum physics, the complex amplitude is called the wave\hyp function of the quantum particle.

We emphasize that the wave\hyp function is not something that can be directly observed but rather can be seen as a mathematical tool that is used to make quantum predictions.
Consider the example of Experiment~\hyperref[exp1]{S1} as explained in the language of wave\hyp functions.
The wave that is associated with the incoming particle is split into two waves each heading towards one of the two detectors.
The amplitudes of the outgoing waves are equal to each other and smaller than the amplitude of the incoming wave by a factor of $\sqrt{2}$.
This means that the respective probabilities of detecting single photons at the two detectors are equal to each other and half as large as the probability of sending a single photon towards the beam\hyp splitter.
The following section presents a complete discussion of single-particle quantum physics in terms of single-particle histories and their associated wave functions. 

\subsubsection{Indistinguishable Histories and the Rules of Single-Particle Quantum Physics}

Other than its physical meaning, most of the properties of waves discussed in Sec.~\ref{subsubsec:classicalwavetheory} are directly applicable to the wave\hyp function. 
Of particular interest is the possibility of interference.
In quantum physics, the wave\hyp functions of different histories taken by a particle can also be added together, and this can result in constructive or destructive interference. 
However, unlike classical waves, wave\hyp functions that overlap do not always interfere.
Instead, the superposition principle can only be applied if the wave\hyp functions are those of fundamentally indistinguishable histories. 
By fundamentally indistinguishable, we mean that there is no way, even in principle, to determine which history was taken by the quantum particles to reach the detectors. 
In other words, photon histories for a particular detection outcome interfere with each other only if it is not possible to determine which history the photons actually took on their way to the detectors.

The interference of indistinguishable histories leads us to an important principle in quantum physics.

\vspace{10pt}
\textit{The indistinguishability principle}.---interference occurs if and only if two or more fundamentally indistinguishable histories contribute toward the same detection outcome.
\vspace{10pt}

As an example, consider Experiment~\hyperref[exp3]{S3}. 
This experiment is done in such a way that the single-particle histories are fundamentally indistinguishable, i.e., no physical system in the setup can gather any information about which path the photon took on its way to the detector.
Thus the indistinguishability principle mandates that interference occurs between the two single-particle histories that lead to the same detector.
In the specific case of a detection at $D_1$, the history that comprises two reflections is fundamentally indistinguishable from the path that involves two transmissions, and both of these histories contribute towards a detection in $D_1$.
Hence, the waves associated with these two histories interfere and, as detailed in the next section (Sec.~\ref{subsec:explanation-single}), give us the detection probabilities that we observe. 

What about the detection probabilities observed in Experiment~\hyperref[exp5]{S5}?
Recall that placing an NDD in one of the paths destroys the interference. This is because the NDD provides information on which path the photon took.
As a consequence, the histories are now distinguishable.
In other words, the histories are no longer fundamentally indistinguishable. 
If there is any way whatsoever to infer which path the photon took, then the indistinguishability principle entails that the amplitudes from different wave\hyp functions cannot be summed, even if those wave\hyp functions represent the same detection outcome.
For these distinguishable histories, we must sum the probabilities for each wave\hyp function, rather than the wave\hyp functions themselves, to obtain the final detection probability. We illustrate this concept with an example as follows.

In Experiments~\hyperref[exp5]{S5} and~\hyperref[exp6]{S6}, the two histories for each detection are distinguishable.
Hence, the two histories act independently, with each history contributing one quarter probability to the photon detection in the given detector (each history passes through two beam\hyp splitters).
Summing the probabilities of $1/4$ from each of the two histories contributing to each detection outcome, we obtain the correct detection probability of $1/2$ at $D_1$ and $1/2$ at $D_2$.

Other than the idea of interference between indistinguishable histories, there is one more important difference to consider when moving from classical wave theory to the quantum physics of wave\hyp functions. 
In classical theory, light can split and travel along two paths but a single photon described by quantum physics cannot. 
When a single photon arrives at a beam\hyp splitter, instead of choosing a single path, quantum physics describes it as proceeding through the apparatus in a superposition of both paths or both single-particle histories. 
This superposition of histories can continue until a path measurement is made, at which point one of the two possible outcomes will result, and we say that the superposition has collapsed.
This collapse precludes the detection of the particle in the other detector.
It is this collapse that leads to the particle-like behavior of some quantum experiments. 

Consider again the example of Experiments~\hyperref[exp5]{S5} and~\hyperref[exp6]{S6}. As a photon leaves the first beam\hyp splitter it enters into a superposition of two paths. 
However, as the photon passes through the NDD, a measurement is made on the photon's path and the indistinguishability between the histories is lost.
After the NDD, the photon is no longer in a superposition of the two histories so the photon must have definitely taken one of the two paths, which eventually leads to the two paths do not interfere but rather contribute to detection probabilities independently.
Thus, concepts of indistinguishability and superposition are closely related and fundamental for explaining many quantum phenomena.

Let us consolidate what we have learned so far about using quantum physics to predict outcomes of experiments involving light.
We gather the operating principles described above into the following basic rules for single-particle quantum experiments. 

\vspace{6pt}
\textit{Rules of single-particle quantum physics}.---
\begin{enumerate}
\item To predict the probability of detecting a particle at a specific detector, consider all the histories of the particle that result in the particle arriving at the detector.
\item 
Each history of a single quantum particle is associated with a wave and its corresponding complex amplitude $a$.
\item 
The amplitude of a history is multiplied by a $1/\sqrt{2}$ factor each time the particle is transmitted through a beam\hyp splitter and by a factor of $i/\sqrt{2}$ each time it is reflected at a beam\hyp splitter. Mirrors leave the amplitude unchanged.
\item A history's amplitude can incur a phase shift relative to another history if it involves a particle taking a longer path compared to the same particle in another history.
\item If two or more fundamentally indistinguishable histories lead to the same detection outcome, then the amplitudes of those histories must be summed to obtain the final detection amplitude.
\item The probability (Pr) of detecting the particle in the chosen detector is proportional to the squared absolute value of the detection amplitude: $\text{Pr}= |a|^2$.
If more than one distinguishable history leads to the same detection, then their probabilities are summed directly.
\item Observing a particle in one path precludes observing it in any other path.
\end{enumerate}
\vspace{6pt}

Equipped with these basic rules, we can now provide an explanation of the single-particle experiments described above.

\subsubsection{Explanation of Experiments~\hyperref[exp1]{S1} through \hyperref[exp6]{S6}}
\label{subsec:explanation-single}

We now explain the results of Experiments~\hyperref[exp1]{S1} through \hyperref[exp6]{S6}.
For each experiment, we predict the pattern of detections using the following procedure: 
Firstly, for each outcome, identify each history (i.e., path) that contributes to the outcome. 
Calculate the associated amplitudes from the incoming amplitude using the rule for beam\hyp splitters and phase shifts. 
Next, apply the indistinguishability principle, i.e., sum the amplitudes from fundamentally indistinguishable paths. 
Finally, compute the probability of each outcome by taking the squared absolute value of the amplitude and sum probabilities from different distinguishable paths.
Let us see this procedure in action.

% BEGIN \input{Figure_MZI.tex}
\begin{figure}
\newcommand*{\figsScale}{1} % beamsplitter and reflectors scale up or down by this ratio
\newcommand*{\RedOpacity}{0.6} % Opacity of the red dots
\begin{subfigure}[t]{0.45\columnwidth}
\includegraphics[width = 0.9\textwidth]{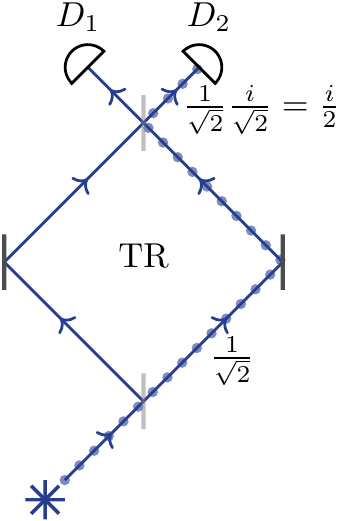}
\caption{}\label{Fig:Exp3a}
\end{subfigure}
\begin{subfigure}[t]{0.45\columnwidth}
\includegraphics[width = 0.9\textwidth]{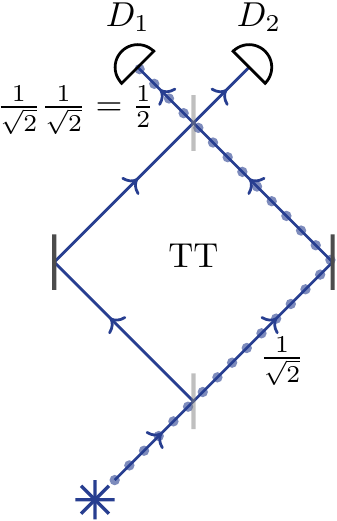}
\caption{}\label{Fig:Exp3b}
\end{subfigure}

\begin{subfigure}[t]{0.45\columnwidth}
\includegraphics[width = 0.9\textwidth]{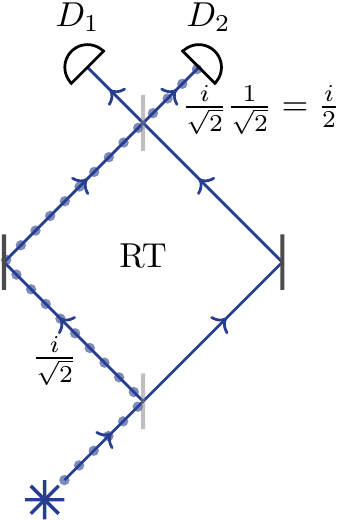}
\caption{}\label{Fig:Exp3c}
\end{subfigure}
\begin{subfigure}[t]{0.45\columnwidth}
\includegraphics[width = 0.94\textwidth]{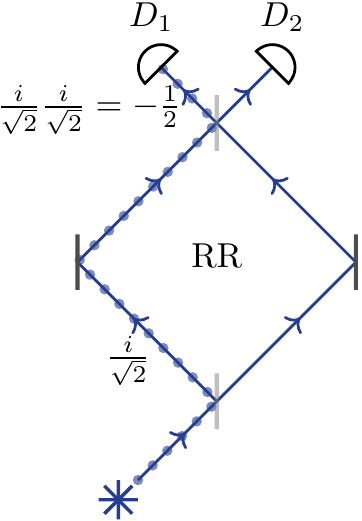}
\caption{}\label{Fig:Exp3d}
\end{subfigure}

\caption{\textbf{The four possible histories for the single photon in Experiment} \hyperref[exp3]{S3}. Subfigures (a) and (c) show the two possible histories for outcome $D_2$ whereas (b) and (d) show the two possible histories for outcome $D_1$. The labels T and R represent `transmitted' and `reflected', respectively. For example, path TR corresponds to a photon transmitted through the first beam-splitter and reflected by the second. The displayed amplitudes can be calculated using the rules described in the main text.
}\label{Fig:SingleParticlePaths}
\end{figure}% END \input{Figure_MZI.tex}

For each outcome of Experiment~\hyperref[exp1]{S1}, there is only one contributing history. For a detection at $D_1$, this history corresponds to a reflection off of the beam-splitter (Rule~1). Likewise, for a detection at $D_2$, this history corresponds to a transmission through the beam-splitter (Rule~1). When a single photon is emitted from the source with unit probability, the incoming amplitude of the photon prior to the beam-splitter is $a=1$ (Rule~2). 
The amplitude of the reflected history is $a(D_1)=1\times i/\sqrt{2}$ whereas the amplitude of transmitted history is $a(D_2)=1\times 1/\sqrt{2}$ (Rule~3).
Since there is only one history contributing to each detection outcome, there are no longer or shorter paths that might have incurred a phase shift (Rule~4).
Moreover, since only one history contributes, we need not sum over any amplitudes (Rule~5).
Hence, the probability of detecting the single photon at $D_1$ is $\mathrm{Pr}(D_1)=|1/\sqrt{2}|^2=1/2$ (Rule~6). 
Likewise, the probability of detecting the photon at $D_2$ is $\mathrm{Pr}(D_2)=|i/\sqrt{2}|^2=1/2$ (Rule~6).
In this way, one can also show that this procedure recovers the result expected for Experiment~\hyperref[exp2]{S2}.
Thus, we can correctly predict the detection probabilities in the first two experiments. Furthermore, Rule~7 allows us to predict the correct detection pattern, including the fact that the detectors do not click simultaneously.
That is, if the photon is observed in one of the paths, then the photon cannot be present in the other path.
More simply, only one of the two (or four) detectors of Experiment \hyperref[exp1]{S1} (or \hyperref[exp2]{S2}) will click for each single photon input into the experimental setup.

For Experiment~\hyperref[exp3]{S3}, there are two possible histories for each of the two detection outcomes.
This is because the photon could take either the left or the right path on its way to either detector.
Let us label each possible history by a sequence of letters `T' and `R' for transmissions and reflections, respectively, at the beam\hyp splitters. 
For outcome $D_1$, the two possible histories are TT and RR. 
That is, to end up in $D_1$, the photon is either transmitted twice or reflected twice (Rule~1). 
For outcome $D_2$, the two possible histories are TR or RT (Rule~1). 
These histories are highlighted in Fig~\ref{Fig:SingleParticlePaths}. 

Rule~2 stipulates that each history is associated with an amplitude. As in \hyperref[exp1]{S1} and \hyperref[exp2]{S2}, the incoming amplitude is $a=1$. Then the amplitude of each history is determined by following Rule~3 for the beam\hyp splitters and mirrors:
  \begin{align*}
		a(\text{TR}) = \frac{1}{\sqrt{2}} \times \frac{i}{\sqrt{2}} &= \frac{i}{2}, \\
    a(\text{TT}) = \frac{1}{\sqrt{2}} \times \frac{1}{\sqrt{2}} &= \frac{1}{2}, \\
		a(\text{RT}) = \frac{i}{\sqrt{2}} \times \frac{1}{\sqrt{2}} &= \frac{i}{{2}}, \\
		a(\text{RR}) = \frac{i}{\sqrt{2}} \times \frac{i}{\sqrt{2}} &= -\frac{1}{{2}}.
	\end{align*}
Rule~4 is not applicable here because we assume that there is no path-length difference between different histories of the same outcome.
In addition, the histories RR and TT are fundamentally indistinguishable since there is no information available that indicates which path the single photons actually took.
Similarly, the histories RT and TR are fundamentally indistinguishable. 
Then by Rule~5 we have the following detection amplitudes:
	\begin{align*}
		a(D_1) &= a(\text{TT}) + a(\text{RR}) = \frac{1}{2} - \frac{1}{2} = 0, \\
  		a(D_2) &= a(\text{TR}) + a(\text{RT}) =\frac{i}{{2}} + \frac{i}{{2}} = i.
	\end{align*}
Next, we use Rule~6 to predict the detection probabilities:
	\begin{align*}
		\mathrm{Pr}(D_1) &= |a(D_1)|^2 = |0|^2=0, \\
		\mathrm{Pr}(D_2) &= |a(D_2)|^2 = |i|^2=1.
	\end{align*}
This gives us the correct probability of detecting photons in $D_1$ and $D_2$ respectively.

To explain Experiment~\hyperref[exp4]{S4}, we must account for the relative phase introduced by the path extension added to the left arm of the apparatus as informed by Rule~4.
In particular, the amplitudes for the histories that include a reflection at the first beam-splitter must accumulate an additional half-cycle phase (180 degrees) due to the extension, so we have:
  \begin{align*}
		a(\text{RT}) = \frac{i}{\sqrt{2}} \times -1\times \frac{1}{\sqrt{2}} &= -\frac{i}{{2}}, \\
		a(\text{RR}) = \frac{i}{\sqrt{2}} \times -1\times \frac{i}{\sqrt{2}} &= \frac{1}{{2}}.
\end{align*}	
On the other hand, the amplitudes for the histories that include a transmission at the first beam\hyp splitter remain the same as in Experiment~\hyperref[exp3]{S3}.
\begin{align*}
a(\text{TR}) = \frac{1}{\sqrt{2}} \times \frac{i}{\sqrt{2}} &= \frac{i}{2}, \\
a(\text{TT}) = \frac{1}{\sqrt{2}} \times \frac{1}{\sqrt{2}} &= \frac{1}{2}.
\end{align*}
We can now see that the summed detection amplitudes are reversed for the two detectors:
	\begin{align*}
		a(D_1) &= a(\text{TT}) + a(\text{RR}) = \frac{1}{2} + \frac{1}{2} = 1, \\
  		a(D_2) &= a(\text{TR}) + a(\text{RT}) =\frac{i}{{2}} - \frac{i}{{2}} = 0.
	\end{align*}
Thus the probabilities are $\mathrm{Pr}(D_1)=1$ and $\mathrm{Pr}(D_2)=0$, and the constructive and destructive interference pattern is swapped.

These same rules also enable an unambiguous explanation of Experiments~\hyperref[exp5]{S5} and~\hyperref[exp6]{S6}.
If the arms of the Mach--Zehnder interferometer are monitored using an NDD, then it is known which path each photon takes. 
In other word, the contributing histories are now distinguishable.
In this case, we can predict the correct outcome probabilities using Rule~6. 

Recall that detection $D_{1}$ resulted from histories involving two transmissions (TT) or two reflections (RR).
The presence of the NDDs makes histories TT and RR distinguishable based on which NDD detects the photon.
As the contributing paths are distinguishable, the probabilities are calculated for the distinguishable outcomes individually and added as described by Rule~6:
\begin{equation*}
\mathrm{Pr}(D_1) = |a(\text{TT})|^2+|a(\text{RR})|^2 =\frac{1}{4}+\frac{1}{4}=\frac{1}{2},
\end{equation*}
Similarly detections in $D_{2}$ result from the histories TR and RT, which are now distinguishable because of the presence of the NDDs.
Thus the probability of detection in $D_{2}$ is obtained by summing the individual probabilities of detection for the cases TR and RT according to 
\begin{equation*}
\mathrm{Pr}(D_2) = |a(\text{TR})|^2+|a(\text{RT})|^2 =\frac{1}{4}+\frac{1}{4}=\frac{1}{2}.
\end{equation*}
Applying Rule~6 gives us the correct (1/2) detection probabilities for each detection. 

Notice that the additional path length of Experiment~\hyperref[exp6]{S6} has no effect on the detection probabilities. 
This is because the eventual detection probabilities do not depend on the complex probability amplitude, but depend only on the individual probabilities [i.e., $|a(\text{TT})|^{2}, |a(\text{TR})|^{2}, |a(\text{RT})|^{2}$ and $|a(\text{RR})|^{2}$], which are independent of any phase acquired in the paths. 
This leads to identical detection probabilities for the two detectors in Experiments~\hyperref[exp5]{S5} and~\hyperref[exp6]{S6}, and no interference is observed in either experiment. 
In summary, the presence of the NDDs destroys the interference and causes the single photons to exhibit particle-like behavior similar to Experiments~\hyperref[exp1]{S1} and~\hyperref[exp2]{S2}.

Let us summarize the quantum description of single-particle experiments. We have demonstrated that a few simple principles of quantum physics can completely explain the particle-like and wave-like behavior of light in many experiments.
This enables us to make correct predictions for the probabilities of detecting quantum particles, even if the physical meaning of the principles may still seem a bit mysterious. 

There are more mysteries and surprises ahead once we consider two or more quantum particles.
Before moving on to the case of two-particle interference, we will introduce the concept of interaction-free measurement, which is important for understanding Hardy's version of Bell's theorem later in the manuscript.

\subsection{Interaction-Free Measurement}
\label{Sec:InteractionFreeMeasurement}

% BEGIN \input{Figure_IFM.tex}
\begin{figure}
\includegraphics[width=0.45\columnwidth]{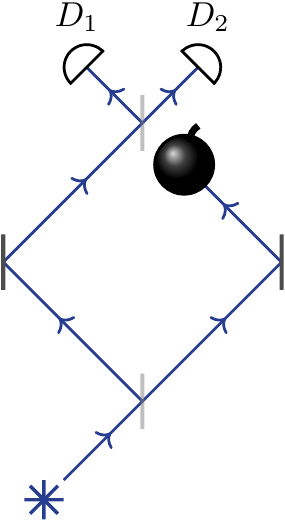}

\caption{\textbf{Interaction-free measurement setup.} A measurement of the photon in $D_{1}$ can successfully detect the presence of the bomb, which has blocked the right path, without any photons from the source hitting it.}\label{Fig:IFM}
\end{figure}
% END \input{Figure_IFM.tex}

Can one infer the presence of something without ever having interacted with it? 
It turns out that single-particle quantum experiments allow for exactly such a possibility.

Usually when we think of measurements using light, we imagine reflecting light off of the object that is being observed and detecting the reflected light.
The so-called ``interaction-free'' measurement is a way to use light to detect the presence of an object based on single-photon interference without the light physically interacting with the object.

The setup for illustrating interaction-free measurement is very similar to the setup of Experiment~\hyperref[exp3]{S3}\m the Mach--Zehnder interferometer. 
We have one single-photon source, two beam\hyp splitters, and two mirrors. However, the right path is blocked midway by an obstacle. To heighten the drama, quantum physicists like to assume this object is a ``bomb'' that will explode if a photon strikes it (see Fig.~\ref{Fig:IFM}).
Can we detect the presence of the bomb in the setup without any explosion occurring?

First, we suppose that there is no bomb in the setup. In this case, the setup is exactly the same as Experiment~\hyperref[exp3]{S3}.
This allows us to predict that an incident single photon will always be detected by detector $D_2$. 

On the other hand, in the case that there is actually a bomb in the right path, the probability of detecting a photon at each detector will change as follows. At the first beam\hyp splitter, an incident photon will choose the right or left path with an equal probability and thus there is a $1/2$ probability that the photon will hit the bomb and make it explode. These are unsuccessful trials from the point of view of interaction-free measurement. However, there is a $1/2$ probability for the photon to take the left path and arrive at the second beam\hyp splitter.
Here, the photon either takes the path leading to $D_1$ or the one leading to $D_2$ with equal likelihood. Thus for an incident photon, either the bomb explodes with $1/2$ probability or one of the two detectors will detect the photon each with an equal probability of $1/4$.

So far, we have seen that if there is no bomb in the setup, then an incident photon always arrives at $D_{2}$; but if there is a bomb, then it could arrive at $D_{1}$ or $D_{2}$. If a single photon is detected in $D_2$, then we cannot say if the bomb was present because there is a chance for a photon to arrive at detector $D_2$ in both cases. However, this is completely different if a photon arrives at $D_1$. In this case, we can infer that the bomb must be present in the setup, because otherwise all of the photons would interfere and none would take the path towards $D_1$. 
Hence, for a single incident photon, there is a $1/4$ probability that we can infer the presence of the bomb in the setup without making the bomb explode.
In this way, quantum physics allows us to detect an object using light without the light ever having interacted with the object.

\subsection{Summary of Single-Particle Experiments Quantum Physics}

Single quantum particles exhibit wave-particle duality, a mixture of particle-like and wave-like behavior. 
Quantum physics explains single-particle behavior by firstly attaching a complex-valued amplitude to each history, then demanding that amplitudes from fundamentally indistinguishable paths be summed, and finally by deriving probabilities of detection as the squared absolute value of the total amplitude. 
This strange single-particle behavior is exemplified by the possibility of performing an interaction-free measurement.
We now turn our attention to the strange behavior of two quantum particles with a focus on entanglement, which encapsulates the second key surprise of quantum physics.

\section{Two-Particle Interference and Entanglement}
\label{Sec:2Particle}

% BEGIN \input{Figure_entanglement.tex}
\begin{figure}
\captionsetup[subfigure]{labelformat=empty}
\newcommand*{\figsScale}{1} % beamsplitter and reflectors scale up or down by this ratio

\begin{subfigure}[b]{0.45\columnwidth}
\includegraphics[width = 0.9\textwidth]{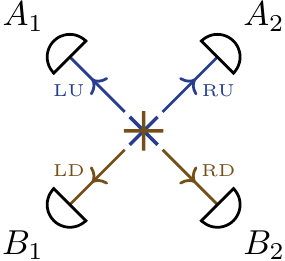}
\caption{(\hyperref[expE1]{E1})}
\label{Fig:E1}
\end{subfigure}
\begin{subfigure}[t]{0.45\columnwidth}
\includegraphics[width = 0.9\textwidth]{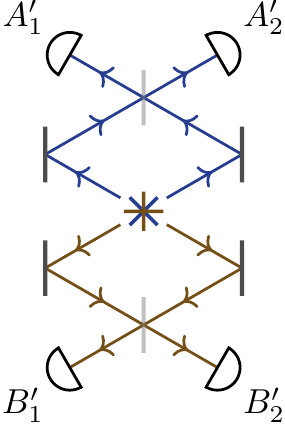}
\caption{(\hyperref[expE2]{E2})}
\label{Fig:E2}
\end{subfigure}%

\begin{subfigure}[t]{0.45\columnwidth}
\includegraphics[width = 0.85\textwidth]{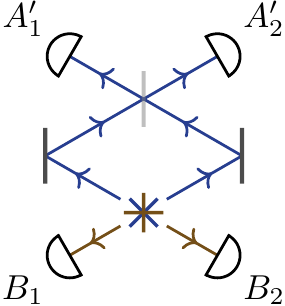}
\caption{(\hyperref[expE3]{E3})}
\label{Fig:E3}
\end{subfigure}
\begin{subfigure}[t]{0.45\columnwidth}
\includegraphics[width = 0.95\textwidth]{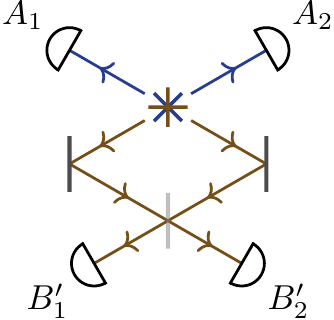}
\caption{(\hyperref[expE4]{E4})}
\label{Fig:E4}
\end{subfigure}%

\captionsetup{width=1.0\columnwidth}
\caption{\textbf{Two-particle interference experiments.} (\hyperref[expE1]{E1}) Direct measurement of the photons when emitted from the entangled-photon source, depicted as the colored star in the center. The source emits two photons back to back.  (\hyperref[expE2]{E2}) the setup for observing two particle interference by merging paths for both Alice and Bob (\hyperref[expE3]{E3}) and (\hyperref[expE4]{E4}) If only one of Alice and Bob merge the paths of the interferometer before detection, then no interference is observed. 
}
\label{Fig:TwoParticle}
\end{figure}

% END \input{Figure_entanglement.tex}
So far, we have only discussed single-particle experiments, which demonstrate wave\hyp particle duality.
In the early days of quantum physics, this strange duality was a source of much debate.
Nonetheless, as quantum physics matured, and found more and more success in explaining and predicting the outcomes of ever-improving experiments, wave\hyp particle duality became an accepted feature of quantum physics.

Making the leap from one to two quantum particles exposes a second, and key, surprise of quantum physics, that of entanglement. 
Entanglement is a property of two or more quantum particles that prohibits the description of each particle as an individual entity. 
The properties of entangled particles depend on each other, but cannot be known before being observed.
That is, a random result is obtained from a measurement of an individual particle. 
This interplay between mutual dependence, or correlation, and individual randomness is the signature of entanglement. 
As we discuss below, these properties make the concept of entanglement counter-intuitive, as well as open to being misunderstood. 

Entanglement has many surprising consequences.
For example, if two particles are entangled, then the outcome of an experiment that is performed on one of the two particles can instantaneously alter the expected outcome of an experiment that is performed on the second, separately (and potentially, distantly) located, particle.
One might imagine that this could be exploited for sending messages faster than light, for example by performing measurement at one location and then looking at the changes in the expected outcomes at a distant position. 
Sending messages faster than light, also called signaling, would violate the special theory of relativity, which mandates no messages can be send faster than the speed of light.
Einstein, who founded the theory of relativity, derisively referred to such instantaneous action as ``spooky action at a distance'', and considered this seemingly-absurd prediction as evidence that quantum physics had to be an incomplete description of reality, as we will describe in more detail in the next section (Sec.~\ref{Sec:EPR}). 
In fact, as we explain in Sec.~\ref{sec:signaling}, quantum physics conforms with locality because it is not possible to use entanglement to send a signal, either faster-than-light or otherwise. 

To explain the concept of entanglement, we now present some simple two-particle experiments that can be understood by generalizing the ideas that we introduced in Sec.~\ref{Sec:OneParticle}.
We also present the quantum physics explanation of these experiments using rules similar to the ones introduced for single particles.
We conclude this section with a short description of why entanglement cannot be used to send messages faster than the speed of light.

\subsection{Two-Particle Experiments and their Quantum Explanation}
The experiments involve the same equipment that we have encountered before, with one addition: a source of entangled pairs of photons.
The source emits two photons, one upwards and one downwards, each either into the left path or the right path.
The two possible paths for the upwards going photon are labeled $LU$ (Left-Up) and $RU$ (Right-Up), whereas those for the downward going photon are $LD$ (Left-Down) and $RD$ (Right-Down) as shown in Fig.~\hyperref[Fig:E1]{7:E1}.
The photons are always emitted in pairs.
Two photons from an emitted pair travel in opposite directions, which could either be $LU$ and $RD$, or be $RU$ and $LD$ with equal likelihood.

The two photons that are emitted by the source exhibit the basic signature of entanglement: the paths that the photons take are simultaneously correlated and random.
The paths that the photons take are correlated because if the upwards going photon takes the $LU$ path, then the other certainly takes the $RD$ path. 
Likewise for the case of one photon taking the $RU$ path, which implies that the other photon takes the $LD$ path.
Moreover, the path that any individual photon takes is not known beforehand, it is completely random.

Entanglement thus describes a special kind of quantum superposition of two or more particles. 
But what exactly is in a superposition?
Two or more particles could be in superpositions over different possible histories, where a history now refers to the collection of the paths taken by all the particles in the experiment.
 
While the aforementioned source of photon pairs is not typically constructed in real life, sources with analogous behavior are well-developed and are routinely used in experiments.
We discuss this in Sec.~\ref{Sec:Experiments}.
It is convenient to imagine that the experiments of Fig.~\ref{Fig:TwoParticle} are being performed by two experimenters, named Alice and Bob. 
Alice is located above the source and receives the upward going photon, while Bob is located below the source, receiving the downward going photon.
The path of the photon traveling towards Alice is depicted in blue and the detectors that are used by Alice are denoted with an `$A$'. 
Similarly, the path of the photon traveling towards Bob is depicted in brown, and his detectors are denoted with a `$B$'.
Our use of colors is not meant to imply anything about the color of the light in the actual photons, but merely to emphasize that there are now two different photons, in contrast to the experiments that we have encountered so far (Sec.~\ref{subsec:singlephotonexperiments}).
Now we examine five different experiments that are performed by Alice and Bob.

\subsubsection{Experiment~E1}
\label{expE1}
First we consider a simple situation in which both Alice and Bob have placed detectors directly in the paths of the photons.
Thus, Alice can detect if the upward going photon is traveling in the left path $LU$ via detector $A_{1}$ or the right path $RU$ via detector $A_{2}$.
Likewise, Bob can distinguish between the downward going photon travelling in the left path $LD$ if detector $B_{1}$ clicks or the right path $RD$ if $B_{2}$ clicks.
This setup is depicted in Fig.~\hyperref[Fig:E1]{7:E1}.

Now we can ask the question, ``What is the detection pattern for Alice's and Bob's detectors?''
That is, how often does Alice detect the photons in $A_{1}$ and $A_{2}$ detectors, and how often does Bob detect photons in $B_{1}$ and $B_{2}$?
More importantly, is there some dependence or correlation between Alice's detection and Bob's?

These questions can be answered promptly based on our description of the entangled photon source.
Alice's left detector $A_{1}$ will click in half the cases and the same for the right detector $A_{2}$.
Similarly Bob's photon detection events will be split half and half into the two detectors $B_{1}$ and $B_{2}$.
The outcomes of Alice's and Bob's detections are correlated: if $A_{1}$ clicks, then $B_{2}$ will surely click.
Similarly, if $A_{2}$ clicks, then so will $B_{1}$.
In other words, the detection events $(A_{1},B_{2})$ and $(A_{2},B_{1})$ both happen with probability 1/2 whereas the events $(A_{1},B_{1})$ and $(A_{2},B_{2})$ happen with zero probability.

This detection pattern can be thought to result from the detectors picking out one out of the two possible histories for the photon pairs.
As the source only emits photons in the $LU$--$RD$ and $RU$--$LD$ directions, these two pairs of directions are the only two histories possible for the photon pairs.
If Alice's photon is detected in $A_{1}$, then this outcome is explained by only one out of the two histories, i.e., the $LU$--$RD$ history, which means that Bob's photon will be detected in $B_{2}$.

Let us note that according to quantum physics it is completely undetermined which of the two possibilities, $LU$--$RD$ or $RU$--$LD$, will occur in any given experiment. 
This is only decided when the measurement is made. 
Yet, as soon as the upper particle is detected in $A_1$, then we know that the lower particle will definitely be detected in $B_2$, and similarly for $A_2$ and $B_1$. 
This observation is closely related to the ``spooky action at a distance'', which was disturbing to Einstein, among others. 
We will come back to this point in more detail in Sec.~\ref{Sec:EPR}.

\subsubsection{Experiment~E2 and the Rules of Many-Particle Quantum Physics}\label{expE2}

% BEGIN \input{Figure_entanglement_1.tex}
\begin{figure}
\newcommand*{\figsScale}{1} % beamsplitter and reflectors scale up or down by this ratio

\begin{subfigure}[t]{0.45\columnwidth}
\includegraphics[width = 0.9\textwidth]{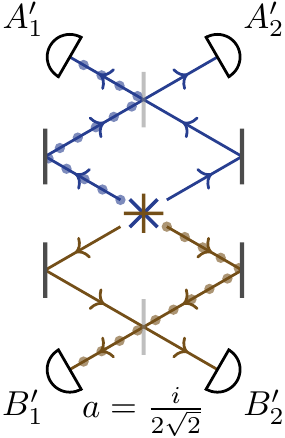}
\caption{}
\label{Fig:EntanglementQuantumEraser_0}
\end{subfigure}%
\begin{subfigure}[t]{0.45\columnwidth}
\includegraphics[width = 0.9\textwidth]{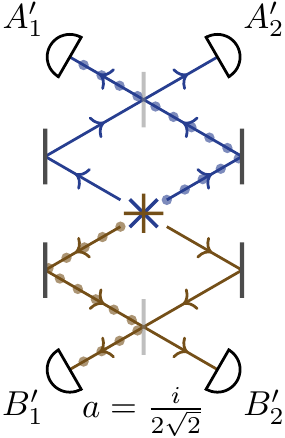}
\caption{}
\label{Fig:EntanglementQuantumEraser_1}
\end{subfigure}%
\caption{\textbf{The two histories for outcome ($A_1^{\prime}$, $B_1^{\prime}$) detection in~\hyperref[expE2]{E2}}. 
These are the two indistinguishable histories that contribute to a detection pattern ($A_1^{\prime}$, $B_1^{\prime}$) in Experiment~\hyperref[expE2]{E2}.
The two histories contribute a probability amplitude $i/2$ and interfere constructively, thus giving a probability $1/2$ of both detectors ($A_1^{\prime}$, $B_1^{\prime}$) clicking.
}\label{Fig:TwoParticleA1B1}

\end{figure}% END \input{Figure_entanglement_1.tex}

To see how interference manifests itself in the two-particle case, let us now imagine a different setup, which we depict in Fig.~\hyperref[Fig:E2]{7:E2}.
In this setup, the upwards going paths are made to meet at a beam\hyp splitter before Alice detects the photons.
Similarly, both of Bob's paths are made to meet at a beam\hyp splitter before the detection is performed.
This seemingly small addition completely changes the pattern of detector clicks.
The resulting pattern can be explained as arising from two-particle interference, which follows the same basic principles as single-particle interference, but in a somewhat more complex way, as we now explain.

In this two-particle experiment, we are interested in Alice's and Bob's detection probabilities and correlations.
In other words, what are the probabilities of four distinct detection events: $(A_{1}^{\prime},B_{1}^{\prime})$, $(A_{1}^{\prime},B_{2}^{\prime})$, $(A_{2}^{\prime},B_{1}^{\prime})$ and $(A_{2}^{\prime},B_{2}^{\prime})$?
We have introduced primes to the detector labels to indicate that these detectors are measuring a different physical property now because the paths now meet at the beam\hyp splitter before arriving at the detector.

To compute the detection probabilities via quantum physics, we can exploit rules similar to those of Sec.~\ref{subsec:QuantumTheory}.
For the many-particle case, we are interested in the probabilities of observing a specific detection outcome, i.e., of observing some specific detectors clicking.
As in the single-particle case we assign an amplitude to each history that contributes to the given detection pattern. 
Furthermore as in the single-particle case, two or more histories will interfere only if they are fundamentally indistinguishable.

Thus, the set of seven rules for single-particle interference are modified to the following for many-particle interference:

\vspace{6pt}
\textit{Rules of many-particle quantum physics}.---
\begin{enumerate}
\item To predict the probability of a specific detection outcome, consider the histories of all the particles that result in the particles arriving at the detectors specified by the outcome.
\item 
Each history of the quantum particles is associated with a wave and its corresponding complex amplitude $a$.
\item 
The amplitude of a history is multiplied by a $1/\sqrt{2}$ factor each time an involved particle is transmitted through a beam\hyp splitter and by a factor of $i/\sqrt{2}$ each time it is reflected at a beam\hyp splitter. Mirrors leave the amplitude unchanged.
\item A history's amplitude can incur a phase shift relative to another history if it involves a particle taking a longer path compared to the same particle in another history.
\item If two or more fundamentally indistinguishable histories lead to the same detection outcome, then the amplitudes of those histories must be summed to obtain the final detection amplitude.
\item The probability (Pr) of observing the given detection outcome is proportional to the squared absolute value of the detection amplitude: $\text{Pr}= |a|^2$.
If more than one distinguishable history leads to the same detection outcome, then their probabilities are summed directly.
\item Observing a particle in one path precludes observing it in any other path.
\end{enumerate}
\vspace{6pt}

It is each to check that these rules include the single-particle version as a special case. 

Note that Rule~2 requires us to specify the probability amplitude of each history depending on the chosen source.
The application of this rule was more straightforward, as each history started with amplitude $a = 1$ but we need to be more careful in the two-particle case. 
In the current experiments, we are using a source of entangled photons.
We can assign probability amplitudes to each history based on the fact the photon pairs are only emitted in the $LU$--$RD$ or $LD$--$RU$ directions with equal likelihood and never in the other two pairs of directions.
Hence, we will assume that the entangled source contributes a factor of $1/\sqrt{2}$ to the amplitude for the histories in which photon pairs are emitted in the $LU$--$RD$ or $LD$--$RU$ directions, and it contributes a factor of zero for histories that involve photons emitted in the $LU$--$RU$ and $LD$--$RD$ directions. 
The fact that we are assigning amplitudes to these two histories implies that they are fundamentally indistinguishable unless detectors are placed in their paths.

Note that it would also be consistent with the probabilities mentioned above to, for example, assign an amplitude of $1/\sqrt{2}$ to $LU$--$RD$ and an amplitude of $-1/\sqrt{2}$ to $LD$--$RU$, since the probabilities are given by the square of the absolute value of the amplitudes. However, this would nevertheless be a physically different source because the histories would interfere in a different way. The particles would still be entangled in this case, but they would be entangled in a different way. 
Here we choose the positive value of the amplitude for simplicity.

We now use these rules to calculate the probabilities of different detection outcomes.
First, let us consider the outcome $(A_1^\prime,B_1^\prime)$.
Following Rule~1, observe that there are two indistinguishable histories leading to this outcome and we depict these in the two parts of Fig.~\ref{Fig:TwoParticleA1B1}.
The first history is the case in which Alice's photon is emitted in direction $LU$ and reflected by her beam\hyp splitter, while Bob's photon is emitted in $RD$ and transmitted by his beam\hyp splitter as depicted in Fig.~\ref{Fig:EntanglementQuantumEraser_0}.
Now, we assign an amplitude to this history (Rule~2).
The amplitude for this history is computed by multiplying the three factors arising respectively from the source, from the transmission in Alice's lab, and from the reflection in Bob's lab (Rule~3).
Thus we have
\begin{align*}
a(LU_A, R_A;RD_B, T_B) = & \frac{1}{\sqrt{2}} \times \frac{1}{\sqrt{2}} \times \frac{i}{\sqrt{2}} \\
= & \frac{i}{2\sqrt{2}}.
\end{align*}
The other possible history involves Alice's photon being emitted along $RU$ and transmitted at the beam\hyp splitter while Bob's photon is emitted along $LD$ and reflected at the beam\hyp splitter shown in Fig.~\ref{Fig:EntanglementQuantumEraser_1}.
Because this history includes contributions from one transmission, one reflection and the factor from the signal, the resultant amplitude for this history is the same as the previous one:
\begin{equation*}
a(RU_A, T_A; LD_B, R_B) = \frac{i}{2\sqrt{2}}.
\end{equation*}
The amplitudes for the two potential histories in which the photon pairs were emitted in the directions $LU$--$RU$ and $LD$--$RD$ are zero, since the source always produces photons going in opposite directions.

For entangled particles, the two histories described above are fundamentally indistinguishable. We have to add the probability amplitudes resulting from these indistinguishable histories to obtain the total probability amplitude for the given detection outcome (Rule~5). Thus, the probability amplitude for the detection outcome~$(A_1^\prime,B_1^\prime)$ is
\begin{align*}
a(A_1^\prime, B_1^\prime) = & a(LU_A, R_A;RD_B, T_B) + a(RU_A, T_A; LD_B, R_B)\\
= & \frac{i}{\sqrt{2}}.
\end{align*}
Squaring this probability amplitude gives us the probability of the detection outcome $A_1^\prime,B_1^\prime$ as (Rule~6)
\begin{equation*}
\operatorname{Pr}(A_1^\prime, B_1^\prime) = \left|a(A_1^{\prime}, B_1^{\prime})\right|^2=\left|\frac{i}{\sqrt{2}}\right|^2=\frac{1}{2}.
\end{equation*}
Thus, the probability of detector $A_{1}^{\prime}$ clicking in Alice's lab and $B_{1}^{\prime}$ clicking in Bob's lab for a given pair of photons is one half.
Since the amplitudes for the two histories were in phase (i.e., had the same complex phase), this outcome is a result of constructive interference of the histories.

Now let us consider the detection outcome $(A_1^\prime,B_2^\prime)$.
Without going through all of the details as above, there are once again two indistinguishable histories leading to this outcome.
Either the particles are emitted in $LU$--$RD$ and both reflected, or they are emitted in $LD$--$RU$ and both transmitted.
These histories respectively have amplitudes of $1/\sqrt{2}\times 1/\sqrt{2}\times 1/\sqrt{2}=1/(2\sqrt{2})$ and $1/\sqrt{2}\times i/\sqrt{2}\times i/\sqrt{2}=-1/(2\sqrt{2})$.
Adding these amplitudes together, we see that these two histories destructively interfere, leading to a total amplitude (and probability) of zero.
Thus, the event in which Alice's detector $A_{1}^{\prime}$ clicks concurrently to Bob's $B_{2}^{\prime}$ has zero probability.

The remaining two cases $(A_2^\prime,B_2^\prime)$ and $(A_2^\prime,B_1^\prime)$ are similar to the two cases described above.
Hence, we can easily see that the respective probabilities of these paths are $1/2$ and $0$.
These results are summarized in Table~\ref{Tab:TwoParticle}.

From Rule~7, we know that if a photon is detected in  $A_{1}^{\prime}$, then it cannot be detected in  $A_{2}^{\prime}$.
This means that in half the cases $A_{1}^{\prime}$- $B_{1}^{\prime}$ outcome is observed and in the other half, $A_{2}^{\prime}$- $B_{2}^{\prime}$ but both outcomes will never be observed simultaneously.
In other words, two photons in the setup will only ever give two detector clicks.

% BEGIN \input{table1.tex}
\begin{table}[]\centering
\renewcommand{\arraystretch}{1.2}
\begin{tabular}{ccccc}
\toprule
Outcome & Contributing history & Amplitude & Total & Prob.\\\midrule
\multirow{2}{*}{$(A_1^\prime,B_1^\prime)$} & $LU_A, R_A; RD_B, T_B$ & $\frac{i}{2\sqrt{2}}$ & \multirow{2}{*}{ $\frac{i}{\sqrt{2}}$} & \multirow{2}{*}{$\frac{1}{2}$} \\ \cline{2-3}
\addlinespace[1mm]
 & $RU_A, T_A; LD_B, R_B$ & $\frac{i}{2\sqrt{2}}$ & & \\ \midrule
\multirow{2}{*}{$(A_1^\prime,B_2^\prime)$} & $LU_A, R_A; RD_B, R_B$ & $\frac{-1}{2\sqrt{2}}$ & \multirow{2}{*}{0} & \multirow{2}{*}{0} \\ \cline{2-3}
\addlinespace[1mm]
 & $RU_A, T_A; LD_B, T_B$ & $\frac{1}{2\sqrt{2}}$ & & \\ \midrule
\multirow{2}{*}{$(A_2^\prime,B_1^\prime)$} & $RU_A, R_A; LD_B, R_B$ & $\frac{-1}{2\sqrt{2}}$ & \multirow{2}{*}{0} & \multirow{2}{*}{0} \\ \cline{2-3}
 \addlinespace[1mm]
& $LU_A, T_A; RD_B, T_B$ & $\frac{1}{2\sqrt{2}}$ & & \\ \midrule
\multirow{2}{*}{$(A_2^\prime,B_2^\prime)$} & $RU_A, R_A; LD_B, T_B$ & $\frac{i}{2\sqrt{2}}$ & \multirow{2}{*}{$\frac{i}{\sqrt{2}}$} & \multirow{2}{*}{$\frac{1}{2}$} \\ \cline{2-3}
\addlinespace[1mm]
& $LU_A, T_A; RD_B, R_B$ & $\frac{i}{2\sqrt{2}}$ & & \\ \bottomrule
\end{tabular}
\caption{Probabilities and amplitudes for the different possible detection outcomes of experiment~\hyperref[expE2]{E2}. 
Of the four possible outcomes, only two have nonzero probability of occurrence.
The remaining two suffer from destructive interference between their individual histories.}
\label{Tab:TwoParticle}
\end{table}% END \input{table1.tex}

We can say that Experiment~\hyperref[expE2]{E2} demonstrates two-particle interference because, out of the four possible detection outcomes, only two outcomes undergo constructive interference and occur with nonzero probability.
The other two outcomes suffer from a destructive interference of their respective histories and have zero probability of occurrence.
In the respects described thus far, two-particle interference is similar to single-particle interference.
However, the cases of one- and two-particle interference are also different in some important respects.

The fundamental difference between one- and two-particle interference is apparent when we consider the experiment from the perspective of the individual experimenters.
For both Alice or Bob individually, their half of the setup looks a lot like a Mach--Zehnder interferometer (as described in Sec.~\ref{Sec:OneParticle}), with the difference that now there is an entangled source where the first beam\hyp splitter was placed in the Mach--Zehnder interferometer.
Recall that if the two paths of a Mach--Zehnder interferometer are balanced, then all the photons are detected in only one of the two detectors.
In contrast, Alice's detector $A_1^\prime$ clicks with 50\% probability, and the same for $A_2^\prime$.
In other words, if Alice is not aware of Bob's outcomes she sees a completely random detection pattern on her detectors $A_1^\prime$ and $A_2^\prime$.
Bob observes the same phenomenon in his lab.
It is not apparent from Alice's or Bob's perspectives individually that any interference is happening whatsoever, since both of their individual outcomes are equally likely to occur.

Furthermore, a distinct signature of single-particle interference was that changing the path lengths in one arm of a Mach--Zehnder interferometer changed the detection probabilities.
Recall that for zero path length difference, all the photons were detected in the bright detector, but as one of the paths was extended, the photon detections would be split between the two detectors and eventually, for a certain path difference, all the photons would be detected in only the dark detector. Let us consider what happens in a two-particle case, i.e., when Alice extends one of the arms of her interferometer.
What are the probabilities of each detector clicking now?
One can verify by simple calculations that it is impossible for Alice to affect the single-particle probabilities by using an extension. 
Thus, no matter how long or short the extension is, the probability of each of her detectors clicking remains fixed at $\frac{1}{2}$.
In this aspect as well, the interference that is happening here is fundamentally different from single-particle interference.
However, one can also show that, while the individual probabilities for Alice and Bob do not change, the joint probabilities listed in Table~\ref{Tab:TwoParticle}, in fact, do change when the length of the paths is no longer the same. 
All of these points show that two-particle interference is distinct from the single-particle case.
In other words, it is the combined two-photon histories that are interfering, not the histories for either photon individually, and the interference manifests in the two-particle detections (i.e., the joint detection probabilities), not in the detection probabilities for individual particles. 
This ensures that the two-particle interference cannot be used for faster-than-light signaling as we detail in Sec.~\ref{sec:signaling}.

Note that, just as for Experiment~\hyperref[expE1]{E1}, the detection pattern of the complete Experiment~\hyperref[expE2]{E2} again shows the dual signatures of entanglement: correlation and randomness.
If detector $A_1^\prime$ clicks, then $B_1^\prime$ must click as well.
Similarly, if $A_2^\prime$ clicks, then $B_2^\prime$ must click as well.
Thus, the clicks of the detectors are perfectly correlated.
Furthermore, the outcomes of measurements performed in one lab are completely random if viewed in isolation.

\subsubsection{Experiment~E3}
\label{expE3}

We saw in section II that single-particle interference only occurs if there are at least two indistinguishable histories that the particle could take to arrive at the given detector. 
Similarly, two-particle interference can only happen if there are at least two indistinguishable histories. This can be illustrated by considering a slightly modified experiment that we depict in Fig.~\hyperref[Fig:E3]{7:E3}.
In this experiment, Alice's paths meet at a beam\hyp splitter before detection, while Bob measures the photons directly.
Alice's detectors use the primed notation introduced previously because she is using a beam\hyp splitter, whereas Bob is not using a beam\hyp splitter, so we do not use primes to denote his detectors.
In this setting, what are the probabilities of each of the four possible pairs of detector clicks?
That is, what are the probabilities of the pairs $(A_1^\prime, B_1)$, $(A_1^\prime, B_2)$, $(A_2^\prime, B_1)$ and $(A_2^\prime, B_2)$ of detectors clicking?

We can easily check that these probabilities are all equal to $\frac{1}{4}$.
This is because only one history contributes to each of the possible pairs of detection outcomes.
Each of these histories picks up a factor of $1/\sqrt{2}$ from the source and either a factor of $1/\sqrt{2}$ or $\mathrm{i}/\sqrt{2}$ from the beam\hyp splitter, thus giving a total probability of $1/4$.

For concreteness, let us consider the history that leads to detections of the photon pairs in the detectors $A_{1}^{\prime}$ and $B_{2}$.
This detection pattern sees contribution from only one history, in which the photons are emitted in the $LU$--$RD$ directions, Bob's detector $B_{2}$ detects Bob's photons, and Alice's photon reflects off from her beam\hyp splitter into detector $A_{1}^{\prime}$.
This gives us a probability amplitude of $1/\sqrt{2} \times i/\sqrt{2} = 1/2$, which means that there is probability $|i/2|^{2} = 1/4$ of detecting photons in the detectors $A_{1}^{\prime}$ and $B_{2}$.
A similar analysis gives the same probability for each of the other three detection patterns.

Observe that in this experiment, only one history contributes to each outcome of the experiment 
This means that there are no collaborating or competing histories, and so there is no constructive or destructive interference.
As a result, all the pairs of detection are equally likely, which is what we see from the above calculations.

\subsubsection{Experiment~E4}
\label{expE4}
What happens instead if Alice detects her photons directly but Bob places a beam\hyp splitter in the paths of his photons before detecting them?
This situation (see Fig.~\hyperref[Fig:E4]{7:E4}) is analogous to the previous Experiment~\hyperref[expE3]{E3} with Alice and Bob having changed their roles.
In this case too, each detection outcome sees contributions from one and only one history, and each pair $(A_{1},B_{1}^{\prime})$, $(A_{1},B_{2}^{\prime})$, $(A_{2},B_{1}^{\prime})$ and $(A_{2},B_{2}^{\prime})$ of detection outcomes are observed with equal probability.
Similar to the case of Experiment~\hyperref[expE3]{E3}, this experiment does not display any interference.

The joint probabilities in Experiments \hyperref[expE3]{E3} and \hyperref[expE4]{E4}, in both of which no interference takes place, are thus very different compared to the joint probabilities for Experiment~\hyperref[expE1]{E1} or \hyperref[expE2]{E2} (see Table~\ref{Tab:TwoParticle}). 
However, in each of these experiments, the individual probabilities on each side are still 1/2 for each detector.
This again highlights how two-particle interference is different from the single-particle case.

\subsubsection{Experiment~E5}
\label{expE5}

% BEGIN \input{Figure_NDD2.tex}
\begin{figure}
\captionsetup[subfigure]{labelformat=empty}
\newcommand*{\figsScale}{1} % beamsplitter and reflectors scale up or down by this ratio

\begin{subfigure}[t]{.4\columnwidth}
\includegraphics[width=\textwidth]{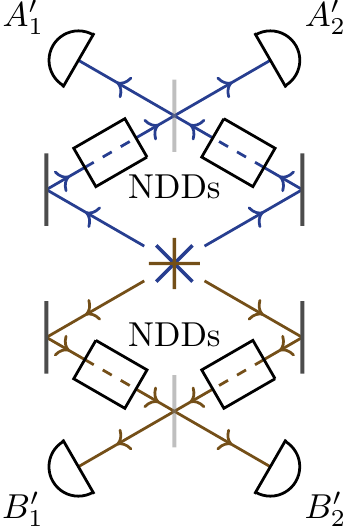}

\caption{(\hyperref[expE5]{E5})}
\label{Fig:NDD3}
\end{subfigure}%

\captionsetup{width=1.0\columnwidth}
\caption{\textbf{Two-particle interference experiment with non-destructive detectors (NDDs).} A non-destructive detector (NDD) is placed in each path of the setup for Experiment (\hyperref[expE2]{E2}). Depending on which NDDs click, one knows which path each photon took.
}
\label{Fig:TwoParticleNDD}
\end{figure}
% END \input{Figure_NDD2.tex}

Two-particle interference, like single-particle interference, can take place only between indistinguishable histories.
Let us now think about an experiment that highlights the role of indistinguishability in the two-particle case.
We consider a modified version of Experiment~\hyperref[expE2]{E2} as depicted in Fig.~\hyperref[Fig:NDD3]{9:E5}. 
Analogous to Experiments~\hyperref[exp5]{S5} and~\hyperref[exp6]{S6}, NDDs are placed in the paths of the two particles.
Now, what is the pattern of Alice's and Bob's detections and what are the correlations that they observe?

The addition of NDDs in the paths makes the different histories distinguishable.
Let us focus on the probability of observing a simultaneous $(A_{1}^{\prime}, B_{1}^{\prime})$ detection.
The two histories that contribute to this detection, i.e., $LU_A, R_A; RD_B, T_B$ and $RU_A, T_A; LD_B, R_B$ (similar to the highlighted histories of Fig.~\ref{Fig:TwoParticleA1B1}) are now distinguishable.
This means that their amplitudes are no longer added together but instead we must follow Rule~6, which suggests that we calculate the individual probabilities of detection from these histories and add them together to obtain the actual detection probabilities.

Specifically, the two histories begin with a $1/\sqrt{2}$ amplitude because of the nature of the source.
Both the histories involve one transmission and one reflection, which change the amplitude to $1/\sqrt{2} \times 1/\sqrt{2} \times i/\sqrt{2} = i/\sqrt{8}$.
Thus, both histories contribute a probability of $|i/\sqrt{8}|^{2} = 1/8$ and these probabilities are added together $1/8 +1/8 = 1/4$ to obtain the final 1/4 probability of $(A_{1}^{\prime}, B_{1}^{\prime})$ detection.
We can check that each of the three other detection patterns, namely $(A_{2}^{\prime}, B_{2}^{\prime})$, $(A_{2}^{\prime}, B_{2}^{\prime})$ and $(A_{2}^{\prime}, B_{2}^{\prime})$, have the same 1/4 likelihood of occurrence. 

This experiment clarifies that indistinguishability is an important prerequisite for two-particle interference and entanglement.
Thus, if indistinguishability is removed by adding NDDs to the paths, then Alice's and Bob's detections are still random but no longer display any correlations.
This means that in the absence of indistinguishability, the second signature of entanglement is lost.

\subsection{A Note on Signaling, Correlations in Everyday Life}
\label{sec:signaling}

At this point, we are ready to revisit the issue of signaling.
Why is Alice not able to use these setups to communicate with Bob faster than the speed of light?
The answer lies in the fact that there is nothing that Alice can do locally to influence Bob's probabilities of each detector clicking from his perspective; all Bob ever sees is a string of totally random clicks with detectors $B_{1}$ and $B_{2}$ clicking randomly with equal likelihood.
Thus, there is no way for Alice to use these setups to send a message of any kind, let alone faster than light.
In fact, it turns out that entanglement alone can never be used to send messages; some other ingredient is always necessary (such as a telephone line, which manifestly will not allow for fast-than-light messaging).

However, there still seems to be a connection between the two particles that we have not explained so far. In Experiments \hyperref[expE1]{E1} and \hyperref[expE2]{E2}, how can they give outcomes that are both perfectly correlated, yet apparently completely random?
In everyday life, we also often experience correlations, but we can usually find reasons for them. For example, let's consider two people, Alice and Bob, who own shirts of two colors, blue and red. If we saw them wear the same color day after day (sometimes blue, sometimes red, but always the same for the two of them), we would strongly suspect that they are either coordinating their choices somehow, or that they are following a pattern that is known to both of them (say blue when it is sunny in the morning and red when it is not).

This leads to the question what is actually happening for the quantum particles. Is there really some kind of ``spooky action at a distance'', as quantum physics seems to suggest, and as we briefly described when talking about Experiment~\hyperref[expE1]{E1}? Or is there some yet-to-be-discovered simple explanation for the correlations, as in our example with the shirts? In 1935, Einstein, Podolsky and Rosen published an influential paper~\cite{Einstein1935} that argued in favor of the latter conclusion. Now we describe their argument in detail.

\section{The Einstein--Podolsky--Rosen Argument and Local Hidden Variables}
\label{Sec:EPR}

The EPR argument~\cite{Einstein1935} questions the completeness of quantum theory.
The basis of EPR's argument is the concept of \textit{elements of reality}, which we first describe before presenting their reasoning in detail.

According to EPR, if the outcome of an experiment is known with certainty, then this outcome must correspond to some element of reality in the theory.
More specifically, if the value of an experimental outcome can be predicted with certainty in a given situation, then the value definitely corresponds to an element of reality.
These experimental outcomes are thought to be real in the sense that they \textit{really have} specific values, even if no measurement is made to ascertain them. 
Later in his life, Einstein likened this notion of `elements of reality' to the reality of the existence of the moon when no one is looking. Quoting physicist Abraham Pais,
\begin{quote}
We often discussed his notions on objective reality. I recall that during one walk Einstein suddenly stopped, turned to me and asked whether I really believed that the moon exists only when I look at it.~\cite{Pais1979}
\end{quote}

EPR define a complete physical theory as one that contains a counterpart to each element of reality. 
According to EPR, a theory is satisfactory if it is complete and correct, in which case the correctness is judged by the degree of agreement between theory and experiment. 
EPR do not question the correctness of quantum physics, but they argue that it is not \textit{complete} in the sense that it does not contain counterparts to all elements of reality.

The EPR argument relies on an experiment similar to that described in the previous section, in which two particles are shot from an entangled source and measured by two distant experimenters.
Although the exact setting that EPR considered was somewhat different, the reasoning can be translated meaningfully into the language of Experiments \hyperref[expE1]{E1}--\hyperref[expE4]{E4}.

\subsection{EPR Argument for Two-Photon Experiments}

What are the elements of reality in the two-photon experiments?
These are the outcomes of the two different kinds of measurements that can be performed, i.e., detecting photons with and without their paths meeting on the beam\hyp splitter.
For concreteness, we focus on the measurements performed by Alice on the upward going photons.
Consider the outcome of a Alice's direct measurement of the photon, i.e., without the paths meeting at a beam\hyp splitter as in Experiments~\hyperref[expE1]{E1} and~\hyperref[expE4]{E4}. 
The outcome of this measurement can be inferred with certainty if Bob measures his photon directly as he did in Experiment~\hyperref[expE1]{E1}.
According to EPR, this means that some element of reality must exist that contains information about this outcome.

Next we consider the outcome of Alice's measurement of her photon after the paths meeting at a beam\hyp splitter.
This is the measurement that she had performed in Experiments~\hyperref[expE2]{E2} and~\hyperref[expE3]{E3}.
As for the case of the direct measurement, the outcome of this measurement performed by Alice can be known with certainty if Bob measures his photon after they meet at a beam\hyp splitter as was done in Experiment~\hyperref[expE2]{E2}.
Thus, there must be another element of reality that contains information about Alice's measurement of the photons after they meet at a beam\hyp splitter.
If the outcomes of both of Alice's measurements correspond to elements of reality, then  according to EPR, any complete physical theory must contain a description of these outcomes.
EPR show that quantum physics does not allow for such a description, and is hence claim that it is not a complete theory.

We can see that quantum physics does not allow for a simultaneous prediction of Alice's two measurements by considering all four experiments \hyperref[expE1]{E1}--\hyperref[expE4]{E4} together.
First, consider Experiments \hyperref[expE1]{E1} and \hyperref[expE3]{E3}, in which Bob measures the same quantity, i.e., the location of the photon without placing a beam\hyp splitter in his paths.
This allows Bob to infer Alice's outcome in those instances when she measures the photons directly as in Experiment~\hyperref[expE1]{E1}, i.e., when she has not placed a beam\hyp splitter in her paths.
However, Bob's direct measurement of the photons gives no information about Alice's outcomes in those cases when Alice places a beam\hyp splitter in her photons' path.
This is clear from the outcome of \hyperref[expE3]{E3}, in which Alice's detections were split randomly between her two detectors.
Likewise, we can see that if Bob measures his photon after placing a beam\hyp splitter in his paths (Experiments \hyperref[expE2]{E2} and \hyperref[expE4]{E4}), then he can infer the outcome of Alice's measurement if she too had placed a beam\hyp splitter, but not of her direct measurement of the photons.

Because Bob can either measure his photon directly or measure them after the paths meeting, yet cannot perform both measurements at the same time, he can infer the results of only one of Alice's measurements.
Thus, even though both of Alice's measurements correspond to elements of reality, quantum physics does not allow a simultaneous prediction of these two measurements.
As quantum physics does not contain counter\hyp parts to all elements of reality, EPR concluded that quantum physics is not a complete description of reality 

EPR end their article with a belief that a complete description of reality does indeed exist even though quantum physics does not provide this description.
Thus, the missing ingredient in quantum physics, according to EPR, is the information about eventual measurement outcomes for all elements of reality. 

\subsection{EPR Argument and Local Hidden Variables}

The modern approach to thinking about the EPR argument and related concepts is in terms of the so-called \textit{local hidden variables}.
Local hidden variables are analogous to EPR's elements of reality but there are subtle differences, which we will get to in Section~\ref{Subsec:PredictionsEPRLHV}.
Here we present the EPR in the language of local hidden variables.

To explain the outcomes of all measurements that can be performed on a particles, one might imagine that each particle carries a cheat sheet, or a list of instructions, with information about measurement outcomes. 
These instructions would be the elements of reality sought by EPR. 
Because these instructions are not a part of quantum physics, and are thus inaccessible to us at least for the moment, they are referred to as local hidden variables in modern discourse.

The \emph{local} part of ``local hidden variables'' refers to the principle that the physical properties of one particle should not be able to change instantaneously depending on anything done to some other distant particle. 
This notion of locality is central to Einstein's theory of relativity, which implies that information cannot travel faster than the speed of light. 
Let us emphasize that this principle is consistent with all the available experimental evidence to date, including experiments on quantum entanglement.

The EPR argument is based on the assumption of locality because Alice's elements of reality are thought to be independent of what measurements Bob performs in his lab and vice versa. 
Assuming that the theory of relativity is correct, locality can be ``enforced'' by, for example, putting Alice and Bob in distant enough laboratories, such that there isn't enough time for information to travel from Bob's lab to Alice's lab, for instance about which measurement that Bob has performed on his photon, before Alice performs hers.
(This point has important consequences for experiments as we describe in Sec.~\ref{Sec:loophole}.)

% BEGIN \input{table2.tex}
\begin{table}
\newcolumntype{C}[1]{>{\hsize=#1\hsize\centering\arraybackslash}X}%
\begin{tabularx}{0.95\columnwidth}{ C{1}C{1}C{1}C{1} }
%\begin{tabular}{cccc}
\toprule
  \multicolumn{2}{C{2}}{Alice} & \multicolumn{2}{C{2}}{Bob} \\
  \multicolumn{2}{C{2}}{Performed measurements:} & \multicolumn{2}{C{2}}{Performed measurements:} \\
  $A_1/A_2$ & $A_1^\prime/A_2^\prime$ & $B_1/B_2$ & $B_1^\prime/B_2^\prime$ \\
  \midrule[\heavyrulewidth]
   \cellcolor{xblue!25}$A_1$ & $A_1^\prime$ & \cellcolor{xblue!25}$B_1$ & $B_1^\prime$ \\
 \cellcolor{xblue!25}$A_1$ & $A_1^\prime$ & \cellcolor{xblue!25}$B_1$ & $B_2^\prime$ \\
 \cellcolor{xblue!25}$A_1$ & $A_2^\prime$ & \cellcolor{xblue!25}$B_1$ & $B_1^\prime$ \\
 \cellcolor{xblue!25}$A_1$ & $A_2^\prime$ & \cellcolor{xblue!25}$B_1$ & $B_2^\prime$ \\
 \cellcolor{xblue!25}$A_2$ & $A_1^\prime$ & \cellcolor{xblue!25}$B_2$ & $B_1^\prime$ \\
 \cellcolor{xblue!25}$A_2$ & $A_1^\prime$ & \cellcolor{xblue!25}$B_2$ & $B_2^\prime$ \\
 \cellcolor{xblue!25}$A_2$ & $A_2^\prime$ & \cellcolor{xblue!25}$B_2$ & $B_1^\prime$ \\
 \cellcolor{xblue!25}$A_2$ & $A_2^\prime$ & \cellcolor{xblue!25}$B_2$ & $B_2^\prime$ \\
 \midrule
 $A_1$ & \cellcolor{xgreen!25}$A_1^\prime$ & $B_2$ & \cellcolor{xgreen!25}$B_2^\prime$ \\
 $A_2$ & \cellcolor{xgreen!25}$A_1^\prime$ & $B_1$ & \cellcolor{xgreen!25}$B_2^\prime$ \\
 $A_1$ & \cellcolor{xgreen!25}$A_2^\prime$ & $B_2$ & \cellcolor{xgreen!25}$B_1^\prime$ \\
 $A_2$ & \cellcolor{xgreen!25}$A_2^\prime$ & $B_1$ & \cellcolor{xgreen!25}$B_1^\prime$ \\ 
 \midrule
 $A_1$ & $A_1^\prime$ & $B_2$ & $B_1^\prime$ \\
 $A_1$ & $A_2^\prime$ & $B_2$ & $B_2^\prime$ \\
 $A_2$ & $A_1^\prime$ & $B_1$ & $B_1^\prime$ \\
 $A_2$ & $A_2^\prime$ & $B_1$ & $B_2^\prime$ \\
 \bottomrule
%\end{tabular}
\end{tabularx}
\caption{\textbf{Possible assignment of local hidden variables to explain experiments \hyperref[expE1]{E1}--\hyperref[expE4]{E4}}: 
The sixteen rows of this table list all possible assignments of the local hidden variables for Alice's and Bob's photons as they leave the source.
The first eight rows are ruled out because experiment \hyperref[expE1]{E1} shows that detectors $A_{1}$ and $B_{1}$ never click simultaneously and likewise for detectors $A_{2}$ and $B_{2}$.
The next four rows are ruled out from the results of Experiment \hyperref[expE2]{E2}, which does not see any simultaneous  $A_{1}^{\prime}$ and $B_{2}^{\prime}$ clicks or simultaneous $A_{2}^{\prime}$ and $B_{1}^{\prime}$ clicks.
The remaining four rows are valid because they do not contradict the individual results of the four experiments. 
In fact, if the values in the bottom four rows are assigned randomly with equal likelihood, we can explain the statistics observed in the four experiments. 
}
\label{tab:HiddenVariablesEntanglement}
\end{table}% END \input{table2.tex}

The question then is: can the outcomes of Experiments \hyperref[expE1]{E1}--\hyperref[expE4]{E4} be explained using local hidden variables?
To explore this question, let's consider a situation in which Alice and Bob can choose whether to use a beam\hyp splitter or not.
Furthermore, our explanation should be able to account for a situation in which they are able to defer their measurement choices until after the source has already emitted its pair of photons.
In other words, Alice and Bob can choose whether or not to place a beam\hyp splitter in their paths before detecting the photons, but after the photon pairs have left the source.

Thus, to explain Experiments \hyperref[expE1]{E1}--\hyperref[expE4]{E4} via local hidden variables, the upward-going photon must leave the source carrying information about its detection outcome if Alice measures using detectors $A_1$ and $A_2$ (without the beam\hyp splitter), and also the outcome if $A_1^\prime$ and $A_2^\prime$ are used instead (with the beam\hyp splitter).
Similarly, the downward\hyp going photon must contain information about its detection outcome for detectors $B_{1}$--$B_{2}$, and also for $B_{1}^{\prime}$--$B_{2}^{\prime}$.
The outcome information that one photon contains can depend on the information that the other photon leaves with. 
Indeed, the information a photon carries about its destiny may even be allowed to change when the photon is in transit; however, locality forbids such a change to depend on anything done to the other photon, for instance on which measurements are performed on the other photon.

Thus, we can imagine that the source writes two symbols in the instruction list of the upward\hyp going photon: one that determines the outcome of the $A_1$--$A_2$ detection (i.e. $A_1$ or $A_2$), and one for the $A_1^\prime$--$A_2^\prime$ detection.
Analogously, the source must also write one unprimed symbol~(i.e. $B_1$ or $B_2$) and one primed symbol~(i.e. $B_1^{\prime}$ or $B_2^{\prime}$) in the instruction list for Bob's photon.
These four symbols are the local hidden variables of the two photons.

Thus, our search for a local hidden variable explanation of Experiments~\hyperref[expE1]{E1}--\hyperref[expE4]{E4} turns into the question of assigning values to the two symbols, one each carried by the two photon pairs leaving the source.
Of course, these values need to be assigned in a way that is consistent with the outcomes of Alice's and Bob's detections with and without their paths meeting at a beam\hyp splitter.

Since each of the four symbols can be assigned to one out of two different outcomes, there are a total of $2 \times 2 \times 2 \times 2 = 16$ possible combinations of values that can be assigned to the two particles.
We tabulate these values in the sixteen rows of Table~\ref{tab:HiddenVariablesEntanglement}.

From the results of our experiments, we know that some of these combinations are not permitted.
For instance, from Experiment~\hyperref[expE1]{E1}, we know that $A_1$ and $B_1$ must never occur simultaneously.
This eliminates the combinations in the first eight rows of Table~\ref{tab:HiddenVariablesEntanglement} as these assignments would have resulted in simultaneous $A_1$--$B_1$ clicks.

Additionally, from Experiment~\hyperref[expE2]{E2}, we know that $A_1^\prime$ and $B_2^\prime$ never click for photons from the same pair and neither do $A_2^\prime$ and $B_1^\prime$.
This eliminates the next four rows of the table, leaving us with only four permissible combinations, i.e., the last four rows.

You may convince yourself that, if one assumes that these last four rows all have equal probability of 1/4, then one can also explain the results of Experiments \hyperref[expE3]{E3} and \hyperref[expE4]{E4}. We conclude that if the source simply produces pairs of particles corresponding to one of these four combinations, with equal probability for each of the four, it completely explains the individual outcomes of the four Experiments \hyperref[expE1]{E1}--\hyperref[expE4]{E4}. This seems to support the vision of EPR, even if we may not know yet how to find the local hidden variables for each individual pair of particles.

But does this mean that local hidden variable theories can adequately replace quantum mechanics in explaining all experiments?
Einstein continued his quest for such a local hidden variable explanation of all quantum phenomena until the end of his life.
This quest was in fact doomed to fail, as shown (after Einstein's death) by John Bell in his celebrated 1964 paper~\cite{Bell1964}.
While local hidden variables can match the outcomes of quantum physics in some cases, as the ones we have discussed so far, they predict radically different outcomes for certain other experiments. In the next section, we turn our attention to an experiment in which no assignment of local hidden variables is consistent with the predictions of quantum physics.

\section{EPR, Bell and Hardy}
\label{Sec:Hardy}

We are now ready to experience the second, and perhaps strangest, surprise of quantum physics.
The story of this surprise began with the EPR paper in which the authors argued that quantum physics is incomplete as a description of reality.
Recall that EPR came to this conclusion by making intuitive assumptions about the nature of reality. Although the authors were questioning the completeness of quantum physics under their assumptions, they did not question the accuracy of either the assumptions or of quantum physics.

Without realizing, EPR had taken a key step towards uncovering a deep contradiction between quantum physics and two seemingly obvious assumptions about the universe, namely realism, or, equivalently, the existence of hidden variables, and locality.
It was John Bell who first realized, to his own surprise, that quantum physics was incompatible with these assumptions.
Thus, either quantum physics or one of two cherished assumptions of physics needed to give way.

In his celebrated 1964 paper, John Bell formalized the EPR assumptions in terms of local hidden variables~\cite{Bell1964}.
Using this formalism, he discovered a contradiction between quantum physics on the one hand and the existence of local hidden variables on the other.
Specifically, Bell proved a theorem which showed that quantum predictions regarding a certain experimental setup are measurably differed from those obtained if local hidden variables were assumed to determine the outcomes of all measurements.
This means that we can perform an experiment whose outcomes would be consistent either with local hidden variables or with quantum physics but not with both. 
This experimental test opened the way for resolving the question raised by EPR one way or the other.
In fact, tests of Bell's theorem may be performed by using a two-photon interference experiment similar to that of Fig.~\hyperref[exp2]{E2}; varying the length of the paths that photon travels as done in Experiment~\hyperref[exp1]{S4}; and finally measuring correlations between Alice's and Bob's detection outcomes.

In 1992, Lucien Hardy proposed a simplified, and perhaps more striking, version of the Bell test~\cite{Hardy1992}.
Understanding Hardy's version is less demanding than Bell's original version because it requires simpler mathematics. 
Furthermore, Hardy's version is amenable to a simple explanation using only the concepts of single-particle and two-particle interference introduced in the previous sections, without having to introduce more mathematically advanced concepts such as state vectors or inequalities.
Here we follow Hardy's logic, cast into the interference language of Secs.~\ref{Sec:OneParticle} and~\ref{Sec:2Particle}, to convey the clash between quantum physics and our intuition about nature. 

\subsection{The Setting for Hardy's Paradox}

\begin{figure}
\captionsetup[subfigure]{labelformat=empty}

\begin{subfigure}[t]{0.45\columnwidth}
\includegraphics[width = 0.95\textwidth]{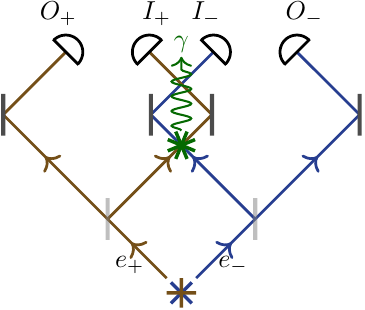}
\caption{(\hyperref[expH1]{H1})}\label{Fig:HardyNoInterfere}
\end{subfigure}
\begin{subfigure}[t]{0.45\columnwidth}
\includegraphics[width = 0.95\textwidth]{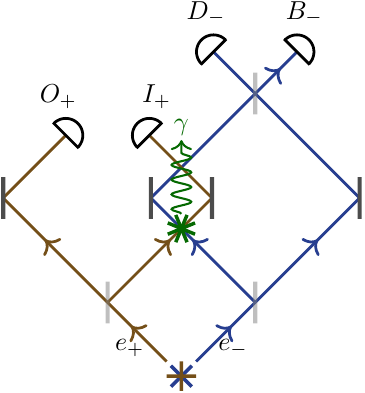}
\caption{(\hyperref[expH2]{H2})}\label{Fig:HardyRightInterfere}
\end{subfigure}
\begin{subfigure}[t]{0.45\columnwidth}
\includegraphics[width = 0.95\textwidth]{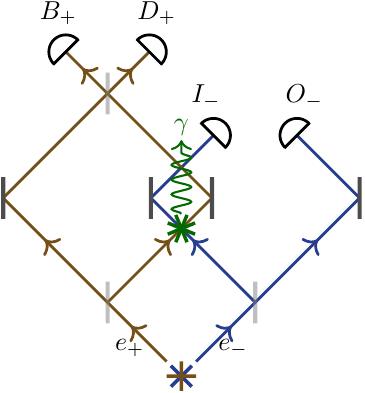}
\caption{(\hyperref[expH3]{H3})}\label{Fig:HardyLeftInterfere}
\end{subfigure}
\begin{subfigure}[t]{0.45\columnwidth}
\includegraphics[width = 0.95\textwidth]{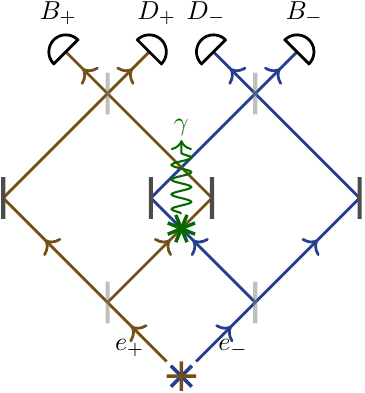}
\caption{(\hyperref[expH4]{H4})}\label{Fig:HardyBothInterfere}
\end{subfigure}
\caption{\textbf{A depiction of Hardy's thought experiment.}
The star at the bottom of the figure depicts a source of electron-positron pairs.
Of a generated pair, the electron, depicted as $e_{-}$, is directed towards the right beamsplitter and the positron, depicted as $e_{+}$, to the left beamsplitter. 
Each of the two particles can take either of the two paths (inner or outer) to reach the beamsplitters at the top of the figure.
The two particles can meet if they both take the respective inner paths, in which case the two will be annihilated and photons of $\gamma$ radiation will be emitted but these not detected by the particle detectors.
The outgoing electrons and positrons are measured at the detectors.
Experiments \hyperref[expH1]{H1} through \hyperref[expH4]{H4} represent four different choices of merging the respective interferometer paths or not for each of the two interferometers.
}\label{Fig:HardyExperiment}
\end{figure}

Hardy's paradox is based on the four experiments depicted in Fig.~\ref{Fig:HardyExperiment}.
We refer to these experiments as~\hyperref[expH1]{H1}, \hyperref[expH2]{H2}, \hyperref[expH3]{H3} and \hyperref[expH4]{H4}.
The experiments rely on the interference of an electron and its antiparticle\footnote{Each fundamental particle is associated with an antiparticle, which has the same mass but opposite charge. When a particle and an antiparticle meet, they annihilate and emit radiation in the form of possibly high-energy photons.}, a positron, each in a Mach-Zehnder interferometer, which we introduced in Sec.~\ref{Sec:OneParticle}.
Even though we are using electrons and positrons, these experiments are similar to the earlier single- and two-particle experiments that relied on photons.

The electron and positron are created (for instance, via a pair-production process~\cite{Hubbell2006}) and shot simultaneously into the two interferometers.
The electron and positron are measured either directly at detectors $I_{\pm}$ and $O_{\pm}$ (placed in the ``inner'' and ``outer'' paths) or at the ``bright'' and ``dark'' detectors $B_{\pm}$ and $D_{\pm}$ after the paths meet at a beam\hyp splitter. 
The minus or plus sign in the subscript refers to either the electron (minus) or the positron (plus).
Furthermore, the detectors are called ``bright'' and ``dark'' for the case with added beam\hyp splitters because, following the discussion in Sec.~\ref{Sec:OneParticle}, if each Mach-Zehnder interferometer was completely on its own, the particles would always be detected at the bright detectors and never at the dark detectors.

% BEGIN \input{Figure_Hardy.tex}
% END \input{Figure_Hardy.tex}

However, the interferometers are arranged such that there is an overlap of the inner trajectory of the electron with the inner trajectory of the positron.
Note that the detectors $I_{\pm}$ in the inner trajectories are downstream from this meeting point.
If both the electron and positron take these ill-fated trajectories, then they meet and annihilate each other, since they are two halves of a particle and antiparticle pair.
They create photons of $\gamma$-radiation in the process, but we assume our detectors are sensitive to electrons and positrons and not photons.
As a result, if the electron and the positron meet at the intersection of their inner paths, nothing is detected at the interferometer outputs.

Let us now consider four experiments in which the electron and positrons are detected either in their inner/outer detectors (i.e., without their paths meeting) or at the bring/dark detectors (i.e., after their paths are merged).

\subsubsection{Experiment~H1}
\label{expH1}

Let us begin by considering the experimental setup \hyperref[expH1]{H1}, which is depicted in Fig.~\hyperref[Fig:HardyNoInterfere]{10:H1}.
In this experiment, the detectors $O_{-}/I_{-}$ are placed directly in the two possible paths of the electron, and detectors $O_{+}/I_{+}$ are placed in the paths of the positron.
In this case, what is the detection pattern?

Because the electron-positron pair will annihilate if the electron and positron meet in the inner path, the pair is never observed in the two inner paths simultaneously.
Hence, the detectors $I_{+}$ and $I_{-}$ never click together in the same run of the experiment.
Other cases of coincident detector clicks might be observed. These are the simultaneous detection at $I_{+}$ and $O_{-}$; at $O_{+}$ and $I_{-}$; or at $O_{+}$ and $O_{-}$.
If this experiment is performed, then these cases, in fact, are observed with equal likelihood.

\subsubsection{Experiment~H2}
\label{expH2}

Next, rather than measuring both the particle and the antiparticle directly, let us consider a situation in which the two paths meet at a beam\hyp splitter for only one interferometer, and for the other interferometer the paths are measured without meeting at a beam\hyp splitter.
Let us first focus on Experiment~\hyperref[expH2]{H2}, in which the right (electron) interferometer paths are combined, as depicted in Fig.~\hyperref[Fig:HardyRightInterfere]{10:H2}.
There are two possible outcomes for the positron in the left interferometer: either the positron can be detected in the inner path or in the outer path.
We can analyze these two possibilities separately.

Suppose that the positron is observed in the outer path (i.e., in detector $O_{+}$).
In this case, what is the effect of the setup on the electron, i.e., what is the likelihood of observing the electron in the detectors $B_{-}$ and $D_{-}$ respectively?
To answer this question, we observe that if the positron took the outer path, then the electron sees a setup very similar to that of the single-particle interference, Experiment~\hyperref[exp3]{S3}.

In this case, quantum physics predicts that the electron is detected only in one of the two detectors, which is the bright detector $B_{-}$.
Thus, if the positron is detected at $O_{+}$, then electrons are detected only at $B_{-}$.
No electrons are observed at the ``dark'' detector $D_{-}$.

Next, let us suppose that the positron is observed in the inner path (i.e., in detector $I_{+}$).
This time, will any electrons be observed in detector $D_{-}$?
That is, what is the likelihood of observing the electron in the detector $B_{-}$ and $D_{-}$?
One may notice that now the situation for the electron is very similar to that of interaction-free measurement (Sec.~\ref{Sec:InteractionFreeMeasurement}).
The original interaction-free measurement argument included a physical obstacle blocking the path of the particle, but in this current case, the obstacle to the particle is its antiparticle meeting the particle in its path.
Because of this obstacle, only electrons that take the outer path are detected.
Following the reasoning of the interaction-free measurement scenario, electrons are detected at both the detectors $B_{-}$ and $D_{-}$ with equal likelihood.

To summarize what we know so far, if the positron is detected in the outer path, then the electrons are detected only at $B_{-}$.
On the other hand, if the positron is observed in the inner path, both $B_{-}$ and $D_{-}$ detect electrons.
Flipping the logic over and focusing on the electron detections, we observe that if electrons are detected in $D_{-}$, then the positron is certainly detected in the inner path.
That is, in case detector $D_{-}$ clicks, then we are certain that $I_{+}$ (and not $O_{+}$) must click as well.

\subsubsection{Experiment~H3}
\label{expH3}

Similar detection patterns would result if the two positron-interferometer beams were combined instead and the electron beams measured without combining (Fig.~\hyperref[Fig:HardyLeftInterfere]{10:H3}).
We call this Experiment~\hyperref[expH3]{H3}.
That is, if the electron is detected in the outer path, then the positron paths can interfere, and all positrons are observed in the bright detector $B_{+}$.
If, alternatively, the electron takes the inner path, then the positron encounters a situation like interaction-free measurement.
In this case, positrons are observed with equal likelihood in both detectors.
Similar to the case of Fig.~\hyperref[Fig:HardyRightInterfere]{10:H2}, observing the positron in $D_{+}$ implies that the electron is detected only in detector~$I_{-}$.

\subsubsection{Experiment~H4}
\label{expH4}

In \hyperref[expH4]{H4}, the final experiment of Hardy's paradox, the paths of both the particle and the antiparticle meet at a beam\hyp splitter before the detectors $B_{+}/D_{+}$ and $B_{-}/D_{-}$ detect the particle and the anti-particle.
The outcome of Experiments \hyperref[expH1]{H1}--\hyperref[expH3]{H3} could be predicted by applying the concepts of single-particle interference and interaction-free measurement in a straight-forward manner.
However, understanding Experiment~\hyperref[expH4]{H4} (Fig.~\hyperref[Fig:HardyBothInterfere]{10:H4}) requires us to carefully analyze our assumptions about the nature of reality.
It is in the context of this setup that the EPR and quantum predictions differ and can be tested in real-life experiments.

\subsection{Predictions from EPR and from Local Hidden Variables}
\label{Subsec:PredictionsEPRLHV}

Before describing the predictions of quantum physics, we will follow a reasoning based on the two seemingly obvious EPR assumptions of realism and locality.
Recall that EPR's first assumption is that if the value of a physical quantity can be predicted with certainty, then there is an element of reality corresponding to this quantity.
EPR's second assumption is that these elements of reality are local, i.e., they do not change instantaneously based on faraway occurrences.

What are the elements of reality in these experiments?
Notice that if the electron is observed in the dark detector $D_{-}$ then the positron is always observed in the inner path.
Consequently, EPR would argue that there is an element of reality attached to the particle that determines which path the positron takes, at least for the case where the electron is detected in $D_{-}$.
Likewise, when the positron is observed in $D_{+}$, then it is known with certainty that the electron will be observed in the inner path.
As a result, EPR would say that there is an element of reality that determines the path taken by the electron, which in this case is the inner path.

Following the reasoning of EPR, these two elements of reality exist and that these elements determine the outcomes of any experiments performed on electron-positron pairs.
What does this mean for Experiment~\hyperref[expH4]{H4}?
According to EPR, if the positron is detected in $D_{+}$ and the electron in $D_{-}$, then the respective elements of reality determine that the electron and the positron both took the inner path.
However, if both of the particles took their respective inner paths, then they must have met and annihilated in each case in which a simultaneous $D_{+}$ and $D_{-}$ measurement would have been expected.
Consequently, the $D_{+}$ and $D_{-}$ detectors should never click simultaneously.

EPR's second assumption, that elements of reality are local, went into the above argument as well.
According to this assumption, there can be no instantaneous communication between the electron and positron because these two particles can be arbitrarily far away during the experiment.
Such instantaneous communication could result in simultaneous $D_{+}$ and $D_{-}$ clicks in the following seemingly contrived, but possible, scenario.
Suppose that the elements of reality also keep track of whether the particle-antiparticle pair is measured at $I_{\pm}$/$O_{\pm}$ or at $D_{\pm}$/$B_{\pm}$, i.e., whether the paths are combined or not.
In this case, as soon as the positron (say) reaches the far end of the interferometer and ``notices'' that its path is combined, it could communicate with the electron to change its properties such that it has some chance of being detected at $D_{-}$.
However, if we assume locality, then this communication is impossible, provided that the measurements on each side are performed sufficiently quickly.
\textit{In summary, if we assume the existence of elements of reality and no instantaneous communication, then simultaneous $D_{+}$ and $D_{-}$ clicks cannot happen.}

% BEGIN \input{table3.tex}
\begin{table}
\newcolumntype{C}[1]{>{\hsize=#1\hsize\centering\arraybackslash}X}%
\begin{tabularx}{0.95\columnwidth}{ C{1}C{1}C{1}C{1} }
  \toprule
  \multicolumn{2}{C{2}}{Positrons} & \multicolumn{2}{C{2}}{Electrons} \\
  \multicolumn{2}{C{2}}{Performed measurements:} & \multicolumn{2}{C{2}}{Performed measurements:} \\
  $I_{+}/O_{+}$ & $B_{+}/D_{+}$ & $I_{-}/O_{-}$ & $B_{-}/D_{-}$ \\
  \midrule[\heavyrulewidth]
 \cellcolor{xgreen!25} $I_{+}$ & $B_{+}$ & \cellcolor{xgreen!25} $I_{-}$ & $B_{-}$ \\
 \cellcolor{xgreen!25} $I_{+}$ & $D_{+}$ & \cellcolor{xgreen!25} $I_{-}$ & $B_{-}$ \\
 \cellcolor{xgreen!25} $I_{+}$ & $B_{+}$ & \cellcolor{xgreen!25} $I_{-}$ & $D_{-}$ \\
 \cellcolor{xgreen!25} $I_{+}$ & $D_{+}$ & \cellcolor{xgreen!25} $I_{-}$ & $D_{-}$ \\
 \midrule
 \cellcolor{xbrown!25}$O_{+}$ & $B_{+}$ & $I_{-}$ & \cellcolor{xblue!25} $D_{-}$ \\
 \cellcolor{xbrown!25}$O_{+}$ & $D_{+}$ & $I_{-}$ & \cellcolor{xblue!25}$D_{-}$ \\
 \cellcolor{xbrown!25}$O_{+}$ & $B_{+}$ & $O_{-}$ & \cellcolor{xblue!25}$D_{-}$ \\
 \midrule
 $I_{+}$ & $\cellcolor{xbrown!25}D_{+}$ & \cellcolor{xblue!25}$O_{-}$ & $B_{-}$ \\
 $I_{+}$ & $\cellcolor{xbrown!25}D_{+}$ & \cellcolor{xblue!25}$O_{-}$ & $D_{-}$ \\
 $O_{+}$ & $\cellcolor{xbrown!25}D_{+}$ & \cellcolor{xblue!25}$O_{-}$ & $D_{-}$ \\
 \midrule
 $O_{+}$ & $B_{+}$ & $I_{-}$ & $B_{-}$ \\
 $O_{+}$ & $D_{+}$ & $I_{-}$ & $B_{-}$ \\
 $I_{+}$ & $B_{+}$ & $O_{-}$ & $B_{-}$ \\
 $I_{+}$ & $B_{+}$ & $O_{-}$ & $D_{-}$ \\
 $O_{+}$ & $B_{+}$ & $O_{-}$ & $B_{-}$ \\
 $O_{+}$ & $D_{+}$ & $O_{-}$ & $B_{-}$ \\
 \bottomrule
\end{tabularx}
\caption{\textbf{Outcomes based on local hidden variable assumption.} The two EPR assumptions are captured by the local hidden variable assumption, which states that the measurement outcomes are known beforehand for each generated electron-positron pair.
In other words, the outcomes of the electron measurements are assumed to be independent of which measurements are performed on the positron and vice versa.
This table lists all possible assignments of the outcomes to the four sets of measurements (inner/outer and bright/dark for the two particles) performed on the pairs.
%There are $2^{4} = 16$ total rows corresponding to the two possible outcomes for each of the four measurements.
The local hidden variable argument rules out simultaneous $D_{-}$-$D_{+}$ clicks as follows.
The first four rows involve inner-inner detections, which result in electron-positron annihilation, and are thus ruled out.
The next three assignments involve simultaneous $D_{-}$-$O_{+}$ detections, which do not arise because of interaction-free measurement.
Analogously, the next three assignments do not arise because of simultaneous $O_{-}$-$D_{+}$ detections, which are ruled out because of interaction free measurements as well.
Only the remaining six measurement outcomes are possible.
These six possibilities do not include any simultaneous $D_{-}$-$D_{+}$ detection.
}
\label{Tab:LHV}
\end{table}
% END \input{table3.tex}

Although these assumptions of realism and locality were first made by EPR, the authors did not realize the full consequences of these assumptions.
It was 29 years before John Bell formalized the EPR argument in terms of local hidden variables, which led him to discover the contradiction between quantum physics and these assumptions and to design decisive tests to settle the EPR versus quantum physics debate.
Bell formalized the two EPR assumptions of realism and locality into the single assumption that local hidden variables (like those presented in Sec.~\ref{Sec:EPR}) can predict all eventual measurement outcomes.

Note that the local hidden variable assumption is stronger than the EPR assumptions.
EPR assume the existence of elements of reality only if predictions with certainty are possible.
In contrast, the local hidden variable assumption states that these elements always exist and underlie all measurement outcomes irrespective of whether these outcomes can be predicted with certainty.
Although stronger than the EPR assumptions, the local hidden variable assumption is in the spirit of Einstein's objections to quantum physics.
Indeed, Einstein's famous words
\begin{quote}
I, at any rate, am convinced that He [God] does not throw dice.
\end{quote}
regarding quantum physics also indicate his inclination towards a hidden-variable approach~\cite{Ballentine1972}.
Einstein himself, after the EPR paper, moved towards phrasing the argument in terms of local hidden variables, without the `prediction with certainty' condition.

The local hidden variable assumption can be stated simply in the context of the Fig.~\ref{Fig:HardyExperiment} setup:
the measurement outcomes of each pair of interfering particles are established at their source.
That is, the information about the eventual measurement outcomes is present in some possibly inaccessible (i.e., hidden) variables, which are set before the particles leave the source.
These hidden variables are local in the sense that outcomes of a particle cannot change based on measurements performed on the other particle.
This independence of outcomes is regarded as locality because the particles could be immensely distant when the measurements are performed on them.
In this case, if the measurements are done quickly, then this independence of hidden variables could only be violated by faster-than-light communication, which is not possible.

We have already seen that the EPR assumptions rule out simultaneous $D_{-}$--$D_{+}$ clicks.
We now show that the local hidden variable assumption leads to the same conclusion.
Because of the hidden-variable assumption, we can list all possible measurement outcomes of the electron-positron pair.
Table~\ref{Tab:LHV} presents this list of possible measurement outcomes.
There are a total of sixteen ($=2\times 2\times 2\times 2$) possible outcomes because each of the four total measurements (namely $I_{-}$/$O_{-}$, $B_{-}$/$D_{-}$, $I_{+}$/$O_{+}$ or $B_{+}$/$D_{+}$) can return one out of two values.

From the analysis of \hyperref[expH1]{H1}, we infer that none of the particle pairs can be assigned outcomes that include simultaneous $I_{-}$--$I_{+}$ events.
This rules out any pair having outcomes depicted in the first four rows of Table~\ref{Tab:LHV}.
Next, we focus on Experiments \hyperref[expH2]{H2} and \hyperref[expH3]{H3}, where the paths were combined for only one of the two interferometers.
As discussed earlier in this section, the logic of interaction-free measurement rules out outcomes which include simultaneous $D_{-}$--$O_{+}$ clicks or simultaneous $O_{-}$--$D_{+}$ clicks.
Thus, rows $5$--$10$ of Table~\ref{Tab:LHV} are eliminated.

This means that only the six remaining measurement outcomes are possible under the local hidden variable assumption.
These six outcomes can be read-off from the bottom six rows of Table~\ref{Tab:LHV}.
Clearly, none of these six rows contains a simultaneous $D_{-}$--$D_{+}$ outcome.
\textit{Thus, the local hidden variable assumption precludes the possibility of simultaneous $D_{-}$--$D_{+}$ clicks.}

% BEGIN \input{Figure_Hardy_Quantum.tex}
\begin{figure}
\begin{subfigure}[t]{0.45\columnwidth}
\includegraphics[width = 0.95\textwidth]{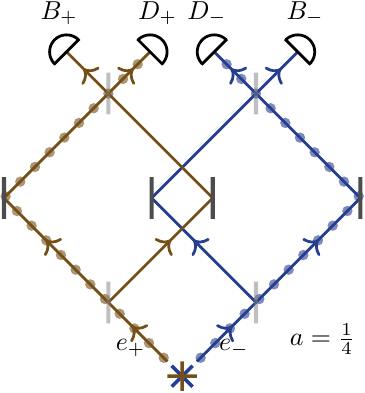}
\caption{}\label{Fig:OuterOuter}
\end{subfigure}
\begin{subfigure}[t]{0.45\columnwidth}
\includegraphics[width = 0.95\textwidth]{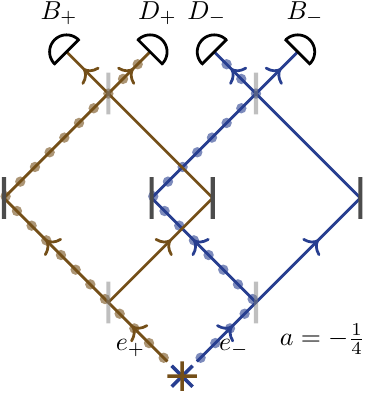}
\caption{}\label{Fig:OuterInner}
\end{subfigure}
\begin{subfigure}[t]{0.45\columnwidth}
\includegraphics[width = 0.95\textwidth]{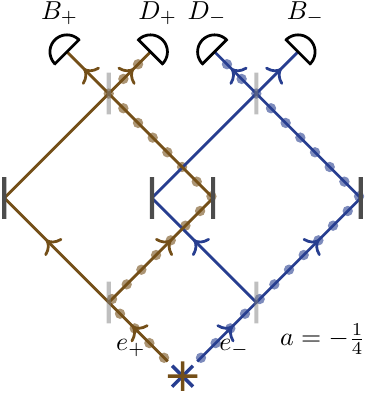}
\caption{}\label{Fig:InnerOuter}
\end{subfigure}
\begin{subfigure}[t]{0.45\columnwidth}
\includegraphics[width = 0.95\textwidth]{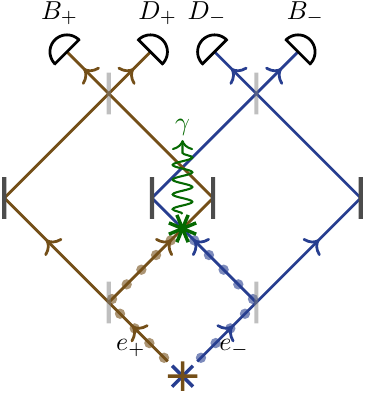}
\caption{}\label{Fig:InnerInner}
\end{subfigure}%
\caption{\textbf{Calculating the probability of simultaneous clicks at $D_{+}$ and $D_{-}$ using quantum theory.}
In quantum mechanics, the particles do not take definite paths but rather see contributions from different histories.
The paths highlighted by the dots in Figures~\ref{Fig:OuterOuter}--\ref{Fig:InnerOuter} represent three histories that result in simultaneous clicks at $D_{+}$ and $D_{-}$.
The history represented in~\ref{Fig:OuterOuter} contributes an amplitude of $a = (1/\sqrt{2})^{4} = 1/4$ because there are four transmission events in total at the four beamsplitters.
Figures~\ref{Fig:OuterInner} and~\ref{Fig:InnerOuter} each contribute an amplitude of $a = (\mathrm{i}/\sqrt{2})^{2}(1/\sqrt{2})^{2} = -1/4$ because of two reflections and two transmissions at the beamsplitters.
The contribution from these histories should be summed and squared to obtain the likelihood $(1/4-1/4-1/4)^{2} = 1/16$ of a simultaneous $D_{+}$-$D_{-}$ detection.
Figure~\ref{Fig:InnerInner} depicts a fourth history that could have resulted in a $D_{+}$-$D_{-}$ detection but does not because of electron-positron annihilation.
}\label{Fig:HardyQuantumPrediction}
\end{figure}% END \input{Figure_Hardy_Quantum.tex}

\subsection{Quantum Predictions}

While the EPR argument, or alternatively the slightly stronger local hidden variable assumption, predicts the absence of simultaneous $D_{+}$ and $D_{-}$ clicks in Experiment~\hyperref[expH4]{H4}, quantum physics predicts that these detectors will indeed click simultaneously in some cases.
We now describe how to determine the prediction of quantum physics based on two-particle interference.
Like we did in Sec.~\ref{Sec:2Particle}, we calculate the prediction of quantum physics by summing over the probability amplitudes of the indistinguishable histories for the two particles and squaring the outcome.

Figs.~\ref{Fig:OuterOuter}--\ref{Fig:InnerOuter} depict the three possible histories of the electron and positron that result in the simultaneous clicking of $D_{-}$ and $D_{+}$. The source contributes a factor of unity to each of the histories since it shoots electron-positron pairs in the two paths in each run of the experiment.
Consider the case of Fig.~\ref{Fig:OuterOuter}, in which both electron and positron take outer paths.
Together the electron and positron undergo a total of four transmissions at the beam\hyp splitters, or a contribution of $1/\sqrt{2}\times 1/\sqrt{2}\times 1/\sqrt{2}\times 1/\sqrt{2} = 1/4$.
The two histories depicted in Figs.~\ref{Fig:OuterInner} and~\ref{Fig:InnerOuter} comprise two reflections and two transmissions.
Thus, their contributions are $-1/4$ each.
There is no possibility of the electron and positron both taking inner paths and being detected since such a case would have resulted in annihilation.
Hence, only the remaining three (outer-outer, inner-outer and outer-inner) histories contribute.
Summing over the contributions and squaring gives a $1/16$ probability of simultaneous $D_{+}$ and $D_{-}$ detection.

To summarize the quantum mechanical prediction, the particles do not take definite paths but see contributions from different histories.
Since there is no inner-inner history, only the other three possibilities contribute giving rise to $1/16$ likelihood of $D_{+}$--$D_{-}$ detections.
Thus, the quantum physics predicts detectors $D_{+}$ and $D_{-}$ clicking simultaneously in $1/16 = 6.25\%$ of all the possible cases.
In contrast, if we assume that the outcomes of experiments are determined by local hidden variables, then there can be no cases of simultaneous detection at $D_{+}$ and $D_{-}$.
Obviously, when Experiment~\hyperref[expH4]{H4} is performed in the lab, either detectors $D_{+}$ and $D_{-}$ sometimes click simultaneously or they don't.
This means that either we must discard the well-established rules of quantum physics or we must give up on the very intuitive idea of local elements of reality.

Years of experimental tests of Bell's original argument and of its many different variants, including the Hardy's variant, agree with the predictions of quantum physics and not with the EPR, or equivalently local hidden variable, predictions.
See Sec.~\ref{Sec:Experiments} for a more detailed account of these experiments.
Thus, we conclude that there are no local elements of reality or local hidden variables. This still leaves us with a choice.
Either there are no elements of reality or hidden variables that determine the outcomes of experiments before they are performed, or nature operates by means of ``non-local'' elements of reality or hidden variables that allow instantaneous communication.

At this stage, one may wonder if and how this strangeness of quantum physics connects with entanglement, which that we described the context of two-particles experiments in Sec.~\ref{Sec:2Particle}.
Indeed, the same entanglement that prevented us from describing two photons as individual entities is behind this strange contradiction between quantum physics and local hidden variables.
This time, though, the entanglement is between the electron and the positron.
Although the electron and the positron started out unentangled when they left their respective sources because there was no correlation between which path each of them is in.
However, correlations were generated subsequently because of the possible annihilation of the electron-positron pair if both particles were in the inner paths of the interferometer.
One can see that this annihilation generates correlations (i.e., mutual dependencies in the particles' paths) as follows: if the electron arrived at its detector after taking the inner path, then we know that the positron took the other path. 
Similarly, if the positron took the inner path, then the electron took the outer path.
This means that the particles are entangled once the histories have crossed the annihilation point.
This entanglement resulted in the detection pattern that led us to conclude that quantum physics is incompatible with local hidden variables.

We can also check that indistinguishability of paths, or histories, is an important condition for observing a departure from local hidden variable predictions.
In our calculations, we obtained the detection probabilities by adding amplitudes from three indistinguishable histories.
If these histories are made distinguishable by adding NDDs in the paths (in analogy to Experiments~\hyperref[exp5]{S5},~\hyperref[exp6]{S6} and~\hyperref[expE5]{E5} before the particles arrive at their respective downstream beam\hyp splitters, then the detection probabilities change.
Although the quantum calculations still predict a small probability of simultaneous detection in $D_{+}$ and $D_{-}$, this probability is not enough to rule out a local hidden variable description if the NDDs are accounted for in the calculations, because the results of the other experiments change as well when NDDs are introduced, since they disturb the interference. 
Remarkably, in the presence of NDDs, quantum physics and local hidden variable theories predict identical probabilities for simultaneous detection in each of the four possible measurement scenarios.
We will see in Sec.~\ref{Sec:Experiments} that this requirement of indistinguishability poses a key challenge in experimentally ruling out local hidden variables and thus verifying quantum physics.

\subsection{Nonlocal Hidden Variables and the Pilot-Wave Theory}

As described above, there are two resolutions to Hardy's paradox: either nature does not operate by means of hidden variables, or these hidden variables are not local.
While most quantum physicists lean towards giving up on the idea of hidden variables entirely~\cite{Schlosshauer2013}, it is possible to construct non-local hidden variable theories that agree with all quantum predictions. The most popular such non-local hidden variable theory is the pilot wave theory, which was introduced by Louis de~Broglie in 1927 and further developed by David Bohm in 1952.
The pilot-wave theory takes both the wave and particle aspects of quantum physics literally: in addition to the particles, there is also a wave that is present everywhere and can interact with the particles.
To match quantum predictions with those of the pilot-wave model, we have to discard locality and necessarily assume that the pilot wave admits nonlocal influences, i.e., it can be influenced instantaneously by distant operations.
Specifically, the measurement made on one particle belonging to an entangled pair instantaneously changes how the pilot wave acts on the other, possibly distant, particle.

Although the pilot wave can change instantaneously based on distant events, it still cannot be used for faster-than-light communication in practice. 
This is because we do not have access to the exact position and location of each particle as these will inevitably be shrouded in randomness.
It turns out that after observing many random events, the average outcomes  remain the same independently of any distant  events.

The pilot-wave theory explains not just the outcomes of experiments \hyperref[expH1]{H1}--\hyperref[expH4]{H4}, but can account for all quantum phenomena.
That is, the predictions of the pilot-wave theory are identical to those of quantum physics.
Hence, it is fair to say that the pilot-wave theory is as correct as quantum physics.
Although we might prefer one over the other as a matter of taste, we cannot tell the difference between their eventual predictions.

\subsection{Summary}

We conclude this section with a summary.
Bell's theorem indicates a deep contradiction between quantum physics and EPR's seemingly obvious assumptions about reality.
Surprisingly, experiments demonstrate quantum physics to be correct and rule out the intuitively appealing possibility of local hidden variables.
This implies that measurement outcomes are either not predetermined at all, or they are determined by nonlocal hidden variables, such as Bohm's pilot wave model.

\section{Experimental Work}
\label{Sec:Experiments}

No theory of physics is meaningful without rigorous verification and testing~\cite{scientificmethod,historyofscience}. 
Throughout history, experimentation has played a key role in shaping and supporting the predictions of physical theories. 
In a theory as radical and counter-intuitive as quantum physics, experimentation has been instrumental not only its validation but also its broad acceptance~\cite{historyofscience,griffiths,gribbin}.
In this section, we discuss some of the key experiments that have justified our support of quantum physics. 

Performing experiments with one or more quantum particles is challenging.
We begin by describing the underlying reason for this challenge, namely quantum decoherence, which destroys interference of single and two particle histories. 
We then introduce some of the properties of photons that enable experiments using these particles.
Next, in support of the reasoning of Secs.~\ref{Sec:OneParticle}--\ref{Sec:Hardy}, we discuss experiments that demonstrate wave\hyp particle duality and the rejection of local hidden variables, the latter via loophole-free tests of Bell's theorem. 
We conclude the section with a short discussion on the technological applications of quantum physics, with focus on information processing.

\subsection{Indistinguishability and the Environment}

Accurate and precise testing of physical theories that lie within the boundaries of day-to-day intuition, such as those that are underpinned by local hidden variables, can be challenging. 
Thus, devising and executing an experiment to test quantum physics, which counters our cherished assumptions about the nature of reality, requires an extra degree of ambition and vigilance. 
Beyond this, an experimental quantum physicist needs to proceed with care due to the fact that quantum particles are easily modified by their environment~\cite{decoherence}. 
This is because it takes only one external, unaccounted, particle to reveal information about a quantum particle that is under test, and hence destroy the indistinguishability of its histories.

To understand this, recall experiments involving NDDs (\hyperref[exp5]{S5}, \hyperref[exp6]{S6}, and \hyperref[expE5]{E5}) and the corresponding experiments in which the NDDs are absent (\hyperref[exp3]{S3} and \hyperref[expE2]{E2}).
These experiments show that quantum interference between histories is observed only when it is not possible to determine which history the photons actually took.
That is, as soon as some other particle obtains information about the histories that each photon took, then the indistinguishability is lost and the interference disappears.

In these examples, the influence comes from the NDD. However, such influences from particles that come from the environment are hard to avoid. That is, as soon as a particle from the environment interacts with a particle under test, a measurement of the environment can reveal the particle's history. This results in a loss of indistinguishability. Remarkably, even if the experimenter does not measure the environment, the different paths are rendered distinguishable. 
These unwanted interactions with the environment, which is referred to as decoherence, is one important reason why it is hard to observe quantum effects.

This loss of indistinguishability from the environment provides a resolution to the famous thought experiment of Schr\"{o}dinger's cat~\citep{schrodinger1935}, which offers the possibility of a cat to be in a quantum superposition of dead and alive.
The existence of the cat in a superposition neglects to account for the fact that the cat will continuously interact with the environment. For example, the cat is breathing air.
%NS: given the comments so far I plan to remove this: or all-pervasive microwave-frequency photons from the early universe~\cite{Durrer2008}. 
This means that, in practice, the cat will only be either dead or alive, and no quantum superposition will be observed. Scientists meticulously isolate their quantum experiments from the environment knowing that it can prevent the observation of quantum strangeness in the laboratory. Such efforts have allowed the observation of wave-particle duality as described below in Sec.~\ref{Sec:ExpHistory}.

\subsection{Photons and their Properties}
\label{Sec:ExpPhotons}

Arguably the most common particle that is used when testing quantum principles is the photon~\cite{zeilinger2010,obrien2014}. Referring to the Hardy experiment of Sec.~\ref{Sec:Hardy}, one should realize that although electron-positron pairs may be produced, they are done so in dedicated, expensive, and bulky particle accelerators that most scientists do not have at their disposal. In contrast, photons can be straightforwardly produced and manipulated, and they interact relatively weakly with each other and the environment. Photons have been employed in many single and two-particle interference experiments that are similar to those described in the single- and two-particles experiments, i.e., Experiments~\hyperref[exp1]{S1}--\hyperref[exp6]{S6}, \hyperref[expE1]{E1}--\hyperref[expE5]{E5}, and \hyperref[expH1]{H1}--\hyperref[expH4]{H4}.

In short, a photon is a particle of the so-called electro-magnetic field as it is described in many texts. Photons are produced when electrically-charged particles oscillate and, in the same way, photons can cause such charged particles to oscillate. This more detailed description of photons was unnecessary in explaining the concepts of wave-particle duality and entanglement of the previous sections, yet becomes useful when attempting to understand the real-life experimental tests of quantum physics that are described in this chapter.

As commonly described in classical physics, polarization denotes the direction of oscillation of the electric component of the light field~\cite{hechtoptics}. As we mentioned, light also has a magnetic component, which is due to the oscillating electric field in conjunction with Faraday's law of induction, but we won't dive into this here~\citep{mandelwolf}. When light is described quantum-mechanically in terms of photons, a similar definition holds, but now there is also the additional possibility of the polarization of light being superposed or entangled~\cite{mandelwolf}. For example, a photon can be polarized at 45$^\circ$ with respect to some reference, e.g. a laboratory bench, which can be equally described by it being in an indistinguishable superposition of being polarized at 0$^\circ$ and 90$^\circ$. 
This situation is completely analogous to that of Experiment~\hyperref[exp3]{S3}, in which the photon was in an indistinguishable superposition of being in the left path and the right path.
Furthermore, pairs of 0$^\circ$-polarized and 90$^\circ$-polarized entangled photons may be produced akin to how the respective pairs of LU-RD and LD-RU entangled photons of Experiment~\hyperref[expE1]{E1} are produced.
Polarization is one of the most widely-employed properties of photons in quantum experiments, including tests of wave-particle duality~\cite{grangier1986} and tests of Bell's theorem~\cite{zeilinger2010,obrien2014,mandel1995,Lundeen2009,freedman,Aspect1982,weihs,boulder,vienna}. This is due to the availability and quality of polarization-sensitive experimental apparatus and the simplicity of polarization-based experiments~\cite{bass}. 

Other widely-employed properties of photons include their path, their energies, or their time of arrival. 
For example, a standard tool in the quantum optics laboratory is the laser, which emits streams of photons that have well-defined energies.
Recall that the experiments that we have discussed in Secs.~\ref{Sec:OneParticle} through~\ref{Sec:Hardy} utilize the path properties of photons.
Path properties have been exploited in the lab for performing loophole-free Bell tests, for example, the Delft experiment that we describe in Sec.~\ref{Sec:loophole} in more detail.
Similarly, the energies and times of arrival have been used for other quantum applications such as quantum communications, which we delve into in more detail in Sec.~\ref{Sec:Applications}.

\subsection{History of Experiments on Wave\hyp Particle Duality}
\label{Sec:ExpHistory}

Before the development of quantum physics, it was widely accepted that light is best described as a wave~\cite{historyofscience}. Although~\citet{newton1704} argued in favor of the interpretation of light as consisting of particles in the 18th century, a famous experiment by Young in 1801 gave experimental support to Huygens' initial idea that light is a wave~\citep{huygens,young}. Drawing from his experience with sound, Young performed an experiment that involved directing a beam of light at a pair of thin slits. The light was then directed to a screen. Upon the screen, Young observed periodic bright and dark bands-- the signature of interference.

His experiment is analogous to the setup of Experiment~\hyperref[exp4]{S4}, except Young's experiment allowed classical light to traverse several different paths, each of a different length. The wave model of light was upheld until Planck proved that light must be absorbed and emitted in discrete amounts of energy in order to explain the spectra of light, i.e., the intensity of light of each color, that is emitted from a hot body~\cite{planck1900}. This idea was followed-up by Einstein, who proposed that the description of light as particles explained the energies and numbers of electrons that are ejected from a metal when bombarded with light. This idea, referred to as the photoelectric effect, earned him a Nobel prize in 1921~\cite{photoelectric}. Experiments by Lenard supported Einstein's explanation of the photoelectric effect in 1907, even though his initial intent was to disprove its existence~\cite{lenard}. 

A 1986 experiment by Grangier, Roger, and Aspect, which used set-ups that are conceptually identical to Experiments~\hyperref[exp1]{S1}, \hyperref[exp3]{S3}, and \hyperref[exp4]{S4}, clearly illustrated wave\hyp particle duality of single photons~\cite{grangier1986}. To understand the experiment, we recall that matter is composed of atoms, which are themselves composed of charged particles, specifically electrons and protons. If a photon impinges on an atom, it can oscillate the atom's constituent charged particles, eliminating the incoming photon. However, this leaves the atom's charged particles oscillating, which can lead to the atom emitting a photon, or many of them that each feature lower energies. 
Thus, by directing a laser beam at a cloud of calcium atoms, Grangier, Roger and Aspect generated a pair of photons, one of which was used for their experiments (i.e.~\hyperref[exp1]{S1}, \hyperref[exp3]{S3}, and \hyperref[exp4]{S4}).

Consistent with the predictions of Experiment~\hyperref[exp1]{S1}, the photon never led to two detectors ($D_1$ and $D_2$) responding simultaneously.
The only exception was a small rate of such coincident detections due to experimental noise. 
This demonstrated the particle character of the photons. To demonstrate their wave character, Grangier and colleagues sent their single photons into a Mach-Zehnder interferometer. Varying its relative path length, they observed a detection rate (at $D_1$ and $D_2$) that varied sinusoidally on a near-zero background, consistent with quantum interference.

As we introduced in Sec.~\ref{Sec:OneParticle}, modern physics now tells us that a photon possesses both wave-like and particle-like characteristics. The wave-like properties are supported by the single-photon version of Young's experiment of Fig.~\hyperref[Fig:Exp3]{S3}, while particle-like properties are revealed by photoelectric experiments, single photon detectors, and Experiments~\hyperref[Fig:Exp1]{S1} and~\hyperref[Fig:Exp2]{S2}. In particular, many experiments have been performed that study the interplay of the wave and particle properties of light, such as the interaction-free measurement of Fig.~\ref{Fig:IFM}. The earliest example of such an experiment is one by~\citet{kwiat1995}. Moreover, the closely-related concept of non-destructive detection is also a widely-studied topic for fundamental studies and practical applications. See, e.g.,~\citet{Grangier1998} and the recent work in~\citet{reiserer2013} as well as the discussion of Sec.~\ref{Sec:Applications}.

In analogy to light, support for a wave\hyp particle duality of matter was initiated by de Broglie~\cite{debroglie}. However, there was previously strong evidence that matter was particle-like. In the 19th century, Dalton deduced the existence of atoms from the proportions of elements that are required for chemical reactions~\cite{dalton}. Furthermore, in 1905, Einstein explained the movements of particles that are suspended in a fluid due to the movement of the fluid's constituent molecules. This phenomena is referred to as Brownian motion after botanist Robert Brown's observation of the trajectory of pollens in water~\cite{brownianmotion}.

The first experimental support for De Broglie's hypothesis that matter can also behave like waves was provided by Davisson and Germer in 1921, who fired electrons at crystalline nickel~\cite{davissongermer}. 
The interaction of the electrons with the periodic array of atoms which constitute the nickel produced alternating bands of high- and low-density electrons in a similar spirit as to how Young's photons interacted with slits to produce periodic bands of light. The first matter-wave experiments that probed the nature of single individual quanta were pioneered by Rauch in 1974, who employed individual neutron particles that interacted with silicon crystal plates to produce interference~\cite{rauch}. Since then, interference of large molecules such as C$_{60}$~\citet{arndt} as well as biologically-significant molecules (e.g. neurotransmitters) have been shown~\citet{geyer}, with a path towards interference of large bio-matter such as proteins~\cite{geyer}. In all of these experiments, many steps were taken to avoid decoherence, e.g. by reducing the amount of radiation from large molecules or the presence of background gases. Indeed, modern quantum physics endorses the existence of matter wave\hyp particle duality in analogy with that of light. 

Beyond quantum duality, a lot of recent scientific interest has focused on the generation of matter that exhibits, in some cases macroscopic~\citep[see for example][]{appel,gross,mcconnell,zarkeshian}, entanglement in order to further probe the boundary of which the rules of classical physics end and those of quantum physics begin, and also to develop new applications.

\subsection{Bell Tests and Hardy's Experiment}
\label{Sec:ExpBell}

Recall from Sec.~\ref{Sec:Hardy} that the predictions of quantum physics for Hardy's experiment are in conflict with the predictions of local hidden variable models. If the electron and positron are governed by local hidden variables, then their annihilation should \textit{never} result in a detection event, i.e. detectors $D_{+}$ and $D_{-}$ of Fig.~\hyperref[expH4]{H4} should not click simultaneously. A healthy skeptic may ask about the role of experimental imperfections in this context. For example, consider the fact that single photon detectors may occasionally produce a supposed detection result, even if no photons are sent to the detector~\cite{detectorreview}. This is called a `dark count'. More importantly, there is an experimental uncertainty, which is referred to as a `standard deviation', that is due to the statistics associated with measurements of any discrete number of quantities, like positrons or photons.

So the question arises: what amount of experimental errors is tolerable in Hardy's or similar experiments to genuinely refute the existence of local hidden variables? This question can be answered by following the approach of Bell~\cite{Bell1964}, as we briefly introduced in Sec.~\ref{Sec:Hardy}. The starting point of this approach is that the Bell test, the Hardy's experiment, and the more experimentally-friendly versions thereof, such as that of Clauser, Horne, Shimony and Holt (CHSH)~\citep{chsh}, all rely not on a single experimental observation, but on the collection of measurements from multiple runs of an experiment. 

Tests of Bell's theorems begin by formulating a mathematical relation that is composed of a set of average values of measurement outcomes of an experiment. 
That is, an experiment is performed repeatedly; the outcomes are recorded; and an average value over these outcomes if performed, for example, on an average, how many events led to simultaneous clicks in two detectors, say $D_{+}$ and $D_{-}$ for the experiment~\hyperref[expH4]{H4}.
Next, this relation is bounded by the assumption that experimental outcomes are predicted by local hidden variables, resulting in a (Bell) inequality~\cite{Bell1964}. For example, consider the table of outcomes based on local hidden variables, Table~\ref{tab:HiddenVariablesEntanglement} (\ref{Tab:LHV}). When an experiment produces results that contradict those that are predicted using local hidden variables, the evaluated relation will surpass the bound, which is referred to as a `violation' of a Bell inequality.

Bell tests are commonly performed using pairs of entangled particles (e.g. photons, electrons, etc.), in which one out of a set of two possible measurements is performed on each particle. For a Bell test using Hardy's experiment, these correspond to either the $I_+/O_+$ or the $B_+/D_+$ experiment for the positron, and analogously for the electron. The results of these measurements, when combined appropriately, can demonstrate a violation of a Bell inequality~\cite{irvine}. Thus, to answer the skeptic, a Bell inequality establishes a limit to the amount of experimental uncertainty that can be tolerated in an experiment in order for it to conclusively disprove the existence of local hidden variables. Often the amount of violation of Bell's inequality is denoted by the number of standard deviations of violation, with more being the most convincing.

Bell's theorem was tested for the first time in 1972, in an experiment by Freedman and Clauser~\cite{freedman}. They used pairs of polarization-entangled photons that were emitted from individual calcium atoms. 
It turns out that the unique properties of calcium atoms led to the entanglement between the polarizations of photons. Yet, despite a violation, and the impressive fact that they used duct tape and spare parts to construct their experiment, there was suspicion that their results could be explained from experimental imperfections. Added doubt was cast by a concurrent experiment by Holt and Pipkin that did not violate a Bell inequality~\cite{holt,pipkin}. Stronger support of quantum predictions was provided in the early 1980s by a series of experiments by Aspect and colleagues~\cite{aspect1981,Aspect1982,aspecttime}. One of their achievements was to improve the calcium-based photon source of Freedman and Clauser to efficiently produce stable polarization-entangled photons. Recall from Sec.~\ref{Sec:ExpHistory} that they also used this source to generate single photons. As a result, good statistical accuracy was achieved, leading to a violation of a Bell inequality by a respectable five standard deviations~\cite{aspecttime}.

\subsection{Closing Loopholes in Bell Tests}
\label{Sec:loophole}

Another achievement of Aspect's 1982 experiment was to address a loophole that was present in all previous experiments~\cite{aspecttime}. A loophole refers to the possibility of explaining the violation of a Bell inequality using a local hidden variable model, and is possible if the assumptions that underpin the formulation of the Bell inequality are not met in an experiment. In order to establish \textit{locality} conditions for a Bell test using two particles, an experimenter must ensure that the choice of measurement that is performed on the first particle cannot even in principle be communicated to the second particle and vice versa~\cite{aspecttime,weihs,loopholesreview}. To envision this, one may imagine that a signal traveling at the speed of light (nothing can travel faster according to the theory of relativity) is emitted by the measurement apparatus for the first particle as soon as the choice of measurement has been made. If this signal were to reach the second particle, then it could, in principle, alter the behavior of the second particle such that its measurement could result in a violation of Bell's theorem. The magnitude of the speed of light implies that for a `locality-loophole free' test of a Bell inequality the measurements must be chosen very quickly.

Aspect's experiment was the first to address the locality loophole by independently and rapidly choosing measurement settings after the photons were emitted~\cite{aspecttime}. Nonetheless, his experiments involved manipulating the measurement settings by using a predictable switching mechanism. (Specifically he deflected photons by using periodically-driven sound waves). This was not a completely satisfactory solution for closing the loophole since, in principle, the source might have somehow `learned' that the settings were predictable, and adjusted the particles' hidden variables accordingly to still violate a Bell inequality. See the discussion of the `fair sampling' loophole below. It was not until 1998 that Weihs and colleagues definitively closed the locality loophole~\cite{weihs}. This was achieved by separating two entangled photons by 400 meters to increase the time available for switching measurements and by using a random number generator to decide the measurement settings. In fact, their random number generator is a device that operates conceptually similar to the setup of Fig.~\hyperref[exp1]{S1}: the probability of detecting a photon at $D_{1}$ or $D_2$ is completely random. The measurement was performed using a robust and high-rate source of polarization-entangled photons that were generated by a process called spontaneous parametric down-conversion, which was well-mastered at the time of the experiment. This process is similarly explained by interaction of light with atoms, yet is less experimentally cumbersome than the calcium-based approaches that we discussed above.

Beyond locality, the so-called \textit{fair sampling} loophole must also be closed~\cite{rowe,loopholesreview}. This refers to the requirement that the particles which are detected in a Bell test should accurately represent all the particles that are generated during the course of the experiment. If particles can somehow decide to not be detected depending on the measurement settings, then the ones that are detected could lead to an apparent violation a Bell inequality in a world that may be completely described by local hidden variables. Thus, it is imperative to detect enough particles in any given experiment to refute a local hidden variable description of reality. Fortunately, for certain Bell inequalities, such as the CH-Eberhard inequality, a total detection efficiency of only at least $2/3$ is needed to close this loophole~\cite{CH, eberhard}. Nonetheless, it is not trivial to achieve such detection efficiencies for photons, especially since photon loss (e.g. due to experimental apparatus) have to be taken into account. For example, the detection efficiency was only a few percent in the Bell experiment by Weihs et al.~\cite{weihs}.

The fair sampling loophole was first closed by Rowe, Wineland (who won the Nobel prize in 2012), and others in 2000 by manipulating a pair of positively-charged beryllium ions~\cite{rowe}. Ions are atoms that not electrically neutral, often due to a lack of electrons. For example, neutral Beryllium atoms have four electrons, while Rowe's ions each contained only three. Without going into detail, the configuration of the charged particles of each ion were entangled or, in other words, the ions were energy-entangled. This was accomplished using a a complex scheme of exciting the ions with laser light in conjunction with the fact that like-ions repel from each other (recall that while like-charged particles repel, differently-charged particles attract). 

Once the ions were entangled, measurements were accomplished by shining laser light on each ion and detecting their emitted photons. In essence, this allows a measurement of the energy of each ion in order to perform the Bell test. The key to closing the loophole came from the fact that Beryllium ions feature a special `cycling' configuration. This referred to the fact that the ions always returned to their original entangled configuration after attempting a measurement (i.e. shining laser light on them and waiting for the detection of a photon). This meant that even if no photons are detected due to loss, the experimenter can try again and again until photons are detected. Thus, all of the energy-entangled ions that are created in the experiment can be measured. Rowe et al's approach allowed the violation of a CHSH-Bell inequality by more than eight standard deviations. Although complex, this experiment illustrates its advantage over the photon-photon entanglement experiments that require low amounts of photon loss.

It took until a 2015 experiment at Delft to close both loopholes simultaneously~\cite{hensen}. This experiment used impurities in diamond, and a clever method to entangle them. (Impurities are collections of charged particles that react to light similarly as atoms or ions do). Two individual impurities, separated by 1.3 kilometers, were independently excited using laser light such that, with equal probability, they could each emit a photon towards a beam\hyp splitter that is situated halfway between the impurities. Due to the fact that each of these photons are indistinguishable, a single photon detection event at the output of the beam\hyp splitter ensured that the impurities became entangled. Specifically, each impurity became energy-entangled similar to that of the ions in Rowe et al's experiment. This so-called `event-ready' scheme allowed the fair sampling loophole to be closed because the detection event heralded the situation in which both impurities had become energy-entangled in their `cycling' configuration~\cite{zukowski,simon}. Thus, similar to Rowe and colleagues, measurements were performed by shining laser light on each impurity and detecting the photon that was emitted, i.e. measuring the energies of each impurity. However, in contrast to Rowe's experiment, locality was enforced due to the separation of the impurities.

The overall very low numbers of photons that were detected (only 245 events were recorded in total) resulted in a violation of the CHSH-Bell inequality by only two standard deviations. This result might have induced a skeptic to doubt the conclusion on purely statistical grounds. However, soon after, experiments in Vienna and Boulder also closed both loopholes~\citep{vienna, boulder}. These experiments utilized polarization-entangled photon pairs that were produced by parametric down-conversion. To ensure locality, the Vienna experiment separated the photons by over 30 meters, while the Boulder experiment had this separation at more than 100 meters. Both experiments utilized a CH-Eberhard-type inequality to ensure fair sampling. The Vienna and Boulder experiments violated this inequality by 11 and 7 standard deviations, respectively, removing any potentially lingering doubts based on statistics. 
Both experiments utilized rapidly-switchable polarization-selective filters, high-speed electronics, and exploited low-loss single-photon detectors. 
After these three experiments, there is now broad consensus in the scientific community that local hidden variables have been definitively ruled out by experiments.

\subsection{Applications}
\label{Sec:Applications}

Besides their fundamental interest, quantum phenomena are also important for practical applications (see e.g.~\citet{laser,transistor,economist,nielsen}). One phenomenon that is important for the semiconductor industry is quantum tunneling~\citep{hund,transistor}, in which an electron can be in a superposition of being transmitted through a charged barrier or not. Global positioning systems rely on clocks which `tick' at a rate defined by the discretized charge configuration of atoms~\citep{gps,atomicclock}.

More directly related to the topics that are discussed in this manuscript, quantum information processing aims to exploit wave\hyp particle duality and entanglement to process information in novel ways~\cite{nielsen}. This idea was first suggested by Manin and separately Feynman, who envisioned using certain controlled quantum particles to simulate complex quantum dynamics and interactions~\cite{manin,feynman}. By the mid 1980s, scientists were beginning to consider the possibility of quantum particles to be able to process and manipulate information in ways that classical objects cannot. For example, in 1985, Deutsch devised a quantum algorithm that, based on the properties of entanglement, is exponentially faster than any possible classical algorithm~\cite{deutsch,nielsencourse}. In 1984, Bennett and Brassard showed that, by exploiting wave\hyp particle duality, a secret key can be established between two parties if information is encoded into a single quantum particle, e.g. by encoding bits using the polarization of a photon~\cite{bb84}. Furthermore in 1993, scientists showed how to transmit quantum properties using entanglement~\cite{bennett}, using a method known as quantum teleportation, with the first experimental demonstration following in 1997~\cite{bouwmeester}. These discoveries spawned the research fields of quantum computing, quantum cryptography, and quantum communication respectively, which are all sub-fields of quantum information science~\cite{nielsen}. Recently, individual impurities in crystals have been shown to be promising as sensors of external stimuli (e.g. heat, electric and magnetic fields), spawning the field of quantum sensing~\cite{degen}. These discoveries have produced a multitude of experiments, including earth-to-satellite quantum communications~\cite{pan}, and have resulted in industrial efforts as well as commercially-available products, e.g. quantum random number generators~\cite{economist}.

\section{Discussion}
\label{Sec:Discussion}
In terms of the disparity between the levels of our practical and philosophical understanding of our theories about the natural world, quantum physics stands out as arguably the starkest example. The theory has resisted attempts at being intuitively internalized by even its discoverers and practicing experts, let alone by the wider human culture.
But some creative physicists and philosophers of the last few decades have demonstrated that an intuitive understanding of quantum physics may not be altogether impossible.

In this manuscript, we followed the lead of these pioneers and aimed to describe two key surprises of quantum physics via simple thought experiments that are based on interference.
First, we described how the interference of a single particle displays the properties of both waves and particles, yet is somehow different from either.
The case of interaction\hyp free measurement, wherein a photon detects the presence of an obstacle without actually interacting with it, exemplifies quantum weirdness in the single\hyp particle case.

Although these single\hyp particle phenomena seem counter\hyp intuitive, it is possible to explain them using the seemingly intuitive concepts of local hidden variables.
The second key surprise of quantum physics is that the outcomes of two\hyp particle interference are incompatible with the notion of local hidden variables.

We followed Hardy's approach~\cite{Hardy1992} to Bell's theorem, and showed how the expected outcomes predicted by quantum interference are fundamentally at odds with the outcomes expected from a local hidden variable approach.
Although it is possible to explain the outcomes of quantum physics using hidden variables, these explanations inevitably involve instantaneous communication between particles.

These two surprising aspects of quantum physics have been verified over decades of experimental effort, beginning with the 1982 experiments of Aspect, Roger and Grangier~\cite{Aspect1982} that demonstrate wave-particle duality of single particle, followed by the first tests of Bell's version of quantum weirdness, and culminating in recent loophole\hyp free tests of the second key surprise.

Looking towards the future, let us mention several significant open questions involving quantum physics. They are all questions regarding its ultimate domain of applicability and relevance.

Quantum physics is an essential part of the framework for the standard model of particle physics, which describes electromagnetism as well as the weak and strong interactions. Despite the efforts of many physicists over several decades, it is not clear yet whether quantum principles also apply to gravity~\cite{Smolin2008}.

A different, if potentially related, question is whether quantum principles apply to macroscopic objects, or whether there is some scale of mass, size or complexity where they cease to be valid. New experiments are constantly being developed to extend the domain in which quantum effects have been demonstrated~\cite{Arndt2014}.
So far quantum physics has withstood all of these challenges.

A fascinating open question is to what extent quantum principles play a role in biological environments~\cite{lambert2013quantum,mohseni2014quantum}.
Many scientists are investigating potential roles for quantum effects in photosynthesis~\cite{engel2007evidence,romero2014quantum} and in the ability of birds (and other animals) to sense magnetic fields~\cite{ritz2000model,hiscock2016quantum}.

Finally, both scientists and philosophers have wondered since the early days of quantum physics whether quantum concepts might help us understand the question of the relationship between mind and matter~\cite{marshall1989consciousness,lockwood1989mind,penrose1994shadows,stapp2011mindful}, with recent new proposals~\cite{fisher2015quantum,kumar2016possible} that are at least partly inspired by the dramatic progress in quantum information science.

\begin{acknowledgments}
C.S.\ would like to thank Rob Thompson for supporting the creation of the ``Quantum Mysteries and Paradoxes'' course that led to this manuscript, many students who took the course for their questions and feedback, and Eleanor Sayre for the suggestion on how to make this manuscript happen. 
All authors would like to thank Phil Langill for piloting this text as a support material for his class. 
We also thank Shreya P.~Kumar, Phil Langill, H\'{e}l\`{e}ne Ollivier, and Borzumehr Toloui for many helpful comments. N.S. is funded by AQT’s Intelligent Quantum Networks and Technologies  (INQNET) research program.   
\end{acknowledgments}
\vspace{1cm}
{\bf Author contributions:}
Section I was written mostly by V.N.\ and C.S., Section II by S.W., P.Z., A.P., I.D., and C.S., Section III by A.D'S.\ and I.D.,
Section IV by I.D.\ and A.D'S., section V by I.D., Section VI by N.S., and Section VII by V.N.\ and C.S. All authors edited and commented on multiple sections. The project was originally conceived by C.S., but grew significantly beyond his initial expectations through the contributions of all the authors.

%merlin.mbs apsrmp4-1.bst 2010-07-25 4.21a (PWD, AO, DPC) hacked
%Control: author (3) reversed first dotless

%\bibliography{main}

\begin{thebibliography}{99}%
\makeatletter
\providecommand \@ifxundefined [1]{%
 \@ifx{#1\undefined}
}%
\providecommand \@ifnum [1]{%
 \ifnum #1\expandafter \@firstoftwo
 \else \expandafter \@secondoftwo
 \fi
}%
\providecommand \@ifx [1]{%
 \ifx #1\expandafter \@firstoftwo
 \else \expandafter \@secondoftwo
 \fi
}%
\providecommand \natexlab [1]{#1}%
\providecommand \enquote  [1]{``#1''}%
\providecommand \bibnamefont  [1]{#1}%
\providecommand \bibfnamefont [1]{#1}%
\providecommand \citenamefont [1]{#1}%
\providecommand \href@noop [0]{\@secondoftwo}%
\providecommand \href [0]{\begingroup \@sanitize@url \@href}%
\providecommand \@href[1]{\@@startlink{#1}\@@href}%
\providecommand \@@href[1]{\endgroup#1\@@endlink}%
\providecommand \@sanitize@url [0]{\catcode `\\12\catcode `\$12\catcode
  `\&12\catcode `\#12\catcode `\^12\catcode `\_12\catcode `\%12\relax}%
\providecommand \@@startlink[1]{}%
\providecommand \@@endlink[0]{}%
\providecommand \url  [0]{\begingroup\@sanitize@url \@url }%
\providecommand \@url [1]{\endgroup\@href {#1}{\urlprefix }}%
\providecommand \urlprefix  [0]{URL }%
\providecommand \Eprint [0]{\href }%
\providecommand \doibase [0]{http://dx.doi.org/}%
\providecommand \selectlanguage [0]{\@gobble}%
\providecommand \bibinfo  [0]{\@secondoftwo}%
\providecommand \bibfield  [0]{\@secondoftwo}%
\providecommand \translation [1]{[#1]}%
\providecommand \BibitemOpen [0]{}%
\providecommand \bibitemStop [0]{}%
\providecommand \bibitemNoStop [0]{.\EOS\space}%
\providecommand \EOS [0]{\spacefactor3000\relax}%
\providecommand \BibitemShut  [1]{\csname bibitem#1\endcsname}%
\let\auto@bib@innerbib\@empty
%</preamble>
\bibitem [{\citenamefont {Appel}\ \emph {et~al.}(2009)\citenamefont {Appel},
  \citenamefont {Windpassinger}, \citenamefont {Oblak}, \citenamefont {Hoff},
  \citenamefont {Kj{\ae}rgaard},\ and\ \citenamefont {Polzik}}]{appel}%
  \BibitemOpen
  \bibfield  {author} {\bibinfo {author} {\bibnamefont {Appel}, \bibfnamefont
  {J{\"u}rgen}}, \bibinfo {author} {\bibfnamefont {Patrick~J.}\ \bibnamefont
  {Windpassinger}}, \bibinfo {author} {\bibfnamefont {Daniel}\ \bibnamefont
  {Oblak}}, \bibinfo {author} {\bibfnamefont {Ulrich~B.}\ \bibnamefont {Hoff}},
  \bibinfo {author} {\bibfnamefont {Niels}\ \bibnamefont {Kj{\ae}rgaard}}, \
  and\ \bibinfo {author} {\bibfnamefont {Eugene~S.}\ \bibnamefont {Polzik}}}
  (\bibinfo {year} {2009}),\ \bibfield  {title} {\enquote {\bibinfo {title}
  {Mesoscopic atomic entanglement for precision measurements beyond the
  standard quantum limit},}\ }\href {10.1073/pnas.0901550106} {\bibfield
  {journal} {\bibinfo  {journal} {Proceedings of the National Academy of
  Sciences}\ }\textbf {\bibinfo {volume} {106}}~(\bibinfo {number} {27}),\
  \bibinfo {pages} {10960}}\BibitemShut {NoStop}%
\bibitem [{\citenamefont {Arndt}\ and\ \citenamefont
  {Hornberger}(2014)}]{Arndt2014}%
  \BibitemOpen
  \bibfield  {author} {\bibinfo {author} {\bibnamefont {Arndt}, \bibfnamefont
  {Markus}}, \ and\ \bibinfo {author} {\bibfnamefont {Klaus}\ \bibnamefont
  {Hornberger}}} (\bibinfo {year} {2014}),\ \bibfield  {title} {\enquote
  {\bibinfo {title} {Testing the limits of quantum mechanical
  superpositions},}\ }\href {https://www.nature.com/articles/nphys2863}
  {\bibfield  {journal} {\bibinfo  {journal} {Nature Physics}\ }\textbf
  {\bibinfo {volume} {10}}~(\bibinfo {number} {4}),\ \bibinfo {pages}
  {271}}\BibitemShut {NoStop}%
\bibitem [{\citenamefont {Arndt}\ \emph {et~al.}(1999)\citenamefont {Arndt},
  \citenamefont {Nairz}, \citenamefont {Vos-Andreae}, \citenamefont {Keller},
  \citenamefont {van~der Zouw},\ and\ \citenamefont {Zeilinger}}]{arndt}%
  \BibitemOpen
  \bibfield  {author} {\bibinfo {author} {\bibnamefont {Arndt}, \bibfnamefont
  {Markus}}, \bibinfo {author} {\bibfnamefont {Olaf}\ \bibnamefont {Nairz}},
  \bibinfo {author} {\bibfnamefont {Julian}\ \bibnamefont {Vos-Andreae}},
  \bibinfo {author} {\bibfnamefont {Claudia}\ \bibnamefont {Keller}}, \bibinfo
  {author} {\bibfnamefont {Gerbrand}\ \bibnamefont {van~der Zouw}}, \ and\
  \bibinfo {author} {\bibfnamefont {Anton}\ \bibnamefont {Zeilinger}}}
  (\bibinfo {year} {1999}),\ \bibfield  {title} {\enquote {\bibinfo {title}
  {Wave-particle duality of {C}$_{60}$ molecules},}\ }\href
  {http://dx.doi.org/10.1038/44348} {\bibfield  {journal} {\bibinfo  {journal}
  {Nature}\ }\textbf {\bibinfo {volume} {401}},\ \bibinfo {pages}
  {680}}\BibitemShut {NoStop}%
\bibitem [{\citenamefont {Aspect}\ \emph
  {et~al.}(1982{\natexlab{a}})\citenamefont {Aspect}, \citenamefont
  {Dalibard},\ and\ \citenamefont {Roger}}]{aspecttime}%
  \BibitemOpen
  \bibfield  {author} {\bibinfo {author} {\bibnamefont {Aspect}, \bibfnamefont
  {Alain}}, \bibinfo {author} {\bibfnamefont {Jean}\ \bibnamefont {Dalibard}},
  \ and\ \bibinfo {author} {\bibfnamefont {G\'erard}\ \bibnamefont {Roger}}}
  (\bibinfo {year} {1982}{\natexlab{a}}),\ \bibfield  {title} {\enquote
  {\bibinfo {title} {{Experimental Test of Bell's Inequalities Using
  Time-Varying Analyzers}},}\ }\href {\doibase 10.1103/PhysRevLett.49.1804}
  {\bibfield  {journal} {\bibinfo  {journal} {Physical Review Letters}\
  }\textbf {\bibinfo {volume} {49}},\ \bibinfo {pages} {1804}}\BibitemShut
  {NoStop}%
\bibitem [{\citenamefont {Aspect}\ \emph {et~al.}(1981)\citenamefont {Aspect},
  \citenamefont {Grangier},\ and\ \citenamefont {Roger}}]{aspect1981}%
  \BibitemOpen
  \bibfield  {author} {\bibinfo {author} {\bibnamefont {Aspect}, \bibfnamefont
  {Alain}}, \bibinfo {author} {\bibfnamefont {Philippe}\ \bibnamefont
  {Grangier}}, \ and\ \bibinfo {author} {\bibfnamefont {G\'erard}\ \bibnamefont
  {Roger}}} (\bibinfo {year} {1981}),\ \bibfield  {title} {\enquote {\bibinfo
  {title} {{Experimental Tests of Realistic Local Theories via Bell's
  Theorem}},}\ }\href {\doibase 10.1103/PhysRevLett.47.460} {\bibfield
  {journal} {\bibinfo  {journal} {Physical Review Letters}\ }\textbf {\bibinfo
  {volume} {47}},\ \bibinfo {pages} {460}}\BibitemShut {NoStop}%
\bibitem [{\citenamefont {Aspect}\ \emph
  {et~al.}(1982{\natexlab{b}})\citenamefont {Aspect}, \citenamefont
  {Grangier},\ and\ \citenamefont {Roger}}]{Aspect1982}%
  \BibitemOpen
  \bibfield  {author} {\bibinfo {author} {\bibnamefont {Aspect}, \bibfnamefont
  {Alain}}, \bibinfo {author} {\bibfnamefont {Philippe}\ \bibnamefont
  {Grangier}}, \ and\ \bibinfo {author} {\bibfnamefont {G\'erard}\ \bibnamefont
  {Roger}}} (\bibinfo {year} {1982}{\natexlab{b}}),\ \bibfield  {title}
  {\enquote {\bibinfo {title} {{Experimental Realization of
  Einstein-Podolsky-Rosen-Bohm Gedankenexperiment: A New Violation of Bell's
  Inequalities}},}\ }\href {\doibase 10.1103/PhysRevLett.49.91} {\bibfield
  {journal} {\bibinfo  {journal} {Physical Review Letters}\ }\textbf {\bibinfo
  {volume} {49}},\ \bibinfo {pages} {91}}\BibitemShut {NoStop}%
\bibitem [{\citenamefont {Ballentine}(1972)}]{Ballentine1972}%
  \BibitemOpen
  \bibfield  {author} {\bibinfo {author} {\bibnamefont {Ballentine},
  \bibfnamefont {Lelslie~E}}} (\bibinfo {year} {1972}),\ \bibfield  {title}
  {\enquote {\bibinfo {title} {{Einstein's Interpretation of Quantum
  Mechanics}},}\ }\href {\doibase 10.1119/1.1987060} {\bibfield  {journal}
  {\bibinfo  {journal} {American Journal of Physics}\ }\textbf {\bibinfo
  {volume} {40}}~(\bibinfo {number} {12}),\ \bibinfo {pages}
  {1763}}\BibitemShut {NoStop}%
\bibitem [{\citenamefont {Bardeen}\ and\ \citenamefont
  {Brattain}(1948)}]{transistor}%
  \BibitemOpen
  \bibfield  {author} {\bibinfo {author} {\bibnamefont {Bardeen}, \bibfnamefont
  {John}}, \ and\ \bibinfo {author} {\bibfnamefont {Walter~H.}\ \bibnamefont
  {Brattain}}} (\bibinfo {year} {1948}),\ \bibfield  {title} {\enquote
  {\bibinfo {title} {{The Transistor, A Semi-Conductor Triode}},}\ }\href
  {https://journals.aps.org/pr/abstract/10.1103/PhysRev.74.230} {\bibfield
  {journal} {\bibinfo  {journal} {Physical Review}\ }\textbf {\bibinfo {volume}
  {74}}~(\bibinfo {number} {2}),\ \bibinfo {pages} {230}}\BibitemShut {NoStop}%
\bibitem [{\citenamefont {Bass}\ \emph {et~al.}(2009)\citenamefont {Bass},
  \citenamefont {DeCusatis}, \citenamefont {Enoch}, \citenamefont
  {Lakshminarayanan}, \citenamefont {Li}, \citenamefont {Macdonald},
  \citenamefont {Mahajan},\ and\ \citenamefont {Van~Stryland}}]{bass}%
  \BibitemOpen
  \bibfield  {author} {\bibinfo {author} {\bibnamefont {Bass}, \bibfnamefont
  {Michael}}, \bibinfo {author} {\bibfnamefont {Casimer}\ \bibnamefont
  {DeCusatis}}, \bibinfo {author} {\bibfnamefont {Jay}\ \bibnamefont {Enoch}},
  \bibinfo {author} {\bibfnamefont {Vasudevan}\ \bibnamefont
  {Lakshminarayanan}}, \bibinfo {author} {\bibfnamefont {Guifang}\ \bibnamefont
  {Li}}, \bibinfo {author} {\bibfnamefont {Carolyn}\ \bibnamefont {Macdonald}},
  \bibinfo {author} {\bibfnamefont {Virendra}\ \bibnamefont {Mahajan}}, \ and\
  \bibinfo {author} {\bibfnamefont {Eric}\ \bibnamefont {Van~Stryland}}}
  (\bibinfo {year} {2009}),\ \href
  {https://www.mhprofessionalresources.com//handbookofoptics/vol1.php} {\emph
  {\bibinfo {title} {Handbook of Optics, Volume I: Geometrical and Physical
  Optics, Polarized Light, Components and Instruments (set)}}}\ (\bibinfo
  {publisher} {McGraw-Hill, Inc.})\BibitemShut {NoStop}%
\bibitem [{\citenamefont {Bell}(1964)}]{Bell1964}%
  \BibitemOpen
  \bibfield  {author} {\bibinfo {author} {\bibnamefont {Bell}, \bibfnamefont
  {John}}} (\bibinfo {year} {1964}),\ \bibfield  {title} {\enquote {\bibinfo
  {title} {{On the Einstein Podolsky Rosen Paradox}},}\ }\href
  {https://doi.org/10.1142/9789812386540_0002} {\bibfield  {journal} {\bibinfo
  {journal} {Physics}\ }\textbf {\bibinfo {volume} {1}}~(\bibinfo {number}
  {3}),\ \bibinfo {pages} {195}}\BibitemShut {NoStop}%
\bibitem [{\citenamefont {Bennett}\ and\ \citenamefont
  {Brassard}(1984)}]{bb84}%
  \BibitemOpen
  \bibfield  {author} {\bibinfo {author} {\bibnamefont {Bennett}, \bibfnamefont
  {Charles~H}}, \ and\ \bibinfo {author} {\bibfnamefont {Gilles}\ \bibnamefont
  {Brassard}}} (\bibinfo {year} {1984}),\ \bibfield  {title} {\enquote
  {\bibinfo {title} {Quantum cryptography: Public key distribution and coin
  tossing},}\ }in\ \href {https://trove.nla.gov.au/version/11518604} {\emph
  {\bibinfo {booktitle} {Proc. IEEE International Conference on Computers,
  Systems and Signal Processing, Bangalore, India, 1984}}},\ p.\ \bibinfo
  {pages} {175}\BibitemShut {NoStop}%
\bibitem [{\citenamefont {Bennett}\ \emph {et~al.}(1993)\citenamefont
  {Bennett}, \citenamefont {Brassard}, \citenamefont {Cr{\'e}peau},
  \citenamefont {Jozsa}, \citenamefont {Peres},\ and\ \citenamefont
  {Wootters}}]{bennett}%
  \BibitemOpen
  \bibfield  {author} {\bibinfo {author} {\bibnamefont {Bennett}, \bibfnamefont
  {Charles~H}}, \bibinfo {author} {\bibfnamefont {Gilles}\ \bibnamefont
  {Brassard}}, \bibinfo {author} {\bibfnamefont {Claude}\ \bibnamefont
  {Cr{\'e}peau}}, \bibinfo {author} {\bibfnamefont {Richard}\ \bibnamefont
  {Jozsa}}, \bibinfo {author} {\bibfnamefont {Asher}\ \bibnamefont {Peres}}, \
  and\ \bibinfo {author} {\bibfnamefont {William~K.}\ \bibnamefont {Wootters}}}
  (\bibinfo {year} {1993}),\ \bibfield  {title} {\enquote {\bibinfo {title}
  {{Teleporting an Unknown Quantum State via Dual Classical and
  Einstein-Podolsky-Rosen Channels}},}\ }\href
  {https://journals.aps.org/prl/abstract/10.1103/PhysRevLett.70.1895}
  {\bibfield  {journal} {\bibinfo  {journal} {Physical Review Letters}\
  }\textbf {\bibinfo {volume} {70}}~(\bibinfo {number} {13}),\ \bibinfo {pages}
  {1895}}\BibitemShut {NoStop}%
\bibitem [{\citenamefont {Billings}(2018)}]{pan}%
  \BibitemOpen
  \bibfield  {author} {\bibinfo {author} {\bibnamefont {Billings},
  \bibfnamefont {Lee}}} (\bibinfo {year} {2018}),\ \href@noop {} {\enquote
  {\bibinfo {title} {{China Shatters ``Spooky Action at a Distance'' Record,
  Preps for Quantum Internet}},}\ }\bibinfo {howpublished}
  {\url{https://www.scientificamerican.com/article/china-shatters-ldquo-spooky-action-at-a-distance-
  rdquo-record-preps-for-quantum-internet/}},\ \bibinfo {note} {[Online;
  accessed 21-February-2018]}\BibitemShut {NoStop}%
\bibitem [{\citenamefont {Bouwmeester}\ \emph {et~al.}(1997)\citenamefont
  {Bouwmeester}, \citenamefont {Pan}, \citenamefont {Mattle}, \citenamefont
  {Eibl}, \citenamefont {Weinfurter},\ and\ \citenamefont
  {Zeilinger}}]{bouwmeester}%
  \BibitemOpen
  \bibfield  {author} {\bibinfo {author} {\bibnamefont {Bouwmeester},
  \bibfnamefont {Dik}}, \bibinfo {author} {\bibfnamefont {Jian-Wei}\
  \bibnamefont {Pan}}, \bibinfo {author} {\bibfnamefont {Klaus}\ \bibnamefont
  {Mattle}}, \bibinfo {author} {\bibfnamefont {Manfred}\ \bibnamefont {Eibl}},
  \bibinfo {author} {\bibfnamefont {Harald}\ \bibnamefont {Weinfurter}}, \ and\
  \bibinfo {author} {\bibfnamefont {Anton}\ \bibnamefont {Zeilinger}}}
  (\bibinfo {year} {1997}),\ \bibfield  {title} {\enquote {\bibinfo {title}
  {Experimental quantum teleportation},}\ }\href
  {https://www.nature.com/articles/37539} {\bibfield  {journal} {\bibinfo
  {journal} {Nature}\ }\textbf {\bibinfo {volume} {390}}~(\bibinfo {number}
  {6660}),\ \bibinfo {pages} {575}}\BibitemShut {NoStop}%
\bibitem [{\citenamefont {Christensen}\ and\ \citenamefont
  {Kwiat}(2017)}]{christensen}%
  \BibitemOpen
  \bibfield  {author} {\bibinfo {author} {\bibnamefont {Christensen},
  \bibfnamefont {Bradley~G}}, \ and\ \bibinfo {author} {\bibfnamefont
  {Paul~G.}\ \bibnamefont {Kwiat}}} (\bibinfo {year} {2017}),\ \enquote
  {\bibinfo {title} {{Nonlocality and Quantum Cakes, Revisited}},}\ in\ \href
  {\doibase 10.1007/978-3-319-38987-5_25} {\emph {\bibinfo {booktitle} {Quantum
  [Un]Speakables II: Half a Century of Bell's Theorem}}},\ \bibinfo {editor}
  {edited by\ \bibinfo {editor} {\bibfnamefont {Reinhold}\ \bibnamefont
  {Bertlmann}}\ and\ \bibinfo {editor} {\bibfnamefont {Anton}\ \bibnamefont
  {Zeilinger}}}\ (\bibinfo  {publisher} {Springer International Publishing})\
  p.\ \bibinfo {pages} {415}\BibitemShut {NoStop}%
\bibitem [{\citenamefont {Clauser}\ and\ \citenamefont {Horne}(1974)}]{CH}%
  \BibitemOpen
  \bibfield  {author} {\bibinfo {author} {\bibnamefont {Clauser}, \bibfnamefont
  {John~F}}, \ and\ \bibinfo {author} {\bibfnamefont {Michael~A.}\ \bibnamefont
  {Horne}}} (\bibinfo {year} {1974}),\ \bibfield  {title} {\enquote {\bibinfo
  {title} {Experimental consequences of objective local theories},}\ }\href
  {https://journals.aps.org/prd/abstract/10.1103/PhysRevD.10.526} {\bibfield
  {journal} {\bibinfo  {journal} {Physical Review D}\ }\textbf {\bibinfo
  {volume} {10}}~(\bibinfo {number} {2}),\ \bibinfo {pages} {526}}\BibitemShut
  {NoStop}%
\bibitem [{\citenamefont {Clauser}\ \emph {et~al.}(1969)\citenamefont
  {Clauser}, \citenamefont {Horne}, \citenamefont {Shimony},\ and\
  \citenamefont {Holt}}]{chsh}%
  \BibitemOpen
  \bibfield  {author} {\bibinfo {author} {\bibnamefont {Clauser}, \bibfnamefont
  {John~F}}, \bibinfo {author} {\bibfnamefont {Michael~A.}\ \bibnamefont
  {Horne}}, \bibinfo {author} {\bibfnamefont {Abner}\ \bibnamefont {Shimony}},
  \ and\ \bibinfo {author} {\bibfnamefont {Richard~A.}\ \bibnamefont {Holt}}}
  (\bibinfo {year} {1969}),\ \bibfield  {title} {\enquote {\bibinfo {title}
  {{Proposed Experiment to Test Local Hidden-Variable Theories}},}\ }\href
  {https://journals.aps.org/prl/abstract/10.1103/PhysRevLett.23.880} {\bibfield
   {journal} {\bibinfo  {journal} {Physical Review Letters}\ }\textbf {\bibinfo
  {volume} {23}}~(\bibinfo {number} {15}),\ \bibinfo {pages} {880}}\BibitemShut
  {NoStop}%
\bibitem [{\citenamefont {Dalton}(1806)}]{dalton}%
  \BibitemOpen
  \bibfield  {author} {\bibinfo {author} {\bibnamefont {Dalton}, \bibfnamefont
  {John}}} (\bibinfo {year} {1806}),\ \href
  {https://doi.org/10.1080/14786440608563325} {\emph {\bibinfo {title} {On the
  absorption of gases by water and other liquids}}}\BibitemShut {NoStop}%
\bibitem [{\citenamefont {Davisson}\ and\ \citenamefont
  {Germer}(1928)}]{davissongermer}%
  \BibitemOpen
  \bibfield  {author} {\bibinfo {author} {\bibnamefont {Davisson},
  \bibfnamefont {Clinton~J}}, \ and\ \bibinfo {author} {\bibfnamefont
  {Lester~H.}\ \bibnamefont {Germer}}} (\bibinfo {year} {1928}),\ \bibfield
  {title} {\enquote {\bibinfo {title} {{Reflection and Refraction of Electrons
  by a Crystal of Nickel}},}\ }\href {https://doi.org/10.1073/pnas.14.8.619}
  {\bibfield  {journal} {\bibinfo  {journal} {Proceedings of the National
  Academy of Sciences}\ }\textbf {\bibinfo {volume} {14}}~(\bibinfo {number}
  {8}),\ \bibinfo {pages} {619}}\BibitemShut {NoStop}%
\bibitem [{\citenamefont {De~Broglie}(1924)}]{debroglie}%
  \BibitemOpen
  \bibfield  {author} {\bibinfo {author} {\bibnamefont {De~Broglie},
  \bibfnamefont {Louis}}} (\bibinfo {year} {1924}),\ \emph {\bibinfo {title}
  {Recherches sur la th{\'e}orie des quanta}},\ \href
  {https://tel.archives-ouvertes.fr/tel-00006807/document} {Ph.D. thesis}\
  (\bibinfo  {school} {Migration-universit{\'e} en cours
  d'affectation})\BibitemShut {NoStop}%
\bibitem [{\citenamefont {Degen}\ \emph {et~al.}(2017)\citenamefont {Degen},
  \citenamefont {Reinhard},\ and\ \citenamefont {Cappellaro}}]{degen}%
  \BibitemOpen
  \bibfield  {author} {\bibinfo {author} {\bibnamefont {Degen}, \bibfnamefont
  {Christian~L}}, \bibinfo {author} {\bibfnamefont {F.}~\bibnamefont
  {Reinhard}}, \ and\ \bibinfo {author} {\bibfnamefont {P.}~\bibnamefont
  {Cappellaro}}} (\bibinfo {year} {2017}),\ \bibfield  {title} {\enquote
  {\bibinfo {title} {Quantum sensing},}\ }\href
  {https://journals.aps.org/rmp/abstract/10.1103/RevModPhys.89.035002}
  {\bibfield  {journal} {\bibinfo  {journal} {Reviews of Modern Physics}\
  }\textbf {\bibinfo {volume} {89}}~(\bibinfo {number} {3}),\ \bibinfo {pages}
  {035002}}\BibitemShut {NoStop}%
\bibitem [{\citenamefont {Derbes}(1996)}]{derbes1996feynman}%
  \BibitemOpen
  \bibfield  {author} {\bibinfo {author} {\bibnamefont {Derbes}, \bibfnamefont
  {David}}} (\bibinfo {year} {1996}),\ \bibfield  {title} {\enquote {\bibinfo
  {title} {{Feynman's derivation of the Schr{\"o}dinger equation}},}\ }\href
  {https://doi.org/10.1119/1.18114} {\bibfield  {journal} {\bibinfo  {journal}
  {American Journal of Physics}\ }\textbf {\bibinfo {volume} {64}}~(\bibinfo
  {number} {7}),\ \bibinfo {pages} {881}}\BibitemShut {NoStop}%
\bibitem [{\citenamefont {Deutsch}(1985)}]{deutsch}%
  \BibitemOpen
  \bibfield  {author} {\bibinfo {author} {\bibnamefont {Deutsch}, \bibfnamefont
  {David}}} (\bibinfo {year} {1985}),\ \bibfield  {title} {\enquote {\bibinfo
  {title} {Quantum theory, the church--turing principle and the universal
  quantum computer},}\ }in\ \href {0.1098/rspa.1985.0070} {\emph {\bibinfo
  {booktitle} {Proceedings of the Royal Society of London A}}},\ Vol.\ \bibinfo
  {volume} {400}\ (\bibinfo {organization} {The Royal Society})\ p.~\bibinfo
  {pages} {97}\BibitemShut {NoStop}%
\bibitem [{\citenamefont {Eberhard}(1993)}]{eberhard}%
  \BibitemOpen
  \bibfield  {author} {\bibinfo {author} {\bibnamefont {Eberhard},
  \bibfnamefont {Philippe~H}}} (\bibinfo {year} {1993}),\ \bibfield  {title}
  {\enquote {\bibinfo {title} {{Background level and counter efficiencies
  required for a loophole-free Einstein-Podolsky-Rosen experiment}},}\ }\href
  {https://journals.aps.org/pra/abstract/10.1103/PhysRevA.47.R747} {\bibfield
  {journal} {\bibinfo  {journal} {Physical Review A}\ }\textbf {\bibinfo
  {volume} {47}}~(\bibinfo {number} {2}),\ \bibinfo {pages} {R747}}\BibitemShut
  {NoStop}%
\bibitem [{\citenamefont {Ede}\ and\ \citenamefont
  {Cormack}(2016)}]{historyofscience}%
  \BibitemOpen
  \bibfield  {author} {\bibinfo {author} {\bibnamefont {Ede}, \bibfnamefont
  {A}}, \ and\ \bibinfo {author} {\bibfnamefont {L.B.}\ \bibnamefont
  {Cormack}}} (\bibinfo {year} {2016}),\ \href
  {https://utorontopress.com/us/a-history-of-science-in-society-9} {\emph
  {\bibinfo {title} {A History of Science in Society: From Philosophy to
  Utility, Third Edition}}}\ (\bibinfo  {publisher} {University of Toronto
  Press})\BibitemShut {NoStop}%
\bibitem [{\citenamefont {Einstein}(1905{\natexlab{a}})}]{brownianmotion}%
  \BibitemOpen
  \bibfield  {author} {\bibinfo {author} {\bibnamefont {Einstein},
  \bibfnamefont {Albert}}} (\bibinfo {year} {1905}{\natexlab{a}}),\ \bibfield
  {title} {\enquote {\bibinfo {title} {{{\"U}ber die von der
  molekularkinetischen Theorie der W{\"a}rme geforderte Bewegung von in
  ruhenden Fl{\"u}ssigkeiten suspendierten Teilchen}},}\ }\href
  {https://doi.org/10.1002/andp.19053220806} {\bibfield  {journal} {\bibinfo
  {journal} {Annalen der Physik}\ }\textbf {\bibinfo {volume} {322}}~(\bibinfo
  {number} {8}),\ \bibinfo {pages} {549}}\BibitemShut {NoStop}%
\bibitem [{\citenamefont {Einstein}(1905{\natexlab{b}})}]{photoelectric}%
  \BibitemOpen
  \bibfield  {author} {\bibinfo {author} {\bibnamefont {Einstein},
  \bibfnamefont {Albert}}} (\bibinfo {year} {1905}{\natexlab{b}}),\ \bibfield
  {title} {\enquote {\bibinfo {title} {{{\"U}ber einen die Erzeugung und
  Verwandlung des Lichtes betreffenden heuristischen Gesichtspunkt}},}\ }\href
  {https://doi.org/10.1002/andp.19053220607} {\bibfield  {journal} {\bibinfo
  {journal} {Annalen der Physik}\ }\textbf {\bibinfo {volume} {322}}~(\bibinfo
  {number} {6}),\ \bibinfo {pages} {132}}\BibitemShut {NoStop}%
\bibitem [{\citenamefont {Einstein}\ \emph {et~al.}(1935)\citenamefont
  {Einstein}, \citenamefont {Podolsky},\ and\ \citenamefont
  {Rosen}}]{Einstein1935}%
  \BibitemOpen
  \bibfield  {author} {\bibinfo {author} {\bibnamefont {Einstein},
  \bibfnamefont {Albert}}, \bibinfo {author} {\bibfnamefont {Boris}\
  \bibnamefont {Podolsky}}, \ and\ \bibinfo {author} {\bibfnamefont {Nathan}\
  \bibnamefont {Rosen}}} (\bibinfo {year} {1935}),\ \bibfield  {title}
  {\enquote {\bibinfo {title} {{Can Quantum-Mechanical Description of Physical
  Reality Be Considered Complete?}}}\ }\href {\doibase 10.1103/PhysRev.47.777}
  {\bibfield  {journal} {\bibinfo  {journal} {Physical. Review}\ }\textbf
  {\bibinfo {volume} {47}},\ \bibinfo {pages} {777}}\BibitemShut {NoStop}%
\bibitem [{\citenamefont {Eisaman}\ \emph {et~al.}(2011)\citenamefont
  {Eisaman}, \citenamefont {Fan}, \citenamefont {Migdall},\ and\ \citenamefont
  {Polyakov}}]{detectorreview}%
  \BibitemOpen
  \bibfield  {author} {\bibinfo {author} {\bibnamefont {Eisaman}, \bibfnamefont
  {Matthew~D}}, \bibinfo {author} {\bibfnamefont {Jingyun}\ \bibnamefont
  {Fan}}, \bibinfo {author} {\bibfnamefont {Alan}\ \bibnamefont {Migdall}}, \
  and\ \bibinfo {author} {\bibfnamefont {Sergey~V.}\ \bibnamefont {Polyakov}}}
  (\bibinfo {year} {2011}),\ \bibfield  {title} {\enquote {\bibinfo {title}
  {{Invited Review Article: Single-photon sources and detectors}},}\ }\href
  {\doibase 10.1063/1.3610677} {\bibfield  {journal} {\bibinfo  {journal}
  {Review of Scientific Instruments}\ }\textbf {\bibinfo {volume}
  {82}}~(\bibinfo {number} {7}),\ \bibinfo {pages} {071101}}\BibitemShut
  {NoStop}%
\bibitem [{\citenamefont {Engel}\ \emph {et~al.}(2007)\citenamefont {Engel},
  \citenamefont {Calhoun}, \citenamefont {Read}, \citenamefont {Ahn},
  \citenamefont {Man{\v{c}}al}, \citenamefont {Cheng}, \citenamefont
  {Blankenship},\ and\ \citenamefont {Fleming}}]{engel2007evidence}%
  \BibitemOpen
  \bibfield  {author} {\bibinfo {author} {\bibnamefont {Engel}, \bibfnamefont
  {Gregory~S}}, \bibinfo {author} {\bibfnamefont {Tessa~R.}\ \bibnamefont
  {Calhoun}}, \bibinfo {author} {\bibfnamefont {Elizabeth~L.}\ \bibnamefont
  {Read}}, \bibinfo {author} {\bibfnamefont {Tae-Kyu}\ \bibnamefont {Ahn}},
  \bibinfo {author} {\bibfnamefont {Tom{\'a}{\v{s}}}\ \bibnamefont
  {Man{\v{c}}al}}, \bibinfo {author} {\bibfnamefont {Yuan-Chung}\ \bibnamefont
  {Cheng}}, \bibinfo {author} {\bibfnamefont {Robert~E.}\ \bibnamefont
  {Blankenship}}, \ and\ \bibinfo {author} {\bibfnamefont {Graham~R.}\
  \bibnamefont {Fleming}}} (\bibinfo {year} {2007}),\ \bibfield  {title}
  {\enquote {\bibinfo {title} {Evidence for wavelike energy transfer through
  quantum coherence in photosynthetic systems},}\ }\href
  {https://www.nature.com/articles/nature05678} {\bibfield  {journal} {\bibinfo
   {journal} {Nature}\ }\textbf {\bibinfo {volume} {446}}~(\bibinfo {number}
  {7137}),\ \bibinfo {pages} {782}}\BibitemShut {NoStop}%
\bibitem [{\citenamefont {Essen}\ and\ \citenamefont
  {Parry}(1955)}]{atomicclock}%
  \BibitemOpen
  \bibfield  {author} {\bibinfo {author} {\bibnamefont {Essen}, \bibfnamefont
  {Louis}}, \ and\ \bibinfo {author} {\bibfnamefont {Jack~V.L.}\ \bibnamefont
  {Parry}}} (\bibinfo {year} {1955}),\ \bibfield  {title} {\enquote {\bibinfo
  {title} {{An Atomic Standard of Frequency and Time Interval: A Caesium
  Resonator}},}\ }\href {https://www.nature.com/articles/176280a0} {\bibfield
  {journal} {\bibinfo  {journal} {Nature}\ }\textbf {\bibinfo {volume}
  {176}}~(\bibinfo {number} {4476}),\ \bibinfo {pages} {280}}\BibitemShut
  {NoStop}%
\bibitem [{\citenamefont {Feynman}(1982)}]{feynman}%
  \BibitemOpen
  \bibfield  {author} {\bibinfo {author} {\bibnamefont {Feynman}, \bibfnamefont
  {Richard~P}}} (\bibinfo {year} {1982}),\ \bibfield  {title} {\enquote
  {\bibinfo {title} {{Simulating Physics with Computers}},}\ }\href
  {https://link.springer.com/article/10.1007%2FBF02650179} {\bibfield
  {journal} {\bibinfo  {journal} {International Journal of Theoretical
  Physics}\ }\textbf {\bibinfo {volume} {21}}~(\bibinfo {number} {6-7}),\
  \bibinfo {pages} {467}}\BibitemShut {NoStop}%
\bibitem [{\citenamefont {Feynman}\ \emph {et~al.}(2010)\citenamefont
  {Feynman}, \citenamefont {Hibbs},\ and\ \citenamefont
  {Styer}}]{feynman2010quantum}%
  \BibitemOpen
  \bibfield  {author} {\bibinfo {author} {\bibnamefont {Feynman}, \bibfnamefont
  {Richard~P}}, \bibinfo {author} {\bibfnamefont {Albert~R.}\ \bibnamefont
  {Hibbs}}, \ and\ \bibinfo {author} {\bibfnamefont {Daniel~F.}\ \bibnamefont
  {Styer}}} (\bibinfo {year} {2010}),\ \href
  {http://store.doverpublications.com/0486477223.html} {\emph {\bibinfo {title}
  {{Quantum Mechanics and Path Integrals}}}}\ (\bibinfo  {publisher} {Dover
  Publications})\BibitemShut {NoStop}%
\bibitem [{\citenamefont {Fisher}(2015)}]{fisher2015quantum}%
  \BibitemOpen
  \bibfield  {author} {\bibinfo {author} {\bibnamefont {Fisher}, \bibfnamefont
  {Matthew~PA}}} (\bibinfo {year} {2015}),\ \bibfield  {title} {\enquote
  {\bibinfo {title} {{Quantum Cognition: The possibility of processing with
  nuclear spins in the brain}},}\ }\href
  {https://doi.org/10.1016/j.aop.2015.08.020} {\bibfield  {journal} {\bibinfo
  {journal} {Annals of Physics}\ }\textbf {\bibinfo {volume} {362}},\ \bibinfo
  {pages} {593}}\BibitemShut {NoStop}%
\bibitem [{\citenamefont {Freedman}\ and\ \citenamefont
  {Clauser}(1972)}]{freedman}%
  \BibitemOpen
  \bibfield  {author} {\bibinfo {author} {\bibnamefont {Freedman},
  \bibfnamefont {Stuart~J}}, \ and\ \bibinfo {author} {\bibfnamefont {John~F.}\
  \bibnamefont {Clauser}}} (\bibinfo {year} {1972}),\ \bibfield  {title}
  {\enquote {\bibinfo {title} {{Experimental Test of Local Hidden-Variable
  Theories}},}\ }\href {https://doi.org/10.1103/PhysRevLett.28.938} {\bibfield
  {journal} {\bibinfo  {journal} {Physical Review Letters}\ }\textbf {\bibinfo
  {volume} {28}}~(\bibinfo {number} {14}),\ \bibinfo {pages} {938}}\BibitemShut
  {NoStop}%
\bibitem [{\citenamefont {Geyer}\ \emph {et~al.}(2016)\citenamefont {Geyer},
  \citenamefont {Sezer}, \citenamefont {Rodewald}, \citenamefont {Mairhofer},
  \citenamefont {D{\"o}rre}, \citenamefont {Haslinger}, \citenamefont
  {Eibenberger}, \citenamefont {Brand},\ and\ \citenamefont {Arndt}}]{geyer}%
  \BibitemOpen
  \bibfield  {author} {\bibinfo {author} {\bibnamefont {Geyer}, \bibfnamefont
  {Philipp}}, \bibinfo {author} {\bibfnamefont {Ugur}\ \bibnamefont {Sezer}},
  \bibinfo {author} {\bibfnamefont {Jonas}\ \bibnamefont {Rodewald}}, \bibinfo
  {author} {\bibfnamefont {Lukas}\ \bibnamefont {Mairhofer}}, \bibinfo {author}
  {\bibfnamefont {Nadine}\ \bibnamefont {D{\"o}rre}}, \bibinfo {author}
  {\bibfnamefont {Philipp}\ \bibnamefont {Haslinger}}, \bibinfo {author}
  {\bibfnamefont {Sandra}\ \bibnamefont {Eibenberger}}, \bibinfo {author}
  {\bibfnamefont {Christian}\ \bibnamefont {Brand}}, \ and\ \bibinfo {author}
  {\bibfnamefont {Markus}\ \bibnamefont {Arndt}}} (\bibinfo {year} {2016}),\
  \bibfield  {title} {\enquote {\bibinfo {title} {Perspectives for quantum
  interference with biomolecules and biomolecular clusters},}\ }\href
  {http://iopscience.iop.org/article/10.1088/0031-8949/91/6/063007/meta}
  {\bibfield  {journal} {\bibinfo  {journal} {Physica Scripta}\ }\textbf
  {\bibinfo {volume} {91}}~(\bibinfo {number} {6}),\ \bibinfo {pages}
  {063007}}\BibitemShut {NoStop}%
\bibitem [{\citenamefont {Gill}(2017)}]{economist}%
  \BibitemOpen
  \bibfield  {author} {\bibinfo {author} {\bibnamefont {Gill}, \bibfnamefont
  {Patrick}}} (\bibinfo {year} {2017}),\ \href
  {https://www.economist.com/news/essays/21717782-quantum-technology-beginning-come-its-own}
  {\enquote {\bibinfo {title} {Technology quarterly: Quantum technology is
  beginning to come into its own},}\ }\bibinfo {howpublished} {The
  Economist}\BibitemShut {NoStop}%
\bibitem [{\citenamefont {Giustina}\ \emph {et~al.}(2015)\citenamefont
  {Giustina}, \citenamefont {Versteegh}, \citenamefont {Wengerowsky},
  \citenamefont {Handsteiner}, \citenamefont {Hochrainer}, \citenamefont
  {Phelan}, \citenamefont {Steinlechner}, \citenamefont {Kofler}, \citenamefont
  {Larsson}, \citenamefont {Abell\'an}, \citenamefont {Amaya}, \citenamefont
  {Pruneri}, \citenamefont {Mitchell}, \citenamefont {Beyer}, \citenamefont
  {Gerrits}, \citenamefont {Lita}, \citenamefont {Shalm}, \citenamefont {Nam},
  \citenamefont {Scheidl}, \citenamefont {Ursin}, \citenamefont {Wittmann},\
  and\ \citenamefont {Zeilinger}}]{vienna}%
  \BibitemOpen
  \bibfield  {author} {\bibinfo {author} {\bibnamefont {Giustina},
  \bibfnamefont {Marissa}}, \bibinfo {author} {\bibfnamefont {Marijn A.~M.}\
  \bibnamefont {Versteegh}}, \bibinfo {author} {\bibfnamefont {S\"oren}\
  \bibnamefont {Wengerowsky}}, \bibinfo {author} {\bibfnamefont {Johannes}\
  \bibnamefont {Handsteiner}}, \bibinfo {author} {\bibfnamefont {Armin}\
  \bibnamefont {Hochrainer}}, \bibinfo {author} {\bibfnamefont {Kevin}\
  \bibnamefont {Phelan}}, \bibinfo {author} {\bibfnamefont {Fabian}\
  \bibnamefont {Steinlechner}}, \bibinfo {author} {\bibfnamefont {Johannes}\
  \bibnamefont {Kofler}}, \bibinfo {author} {\bibfnamefont {Jan-\AA{}ke}\
  \bibnamefont {Larsson}}, \bibinfo {author} {\bibfnamefont {Carlos}\
  \bibnamefont {Abell\'an}}, \bibinfo {author} {\bibfnamefont {Waldimar}\
  \bibnamefont {Amaya}}, \bibinfo {author} {\bibfnamefont {Valerio}\
  \bibnamefont {Pruneri}}, \bibinfo {author} {\bibfnamefont {Morgan~W.}\
  \bibnamefont {Mitchell}}, \bibinfo {author} {\bibfnamefont {J\"orn}\
  \bibnamefont {Beyer}}, \bibinfo {author} {\bibfnamefont {Thomas}\
  \bibnamefont {Gerrits}}, \bibinfo {author} {\bibfnamefont {Adriana~E.}\
  \bibnamefont {Lita}}, \bibinfo {author} {\bibfnamefont {Lynden~K.}\
  \bibnamefont {Shalm}}, \bibinfo {author} {\bibfnamefont {Sae~Woo}\
  \bibnamefont {Nam}}, \bibinfo {author} {\bibfnamefont {Thomas}\ \bibnamefont
  {Scheidl}}, \bibinfo {author} {\bibfnamefont {Rupert}\ \bibnamefont {Ursin}},
  \bibinfo {author} {\bibfnamefont {Bernhard}\ \bibnamefont {Wittmann}}, \ and\
  \bibinfo {author} {\bibfnamefont {Anton}\ \bibnamefont {Zeilinger}}}
  (\bibinfo {year} {2015}),\ \bibfield  {title} {\enquote {\bibinfo {title}
  {{Significant-Loophole-Free Test of Bell's Theorem with Entangled
  Photons}},}\ }\href {\doibase 10.1103/PhysRevLett.115.250401} {\bibfield
  {journal} {\bibinfo  {journal} {Physical Review Letters}\ }\textbf {\bibinfo
  {volume} {115}},\ \bibinfo {pages} {250401}}\BibitemShut {NoStop}%
\bibitem [{\citenamefont {Grangier}\ \emph {et~al.}(1998)\citenamefont
  {Grangier}, \citenamefont {Levenson},\ and\ \citenamefont
  {Poizat}}]{Grangier1998}%
  \BibitemOpen
  \bibfield  {author} {\bibinfo {author} {\bibnamefont {Grangier},
  \bibfnamefont {Philippe}}, \bibinfo {author} {\bibfnamefont {Juan~Ariel}\
  \bibnamefont {Levenson}}, \ and\ \bibinfo {author} {\bibfnamefont
  {Jean-Philippe}\ \bibnamefont {Poizat}}} (\bibinfo {year} {1998}),\ \bibfield
   {title} {\enquote {\bibinfo {title} {Quantum non-demolition measurements in
  optics},}\ }\href {https://www.nature.com/articles/25059} {\bibfield
  {journal} {\bibinfo  {journal} {Nature}\ }\textbf {\bibinfo {volume}
  {396}}~(\bibinfo {number} {6711}),\ \bibinfo {pages} {537}}\BibitemShut
  {NoStop}%
\bibitem [{\citenamefont {Grangier}\ \emph {et~al.}(1986)\citenamefont
  {Grangier}, \citenamefont {Roger},\ and\ \citenamefont
  {Aspect}}]{grangier1986}%
  \BibitemOpen
  \bibfield  {author} {\bibinfo {author} {\bibnamefont {Grangier},
  \bibfnamefont {Philippe}}, \bibinfo {author} {\bibfnamefont {Gerard}\
  \bibnamefont {Roger}}, \ and\ \bibinfo {author} {\bibfnamefont {Alain}\
  \bibnamefont {Aspect}}} (\bibinfo {year} {1986}),\ \bibfield  {title}
  {\enquote {\bibinfo {title} {{Experimental Evidence for a Photon
  Anticorrelation Effect on a Beam Splitter: A New Light on Single-Photon
  Interferences}},}\ }\href
  {http://iopscience.iop.org/article/10.1209/0295-5075/1/4/004/meta} {\bibfield
   {journal} {\bibinfo  {journal} {EPL (Europhysics Letters)}\ }\textbf
  {\bibinfo {volume} {1}}~(\bibinfo {number} {4}),\ \bibinfo {pages}
  {173}}\BibitemShut {NoStop}%
\bibitem [{\citenamefont {Gribbin}(1984)}]{gribbin}%
  \BibitemOpen
  \bibfield  {author} {\bibinfo {author} {\bibnamefont {Gribbin}, \bibfnamefont
  {John}}} (\bibinfo {year} {1984}),\ \href@noop {} {\emph {\bibinfo {title}
  {In search of Schrodinger's cat: Quantum physics and reality}}}\ (\bibinfo
  {publisher} {Bantam})\BibitemShut {NoStop}%
\bibitem [{\citenamefont {Griffiths}(2016)}]{griffiths}%
  \BibitemOpen
  \bibfield  {author} {\bibinfo {author} {\bibnamefont {Griffiths},
  \bibfnamefont {David~J}}} (\bibinfo {year} {2016}),\ \href@noop {} {\emph
  {\bibinfo {title} {Introduction to quantum mechanics}}}\ (\bibinfo
  {publisher} {Cambridge University Press})\BibitemShut {NoStop}%
\bibitem [{\citenamefont {Gross}\ \emph {et~al.}(2012)\citenamefont {Gross},
  \citenamefont {Zibold}, \citenamefont {Nicklas}, \citenamefont {Esteve},\
  and\ \citenamefont {Oberthaler}}]{gross}%
  \BibitemOpen
  \bibfield  {author} {\bibinfo {author} {\bibnamefont {Gross}, \bibfnamefont
  {Christian}}, \bibinfo {author} {\bibfnamefont {Tilman}\ \bibnamefont
  {Zibold}}, \bibinfo {author} {\bibfnamefont {Eike}\ \bibnamefont {Nicklas}},
  \bibinfo {author} {\bibfnamefont {Jerome}\ \bibnamefont {Esteve}}, \ and\
  \bibinfo {author} {\bibfnamefont {Markus~K.}\ \bibnamefont {Oberthaler}}}
  (\bibinfo {year} {2012}),\ \bibfield  {title} {\enquote {\bibinfo {title}
  {Nonlinear atom interferometer surpasses classical precision limit},}\ }\href
  {https://www.nature.com/articles/nature08919} {\bibfield  {journal} {\bibinfo
   {journal} {Nature}\ }\textbf {\bibinfo {volume} {464}}~(\bibinfo {number}
  {7292}),\ \bibinfo {pages} {1165}}\BibitemShut {NoStop}%
\bibitem [{\citenamefont {Hardy}(1992)}]{Hardy1992}%
  \BibitemOpen
  \bibfield  {author} {\bibinfo {author} {\bibnamefont {Hardy}, \bibfnamefont
  {Lucien}}} (\bibinfo {year} {1992}),\ \bibfield  {title} {\enquote {\bibinfo
  {title} {{Quantum mechanics, local realistic theories, and Lorentz-invariant
  realistic theories}},}\ }\href {\doibase 10.1103/PhysRevLett.68.2981}
  {\bibfield  {journal} {\bibinfo  {journal} {Physical Review Letters}\
  }\textbf {\bibinfo {volume} {68}},\ \bibinfo {pages} {2981}}\BibitemShut
  {NoStop}%
\bibitem [{\citenamefont {Hecht}(2016)}]{hechtoptics}%
  \BibitemOpen
  \bibfield  {author} {\bibinfo {author} {\bibnamefont {Hecht}, \bibfnamefont
  {Eugene}}} (\bibinfo {year} {2016}),\ \href
  {https://www.pearson.com/us/higher-education/program/Hecht-Optics-5th-Edition/PGM45350.html}
  {\emph {\bibinfo {title} {Optics}}}\ (\bibinfo  {publisher} {Pearson
  Education})\BibitemShut {NoStop}%
\bibitem [{\citenamefont {Hensen}\ \emph {et~al.}(2015)\citenamefont {Hensen},
  \citenamefont {Bernien}, \citenamefont {Dr{\'e}au}, \citenamefont {Reiserer},
  \citenamefont {Kalb}, \citenamefont {Blok}, \citenamefont {Ruitenberg},
  \citenamefont {Vermeulen}, \citenamefont {Schouten}, \citenamefont
  {Abell{\'a}n}, \citenamefont {Amaya}, \citenamefont {Pruneri}, \citenamefont
  {Mitchell}, \citenamefont {Markham}, \citenamefont {Twitchen}, \citenamefont
  {Elkouss}, \citenamefont {Wehner}, \citenamefont {Taminiau},\ and\
  \citenamefont {Hanson}}]{hensen}%
  \BibitemOpen
  \bibfield  {author} {\bibinfo {author} {\bibnamefont {Hensen}, \bibfnamefont
  {B}}, \bibinfo {author} {\bibfnamefont {H.}~\bibnamefont {Bernien}}, \bibinfo
  {author} {\bibfnamefont {A.~E.}\ \bibnamefont {Dr{\'e}au}}, \bibinfo {author}
  {\bibfnamefont {A.}~\bibnamefont {Reiserer}}, \bibinfo {author}
  {\bibfnamefont {N.}~\bibnamefont {Kalb}}, \bibinfo {author} {\bibfnamefont
  {M.~S.}\ \bibnamefont {Blok}}, \bibinfo {author} {\bibfnamefont
  {J.}~\bibnamefont {Ruitenberg}}, \bibinfo {author} {\bibfnamefont {R.~F.~L.}\
  \bibnamefont {Vermeulen}}, \bibinfo {author} {\bibfnamefont {R.~N.}\
  \bibnamefont {Schouten}}, \bibinfo {author} {\bibfnamefont {C.}~\bibnamefont
  {Abell{\'a}n}}, \bibinfo {author} {\bibfnamefont {W.}~\bibnamefont {Amaya}},
  \bibinfo {author} {\bibfnamefont {V.}~\bibnamefont {Pruneri}}, \bibinfo
  {author} {\bibfnamefont {M.~W.}\ \bibnamefont {Mitchell}}, \bibinfo {author}
  {\bibfnamefont {M.}~\bibnamefont {Markham}}, \bibinfo {author} {\bibfnamefont
  {D.~J.}\ \bibnamefont {Twitchen}}, \bibinfo {author} {\bibfnamefont
  {D.}~\bibnamefont {Elkouss}}, \bibinfo {author} {\bibfnamefont
  {S.}~\bibnamefont {Wehner}}, \bibinfo {author} {\bibfnamefont {T.~H.}\
  \bibnamefont {Taminiau}}, \ and\ \bibinfo {author} {\bibfnamefont
  {R.}~\bibnamefont {Hanson}}} (\bibinfo {year} {2015}),\ \bibfield  {title}
  {\enquote {\bibinfo {title} {Loophole-free bell inequality violation using
  electron spins separated by 1.3 kilometres},}\ }\href
  {https://www.nature.com/articles/nature15759} {\bibfield  {journal} {\bibinfo
   {journal} {Nature}\ }\textbf {\bibinfo {volume} {526}}~(\bibinfo {number}
  {7575}),\ \bibinfo {pages} {682}}\BibitemShut {NoStop}%
\bibitem [{\citenamefont {Hiscock}\ \emph {et~al.}(2016)\citenamefont
  {Hiscock}, \citenamefont {Worster}, \citenamefont {Kattnig}, \citenamefont
  {Steers}, \citenamefont {Jin}, \citenamefont {Manolopoulos}, \citenamefont
  {Mouritsen},\ and\ \citenamefont {Hore}}]{hiscock2016quantum}%
  \BibitemOpen
  \bibfield  {author} {\bibinfo {author} {\bibnamefont {Hiscock}, \bibfnamefont
  {Hamish~G}}, \bibinfo {author} {\bibfnamefont {Susannah}\ \bibnamefont
  {Worster}}, \bibinfo {author} {\bibfnamefont {Daniel~R.}\ \bibnamefont
  {Kattnig}}, \bibinfo {author} {\bibfnamefont {Charlotte}\ \bibnamefont
  {Steers}}, \bibinfo {author} {\bibfnamefont {Ye}~\bibnamefont {Jin}},
  \bibinfo {author} {\bibfnamefont {David~E.}\ \bibnamefont {Manolopoulos}},
  \bibinfo {author} {\bibfnamefont {Henrik}\ \bibnamefont {Mouritsen}}, \ and\
  \bibinfo {author} {\bibfnamefont {P.J.}\ \bibnamefont {Hore}}} (\bibinfo
  {year} {2016}),\ \bibfield  {title} {\enquote {\bibinfo {title} {The quantum
  needle of the avian magnetic compass},}\ }\href
  {https://doi.org/10.1073/pnas.1600341113} {\bibfield  {journal} {\bibinfo
  {journal} {Proceedings of the National Academy of Sciences}\ }\textbf
  {\bibinfo {volume} {113}}~(\bibinfo {number} {17}),\ \bibinfo {pages}
  {4634}}\BibitemShut {NoStop}%
\bibitem [{\citenamefont {Holt}(1973)}]{holt}%
  \BibitemOpen
  \bibfield  {author} {\bibinfo {author} {\bibnamefont {Holt}, \bibfnamefont
  {Richard~A}}} (\bibinfo {year} {1973}),\ \href@noop {} {Ph.D. thesis}\
  (\bibinfo  {school} {Harvard})\BibitemShut {NoStop}%
\bibitem [{\citenamefont {Hubbell}(2006)}]{Hubbell2006}%
  \BibitemOpen
  \bibfield  {author} {\bibinfo {author} {\bibnamefont {Hubbell}, \bibfnamefont
  {John~H}}} (\bibinfo {year} {2006}),\ \bibfield  {title} {\enquote {\bibinfo
  {title} {Electron {\textendash} positron pair production by photons: A
  historical overview},}\ }\href {\doibase 10.1016/j.radphyschem.2005.10.008}
  {\bibfield  {journal} {\bibinfo  {journal} {Radiation Physics Chemistry}\
  }\textbf {\bibinfo {volume} {75}}~(\bibinfo {number} {6}),\ \bibinfo {pages}
  {614}}\BibitemShut {NoStop}%
\bibitem [{\citenamefont {Hund}(1927)}]{hund}%
  \BibitemOpen
  \bibfield  {author} {\bibinfo {author} {\bibnamefont {Hund}, \bibfnamefont
  {Friedrich}}} (\bibinfo {year} {1927}),\ \bibfield  {title} {\enquote
  {\bibinfo {title} {Zur deutung der molekelspektren. i},}\ }\href
  {https://link.springer.com/article/10.1007/BF01400234} {\bibfield  {journal}
  {\bibinfo  {journal} {Zeitschrift f{\"u}r Physik}\ }\textbf {\bibinfo
  {volume} {40}}~(\bibinfo {number} {10}),\ \bibinfo {pages} {742}}\BibitemShut
  {NoStop}%
\bibitem [{\citenamefont {Huygens}(1690)}]{huygens}%
  \BibitemOpen
  \bibfield  {author} {\bibinfo {author} {\bibnamefont {Huygens}, \bibfnamefont
  {Christian}}} (\bibinfo {year} {1690}),\ \href
  {http://digitalcollections.library.cmu.edu/awweb/awarchive?type=file&item=719030}
  {\emph {\bibinfo {title} {Trait\'{e} de la Lumi\`{e}re}}}\ (\bibinfo
  {publisher} {Leiden: Pieter van der Aa})\BibitemShut {NoStop}%
\bibitem [{\citenamefont {Irvine}\ \emph {et~al.}(2005)\citenamefont {Irvine},
  \citenamefont {Hodelin}, \citenamefont {Simon},\ and\ \citenamefont
  {Bouwmeester}}]{irvine}%
  \BibitemOpen
  \bibfield  {author} {\bibinfo {author} {\bibnamefont {Irvine}, \bibfnamefont
  {William T~M}}, \bibinfo {author} {\bibfnamefont {Juan~F.}\ \bibnamefont
  {Hodelin}}, \bibinfo {author} {\bibfnamefont {Christoph}\ \bibnamefont
  {Simon}}, \ and\ \bibinfo {author} {\bibfnamefont {Dirk}\ \bibnamefont
  {Bouwmeester}}} (\bibinfo {year} {2005}),\ \bibfield  {title} {\enquote
  {\bibinfo {title} {{Realization of Hardy's Thought Experiment with
  Photons}},}\ }\href
  {https://journals.aps.org/prl/abstract/10.1103/PhysRevLett.95.030401}
  {\bibfield  {journal} {\bibinfo  {journal} {Physical Review Letters}\
  }\textbf {\bibinfo {volume} {95}}~(\bibinfo {number} {3}),\ \bibinfo {pages}
  {030401}}\BibitemShut {NoStop}%
\bibitem [{\citenamefont {Kumar}\ \emph {et~al.}(2016)\citenamefont {Kumar},
  \citenamefont {Boone}, \citenamefont {Tuszy{\'n}ski}, \citenamefont
  {Barclay},\ and\ \citenamefont {Simon}}]{kumar2016possible}%
  \BibitemOpen
  \bibfield  {author} {\bibinfo {author} {\bibnamefont {Kumar}, \bibfnamefont
  {Sourabh}}, \bibinfo {author} {\bibfnamefont {Kristine}\ \bibnamefont
  {Boone}}, \bibinfo {author} {\bibfnamefont {Jack}\ \bibnamefont
  {Tuszy{\'n}ski}}, \bibinfo {author} {\bibfnamefont {Paul}\ \bibnamefont
  {Barclay}}, \ and\ \bibinfo {author} {\bibfnamefont {Christoph}\ \bibnamefont
  {Simon}}} (\bibinfo {year} {2016}),\ \bibfield  {title} {\enquote {\bibinfo
  {title} {Possible existence of optical communication channels in the
  brain},}\ }\href {https://www.nature.com/articles/srep36508} {\bibfield
  {journal} {\bibinfo  {journal} {Scientific Reports}\ }\textbf {\bibinfo
  {volume} {6}},\ \bibinfo {pages} {36508}}\BibitemShut {NoStop}%
\bibitem [{\citenamefont {Kwiat}\ \emph {et~al.}(1995)\citenamefont {Kwiat},
  \citenamefont {Weinfurter}, \citenamefont {Herzog}, \citenamefont
  {Zeilinger},\ and\ \citenamefont {Kasevich}}]{kwiat1995}%
  \BibitemOpen
  \bibfield  {author} {\bibinfo {author} {\bibnamefont {Kwiat}, \bibfnamefont
  {Paul}}, \bibinfo {author} {\bibfnamefont {Harald}\ \bibnamefont
  {Weinfurter}}, \bibinfo {author} {\bibfnamefont {Thomas}\ \bibnamefont
  {Herzog}}, \bibinfo {author} {\bibfnamefont {Anton}\ \bibnamefont
  {Zeilinger}}, \ and\ \bibinfo {author} {\bibfnamefont {Mark~A.}\ \bibnamefont
  {Kasevich}}} (\bibinfo {year} {1995}),\ \bibfield  {title} {\enquote
  {\bibinfo {title} {Interaction-free measurement},}\ }\href {\doibase
  10.1103/PhysRevLett.74.4763} {\bibfield  {journal} {\bibinfo  {journal}
  {Physical Review Letters}\ }\textbf {\bibinfo {volume} {74}},\ \bibinfo
  {pages} {4763}}\BibitemShut {NoStop}%
\bibitem [{\citenamefont {Kwiat}\ and\ \citenamefont
  {Hardy}(2000)}]{Kwiat2000}%
  \BibitemOpen
  \bibfield  {author} {\bibinfo {author} {\bibnamefont {Kwiat}, \bibfnamefont
  {Paul~G}}, \ and\ \bibinfo {author} {\bibfnamefont {Lucien}\ \bibnamefont
  {Hardy}}} (\bibinfo {year} {2000}),\ \bibfield  {title} {\enquote {\bibinfo
  {title} {The mystery of the quantum cakes},}\ }\href
  {https://doi.org/10.1119/1.19369} {\bibfield  {journal} {\bibinfo  {journal}
  {American Journal of Physics}\ }\textbf {\bibinfo {volume} {68}}~(\bibinfo
  {number} {1}),\ \bibinfo {pages} {33}}\BibitemShut {NoStop}%
\bibitem [{\citenamefont {Lambert}\ \emph {et~al.}(2013)\citenamefont
  {Lambert}, \citenamefont {Chen}, \citenamefont {Cheng}, \citenamefont {Li},
  \citenamefont {Chen},\ and\ \citenamefont {Nori}}]{lambert2013quantum}%
  \BibitemOpen
  \bibfield  {author} {\bibinfo {author} {\bibnamefont {Lambert}, \bibfnamefont
  {Neill}}, \bibinfo {author} {\bibfnamefont {Yueh-Nan}\ \bibnamefont {Chen}},
  \bibinfo {author} {\bibfnamefont {Yuan-Chung}\ \bibnamefont {Cheng}},
  \bibinfo {author} {\bibfnamefont {Che-Ming}\ \bibnamefont {Li}}, \bibinfo
  {author} {\bibfnamefont {Guang-Yin}\ \bibnamefont {Chen}}, \ and\ \bibinfo
  {author} {\bibfnamefont {Franco}\ \bibnamefont {Nori}}} (\bibinfo {year}
  {2013}),\ \bibfield  {title} {\enquote {\bibinfo {title} {Quantum biology},}\
  }\href {https://www.nature.com/articles/nphys2474} {\bibfield  {journal}
  {\bibinfo  {journal} {Nature Physics}\ }\textbf {\bibinfo {volume}
  {9}}~(\bibinfo {number} {1}),\ \bibinfo {pages} {10}}\BibitemShut {NoStop}%
\bibitem [{\citenamefont {Larsson}(2014)}]{loopholesreview}%
  \BibitemOpen
  \bibfield  {author} {\bibinfo {author} {\bibnamefont {Larsson}, \bibfnamefont
  {Jan-{\AA}ke}}} (\bibinfo {year} {2014}),\ \bibfield  {title} {\enquote
  {\bibinfo {title} {Loopholes in bell inequality tests of local realism},}\
  }\href {http://iopscience.iop.org/article/10.1088/1751-8113/47/42/424003}
  {\bibfield  {journal} {\bibinfo  {journal} {Journal of Physics A:
  Mathematical and Theoretical}\ }\textbf {\bibinfo {volume} {47}}~(\bibinfo
  {number} {42}),\ \bibinfo {pages} {424003}}\BibitemShut {NoStop}%
\bibitem [{\citenamefont {Lenard}(1902)}]{lenard}%
  \BibitemOpen
  \bibfield  {author} {\bibinfo {author} {\bibnamefont {Lenard}, \bibfnamefont
  {Philipp}}} (\bibinfo {year} {1902}),\ \bibfield  {title} {\enquote {\bibinfo
  {title} {{Ueber die lichtelektrische Wirkung}},}\ }\href
  {https://doi.org/10.1002/andp.19023130510} {\bibfield  {journal} {\bibinfo
  {journal} {Annalen der Physik}\ }\textbf {\bibinfo {volume} {313}}~(\bibinfo
  {number} {5}),\ \bibinfo {pages} {149}}\BibitemShut {NoStop}%
\bibitem [{\citenamefont {Lockwood}(1989)}]{lockwood1989mind}%
  \BibitemOpen
  \bibfield  {author} {\bibinfo {author} {\bibnamefont {Lockwood},
  \bibfnamefont {Michael}}} (\bibinfo {year} {1989}),\ \href@noop {} {\emph
  {\bibinfo {title} {Mind, brain and the quantum: The compound'I.'}}}\
  (\bibinfo  {publisher} {Basil Blackwell})\BibitemShut {NoStop}%
\bibitem [{\citenamefont {Lundeen}\ and\ \citenamefont
  {Steinberg}(2009)}]{Lundeen2009}%
  \BibitemOpen
  \bibfield  {author} {\bibinfo {author} {\bibnamefont {Lundeen}, \bibfnamefont
  {Jeff~S}}, \ and\ \bibinfo {author} {\bibfnamefont {Aephraim~M.}\
  \bibnamefont {Steinberg}}} (\bibinfo {year} {2009}),\ \bibfield  {title}
  {\enquote {\bibinfo {title} {{Experimental Joint Weak Measurement on a Photon
  Pair as a Probe of Hardy's Paradox}},}\ }\href
  {http://dx.doi.org/10.1103/PhysRevLett.102.020404} {\bibfield  {journal}
  {\bibinfo  {journal} {Physical Review Letters}\ }\textbf {\bibinfo {volume}
  {102}}~(\bibinfo {number} {2}),\ \bibinfo {pages} {020404}}\BibitemShut
  {NoStop}%
\bibitem [{\citenamefont {Mandel}\ and\ \citenamefont
  {Wolf}(1995)}]{mandelwolf}%
  \BibitemOpen
  \bibfield  {author} {\bibinfo {author} {\bibnamefont {Mandel}, \bibfnamefont
  {Leonard}}, \ and\ \bibinfo {author} {\bibfnamefont {Emil}\ \bibnamefont
  {Wolf}}} (\bibinfo {year} {1995}),\ \href
  {http://www.cambridge.org/us/academic/subjects/physics/optics-optoelectronics-and-photonics/optical-coherence-and-quantum-optics?format=HB&isbn=9780521417112#ku5a5cGzuIhWODKF.97}
  {\emph {\bibinfo {title} {{Optical Coherence and Quantum Optics}}}}\
  (\bibinfo  {publisher} {Cambridge university press})\BibitemShut {NoStop}%
\bibitem [{\citenamefont {Manin~Yu}(1980)}]{manin}%
  \BibitemOpen
  \bibfield  {author} {\bibinfo {author} {\bibnamefont {Manin~Yu},
  \bibfnamefont {I}}} (\bibinfo {year} {1980}),\ \href
  {https://web.archive.org/web/20130510173823/http://publ.lib.ru/ARCHIVES/M/MANIN_Yuriy_Ivanovich/Manin_Yu.I._Vychislimoe_i_nevychislimoe.%281980%29.%5Bdjv%5D.zip}
  {\emph {\bibinfo {title} {Vychislimoe i Nevychislimoe}}}\ (\bibinfo
  {publisher} {Computable and Noncomputable)(Moscow: Sov. Radio,
  1980)})\BibitemShut {NoStop}%
\bibitem [{\citenamefont {Marshall}(1989)}]{marshall1989consciousness}%
  \BibitemOpen
  \bibfield  {author} {\bibinfo {author} {\bibnamefont {Marshall},
  \bibfnamefont {IN}}} (\bibinfo {year} {1989}),\ \bibfield  {title} {\enquote
  {\bibinfo {title} {{Consciousness and Bose-Einstein condensates}},}\
  }\href@noop {} {\bibfield  {journal} {\bibinfo  {journal} {New Ideas in
  Psychology}\ }\textbf {\bibinfo {volume} {7}}~(\bibinfo {number} {1}),\
  \bibinfo {pages} {73}}\BibitemShut {NoStop}%
\bibitem [{\citenamefont {Maxwell}(1865)}]{Maxwell1865}%
  \BibitemOpen
  \bibfield  {author} {\bibinfo {author} {\bibnamefont {Maxwell}, \bibfnamefont
  {James~Clerk}}} (\bibinfo {year} {1865}),\ \bibfield  {title} {\enquote
  {\bibinfo {title} {Viii.~{A Dynamical Theory of the Electromagnetic
  Field}},}\ }\href
  {http://rstl.royalsocietypublishing.org/content/155/459.full.pdf+html}
  {\bibfield  {journal} {\bibinfo  {journal} {Philosophical Transactions of the
  Royal Society of London}\ }\textbf {\bibinfo {volume} {155}},\ \bibinfo
  {pages} {459}}\BibitemShut {NoStop}%
\bibitem [{\citenamefont {McConnell}\ \emph {et~al.}(2015)\citenamefont
  {McConnell}, \citenamefont {Zhang}, \citenamefont {Hu}, \citenamefont
  {{\'C}uk},\ and\ \citenamefont {Vuleti{\'c}}}]{mcconnell}%
  \BibitemOpen
  \bibfield  {author} {\bibinfo {author} {\bibnamefont {McConnell},
  \bibfnamefont {Robert}}, \bibinfo {author} {\bibfnamefont {Hao}\ \bibnamefont
  {Zhang}}, \bibinfo {author} {\bibfnamefont {Jiazhong}\ \bibnamefont {Hu}},
  \bibinfo {author} {\bibfnamefont {Senka}\ \bibnamefont {{\'C}uk}}, \ and\
  \bibinfo {author} {\bibfnamefont {Vladan}\ \bibnamefont {Vuleti{\'c}}}}
  (\bibinfo {year} {2015}),\ \bibfield  {title} {\enquote {\bibinfo {title}
  {Entanglement with negative wigner function of almost 3,000 atoms heralded by
  one photon},}\ }\href {https://www.nature.com/articles/nature14293}
  {\bibfield  {journal} {\bibinfo  {journal} {Nature}\ }\textbf {\bibinfo
  {volume} {519}}~(\bibinfo {number} {7544}),\ \bibinfo {pages}
  {439}}\BibitemShut {NoStop}%
\bibitem [{\citenamefont {Mohseni}\ \emph {et~al.}(2014)\citenamefont
  {Mohseni}, \citenamefont {Omar}, \citenamefont {Engel},\ and\ \citenamefont
  {Plenio}}]{mohseni2014quantum}%
  \BibitemOpen
  \bibfield  {author} {\bibinfo {author} {\bibnamefont {Mohseni}, \bibfnamefont
  {Masoud}}, \bibinfo {author} {\bibfnamefont {Yasser}\ \bibnamefont {Omar}},
  \bibinfo {author} {\bibfnamefont {Gregory~S.}\ \bibnamefont {Engel}}, \ and\
  \bibinfo {author} {\bibfnamefont {Martin~B.}\ \bibnamefont {Plenio}}}
  (\bibinfo {year} {2014}),\ \href
  {http://www.cambridge.org/gb/academic/subjects/physics/biological-physics-and-soft-matter-physics/quantum-effects-biology?format=HB#eWIBSOWCadWM8lm2.97}
  {\emph {\bibinfo {title} {Quantum effects in biology}}}\ (\bibinfo
  {publisher} {Cambridge University Press})\BibitemShut {NoStop}%
\bibitem [{\citenamefont {Newton}(1704)}]{newton1704}%
  \BibitemOpen
  \bibfield  {author} {\bibinfo {author} {\bibnamefont {Newton}, \bibfnamefont
  {Isaac}}} (\bibinfo {year} {1704}),\ \href
  {https://archive.org/details/opticksortreatis1730newt} {\emph {\bibinfo
  {title} {Opticks: Or a treatise of the reflexions, refractions, inflexions
  and colours of light}}}\BibitemShut {NoStop}%
\bibitem [{\citenamefont {Nielsen}(2018)}]{nielsencourse}%
  \BibitemOpen
  \bibfield  {author} {\bibinfo {author} {\bibnamefont {Nielsen}, \bibfnamefont
  {Michael}}} (\bibinfo {year} {2018}),\ \href@noop {} {\enquote {\bibinfo
  {title} {Quantum computing for the determined},}\ }\bibinfo {howpublished}
  {\url{http://michaelnielsen.org/blog/quantum-computing-for-the-determined/}},\
  \bibinfo {note} {[Online; accessed 06-March-2018]}\BibitemShut {NoStop}%
\bibitem [{\citenamefont {Nielsen}\ and\ \citenamefont
  {Chuang}(2010)}]{nielsen}%
  \BibitemOpen
  \bibfield  {author} {\bibinfo {author} {\bibnamefont {Nielsen}, \bibfnamefont
  {Michael~A}}, \ and\ \bibinfo {author} {\bibfnamefont {Isaac~L}\ \bibnamefont
  {Chuang}}} (\bibinfo {year} {2010}),\ \href
  {http://www.cambridge.org/gb/academic/subjects/physics/quantum-physics-quantum-information-and-quantum-computation/quantum-computation-and-quantum-information-10th-anniversary-edition?format=HB&isbn=9781107002173#T7GlDd9PdRXerE8K.97}
  {\enquote {\bibinfo {title} {Quantum computation and quantum information},}\
  }\BibitemShut {NoStop}%
\bibitem [{\citenamefont {Oxford}(2018)}]{scientificmethod}%
  \BibitemOpen
  \bibfield  {author} {\bibinfo {author} {\bibnamefont {Oxford},}} (\bibinfo
  {year} {2018}),\ \href@noop {} {\enquote {\bibinfo {title} {{Scientific
  method}},}\ }\bibinfo {howpublished}
  {\url{http://www.oxfordreference.com/view/10.1093/oi/authority.20110803100447727}},\
  \bibinfo {note} {[Online; accessed 20-February-2018]}\BibitemShut {NoStop}%
\bibitem [{\citenamefont {Pais}(1979)}]{Pais1979}%
  \BibitemOpen
  \bibfield  {author} {\bibinfo {author} {\bibnamefont {Pais}, \bibfnamefont
  {A}}} (\bibinfo {year} {1979}),\ \bibfield  {title} {\enquote {\bibinfo
  {title} {Einstein and the quantum theory},}\ }\href {\doibase
  10.1103/RevModPhys.51.863} {\bibfield  {journal} {\bibinfo  {journal}
  {Reviews of Modern Physics}\ }\textbf {\bibinfo {volume} {51}},\ \bibinfo
  {pages} {863}}\BibitemShut {NoStop}%
\bibitem [{\citenamefont {Penrose}(1994)}]{penrose1994shadows}%
  \BibitemOpen
  \bibfield  {author} {\bibinfo {author} {\bibnamefont {Penrose}, \bibfnamefont
  {Roger}}} (\bibinfo {year} {1994}),\ \href
  {https://global.oup.com/academic/product/shadows-of-the-mind-9780195106466?cc=us&lang=en&}
  {\emph {\bibinfo {title} {Shadows of the Mind}}},\ Vol.~\bibinfo {volume}
  {4}\ (\bibinfo  {publisher} {Oxford University Press Oxford})\BibitemShut
  {NoStop}%
\bibitem [{\citenamefont {Pipkin}(1979)}]{pipkin}%
  \BibitemOpen
  \bibfield  {author} {\bibinfo {author} {\bibnamefont {Pipkin}, \bibfnamefont
  {Francis~M}}} (\bibinfo {year} {1979}),\ \bibfield  {title} {\enquote
  {\bibinfo {title} {{Atomic Physics Tests of the Basic Concepts in Quantum
  Mechanics}},}\ }in\ \href
  {https://www.sciencedirect.com/science/article/pii/S006521990860130X} {\emph
  {\bibinfo {booktitle} {Advances in Atomic and Molecular Physics}}},\
  Vol.~\bibinfo {volume} {14}\ (\bibinfo  {publisher} {Elsevier})\ p.\ \bibinfo
  {pages} {281}\BibitemShut {NoStop}%
\bibitem [{\citenamefont {Planck}(1900)}]{planck1900}%
  \BibitemOpen
  \bibfield  {author} {\bibinfo {author} {\bibnamefont {Planck}, \bibfnamefont
  {Max Karl Ernst~Ludwig}}} (\bibinfo {year} {1900}),\ \bibfield  {title}
  {\enquote {\bibinfo {title} {Zur theorie des gesetzes der energieverteilung
  im normalspectrum},}\ }\href
  {https://grundpraktikum.physik.uni-saarland.de/scripts/Planck_1.pdf}
  {\bibfield  {journal} {\bibinfo  {journal} {Verhandlungen der Deutschen
  Physikalischen Gesellschaft}\ }\textbf {\bibinfo {volume} {2}},\ \bibinfo
  {pages} {237}}\BibitemShut {NoStop}%
\bibitem [{\citenamefont {Rauch}\ \emph {et~al.}(1974)\citenamefont {Rauch},
  \citenamefont {Treimer},\ and\ \citenamefont {Bonse}}]{rauch}%
  \BibitemOpen
  \bibfield  {author} {\bibinfo {author} {\bibnamefont {Rauch}, \bibfnamefont
  {Helmut}}, \bibinfo {author} {\bibfnamefont {Wolfgang}\ \bibnamefont
  {Treimer}}, \ and\ \bibinfo {author} {\bibfnamefont {Ulrich}\ \bibnamefont
  {Bonse}}} (\bibinfo {year} {1974}),\ \bibfield  {title} {\enquote {\bibinfo
  {title} {Test of a single crystal neutron interferometer},}\ }\href
  {https://www.sciencedirect.com/science/article/pii/0375960174901327}
  {\bibfield  {journal} {\bibinfo  {journal} {Physics Letters A}\ }\textbf
  {\bibinfo {volume} {47}}~(\bibinfo {number} {5}),\ \bibinfo {pages}
  {369}}\BibitemShut {NoStop}%
\bibitem [{\citenamefont {Raymer}(2017)}]{Raymer2017}%
  \BibitemOpen
  \bibfield  {author} {\bibinfo {author} {\bibnamefont {Raymer}, \bibfnamefont
  {Michael}}} (\bibinfo {year} {2017}),\ \href@noop {} {\emph {\bibinfo {title}
  {Quantum Physics: What Everyone Needs to Know}}}\ (\bibinfo  {publisher}
  {Oxford University Press})\BibitemShut {NoStop}%
\bibitem [{\citenamefont {Reiserer}\ \emph {et~al.}(2013)\citenamefont
  {Reiserer}, \citenamefont {Ritter},\ and\ \citenamefont
  {Rempe}}]{reiserer2013}%
  \BibitemOpen
  \bibfield  {author} {\bibinfo {author} {\bibnamefont {Reiserer},
  \bibfnamefont {Andreas}}, \bibinfo {author} {\bibfnamefont {Stephan}\
  \bibnamefont {Ritter}}, \ and\ \bibinfo {author} {\bibfnamefont {Gerhard}\
  \bibnamefont {Rempe}}} (\bibinfo {year} {2013}),\ \bibfield  {title}
  {\enquote {\bibinfo {title} {{Nondestructive Detection of an Optical
  Photon}},}\ }\href {http://science.sciencemag.org/content/342/6164/1349}
  {\bibinfo  {journal} {Science}\ ,\ \bibinfo {pages} {1246164}}\BibitemShut
  {NoStop}%
\bibitem [{\citenamefont {Ritz}\ \emph {et~al.}(2000)\citenamefont {Ritz},
  \citenamefont {Adem},\ and\ \citenamefont {Schulten}}]{ritz2000model}%
  \BibitemOpen
\bibfield  {journal} {  }\bibfield  {author} {\bibinfo {author} {\bibnamefont
  {Ritz}, \bibfnamefont {Thorsten}}, \bibinfo {author} {\bibfnamefont {Salih}\
  \bibnamefont {Adem}}, \ and\ \bibinfo {author} {\bibfnamefont {Klaus}\
  \bibnamefont {Schulten}}} (\bibinfo {year} {2000}),\ \bibfield  {title}
  {\enquote {\bibinfo {title} {{A Model for Photoreceptor-Based
  Magnetoreception in Birds}},}\ }\href
  {http://www.cell.com/biophysj/fulltext/S0006-3495(00)76629-X} {\bibfield
  {journal} {\bibinfo  {journal} {Biophysical Journal}\ }\textbf {\bibinfo
  {volume} {78}}~(\bibinfo {number} {2}),\ \bibinfo {pages} {707}}\BibitemShut
  {NoStop}%
\bibitem [{\citenamefont {Romero}\ \emph {et~al.}(2014)\citenamefont {Romero},
  \citenamefont {Augulis}, \citenamefont {Novoderezhkin}, \citenamefont
  {Ferretti}, \citenamefont {Thieme}, \citenamefont {Zigmantas},\ and\
  \citenamefont {Van~Grondelle}}]{romero2014quantum}%
  \BibitemOpen
  \bibfield  {author} {\bibinfo {author} {\bibnamefont {Romero}, \bibfnamefont
  {Elisabet}}, \bibinfo {author} {\bibfnamefont {Ramunas}\ \bibnamefont
  {Augulis}}, \bibinfo {author} {\bibfnamefont {Vladimir~I.}\ \bibnamefont
  {Novoderezhkin}}, \bibinfo {author} {\bibfnamefont {Marco}\ \bibnamefont
  {Ferretti}}, \bibinfo {author} {\bibfnamefont {Jos}\ \bibnamefont {Thieme}},
  \bibinfo {author} {\bibfnamefont {Donatas}\ \bibnamefont {Zigmantas}}, \ and\
  \bibinfo {author} {\bibfnamefont {Rienk}\ \bibnamefont {Van~Grondelle}}}
  (\bibinfo {year} {2014}),\ \bibfield  {title} {\enquote {\bibinfo {title}
  {Quantum coherence in photosynthesis for efficient solar-energy
  conversion},}\ }\href {https://www.nature.com/articles/nphys3017} {\bibfield
  {journal} {\bibinfo  {journal} {Nature Physics}\ }\textbf {\bibinfo {volume}
  {10}}~(\bibinfo {number} {9}),\ \bibinfo {pages} {676}}\BibitemShut {NoStop}%
\bibitem [{\citenamefont {Rowe}\ \emph {et~al.}(2001)\citenamefont {Rowe},
  \citenamefont {Kielpinski}, \citenamefont {Meyer}, \citenamefont {Sackett},
  \citenamefont {Monroe},\ and\ \citenamefont {Wineland}}]{rowe}%
  \BibitemOpen
  \bibfield  {author} {\bibinfo {author} {\bibnamefont {Rowe}, \bibfnamefont
  {Mary~A}}, \bibinfo {author} {\bibfnamefont {David}\ \bibnamefont
  {Kielpinski}}, \bibinfo {author} {\bibfnamefont {Volker}\ \bibnamefont
  {Meyer}}, \bibinfo {author} {\bibfnamefont {Wayne~M.}\ \bibnamefont
  {Sackett}, \bibfnamefont {Charles A .and~Itano}}, \bibinfo {author}
  {\bibfnamefont {Christopher}\ \bibnamefont {Monroe}}, \ and\ \bibinfo
  {author} {\bibfnamefont {David~J.}\ \bibnamefont {Wineland}}} (\bibinfo
  {year} {2001}),\ \bibfield  {title} {\enquote {\bibinfo {title} {Experimental
  violation of a bell's inequality with efficient detection},}\ }\href
  {https://www.nature.com/articles/35057215} {\bibfield  {journal} {\bibinfo
  {journal} {Nature}\ }\textbf {\bibinfo {volume} {409}}~(\bibinfo {number}
  {6822}),\ \bibinfo {pages} {791}}\BibitemShut {NoStop}%
\bibitem [{\citenamefont {Rudolph}(2017)}]{Rudolph2017}%
  \BibitemOpen
  \bibfield  {author} {\bibinfo {author} {\bibnamefont {Rudolph}, \bibfnamefont
  {Terry}}} (\bibinfo {year} {2017}),\ \href
  {https://www.amazon.com/Q-Quantum-Terry-Rudolph/dp/0999063502} {\emph
  {\bibinfo {title} {Q is for Quantum}}}\ (\bibinfo  {publisher} {Terry
  Rudolph})\BibitemShut {NoStop}%
\bibitem [{\citenamefont {Scarani}(2006)}]{Scarani2006}%
  \BibitemOpen
  \bibfield  {author} {\bibinfo {author} {\bibnamefont {Scarani}, \bibfnamefont
  {Valerio}}} (\bibinfo {year} {2006}),\ \href
  {https://global.oup.com/academic/product/quantum-physics-a-first-encounter-9780198570479?cc=us&lang=en&}
  {\emph {\bibinfo {title} {{Quantum Physics: A First Encounter Interference,
  Entanglement, and Reality}}}}\ (\bibinfo  {publisher} {Oxford University
  Press},\ \bibinfo {address} {Oxford})\BibitemShut {NoStop}%
\bibitem [{\citenamefont {Scarani}\ \emph {et~al.}(2010)\citenamefont
  {Scarani}, \citenamefont {Lynn},\ and\ \citenamefont {Liu}}]{Scarani2010}%
  \BibitemOpen
  \bibfield  {author} {\bibinfo {author} {\bibnamefont {Scarani}, \bibfnamefont
  {Valerio}}, \bibinfo {author} {\bibfnamefont {Chua}\ \bibnamefont {Lynn}}, \
  and\ \bibinfo {author} {\bibfnamefont {Shiyang}\ \bibnamefont {Liu}}}
  (\bibinfo {year} {2010}),\ \href
  {https://www.worldscientific.com/worldscibooks/10.1142/7965} {\emph {\bibinfo
  {title} {Six Quantum Pieces: A First Course in Quantum Physics}}}\ (\bibinfo
  {publisher} {World Scientific})\BibitemShut {NoStop}%
\bibitem [{\citenamefont {Schawlow}\ and\ \citenamefont
  {Townes}(1958)}]{laser}%
  \BibitemOpen
  \bibfield  {author} {\bibinfo {author} {\bibnamefont {Schawlow},
  \bibfnamefont {Arthur~L}}, \ and\ \bibinfo {author} {\bibfnamefont
  {Charles~H.}\ \bibnamefont {Townes}}} (\bibinfo {year} {1958}),\ \bibfield
  {title} {\enquote {\bibinfo {title} {Infrared and optical masers},}\ }\href
  {\doibase 10.1103/PhysRev.112.1940} {\bibfield  {journal} {\bibinfo
  {journal} {Physical Review}\ }\textbf {\bibinfo {volume} {112}},\ \bibinfo
  {pages} {1940}}\BibitemShut {NoStop}%
\bibitem [{\citenamefont {Schlosshauer}\ \emph {et~al.}(2013)\citenamefont
  {Schlosshauer}, \citenamefont {Kofler},\ and\ \citenamefont
  {Zeilinger}}]{Schlosshauer2013}%
  \BibitemOpen
  \bibfield  {author} {\bibinfo {author} {\bibnamefont {Schlosshauer},
  \bibfnamefont {Maximilian}}, \bibinfo {author} {\bibfnamefont {Johannes}\
  \bibnamefont {Kofler}}, \ and\ \bibinfo {author} {\bibfnamefont {Anton}\
  \bibnamefont {Zeilinger}}} (\bibinfo {year} {2013}),\ \bibfield  {title}
  {\enquote {\bibinfo {title} {The interpretation of quantum mechanics: from
  disagreement to consensus?}}\ }\href {\doibase 10.1002/andp.201300722}
  {\bibfield  {journal} {\bibinfo  {journal} {Annalen der Physik}\ }\textbf
  {\bibinfo {volume} {525}}~(\bibinfo {number} {4}),\ \bibinfo {pages}
  {A51}}\BibitemShut {NoStop}%
\bibitem [{\citenamefont {Schr{\"o}dinger}(1935)}]{schrodinger1935}%
  \BibitemOpen
  \bibfield  {author} {\bibinfo {author} {\bibnamefont {Schr{\"o}dinger},
  \bibfnamefont {E}}} (\bibinfo {year} {1935}),\ \bibfield  {title} {\enquote
  {\bibinfo {title} {{Die gegenw{\"a}rtige Situation in der
  Quantenmechanik}},}\ }\href {\doibase 10.1007/BF01491891} {\bibfield
  {journal} {\bibinfo  {journal} {Naturwissenschaften}\ }\textbf {\bibinfo
  {volume} {23}}~(\bibinfo {number} {48}),\ \bibinfo {pages} {807}}\BibitemShut
  {NoStop}%
\bibitem [{\citenamefont {Shadbolt}\ \emph {et~al.}(2014)\citenamefont
  {Shadbolt}, \citenamefont {Mathews}, \citenamefont {Laing},\ and\
  \citenamefont {O'Brien}}]{obrien2014}%
  \BibitemOpen
  \bibfield  {author} {\bibinfo {author} {\bibnamefont {Shadbolt},
  \bibfnamefont {Peter}}, \bibinfo {author} {\bibfnamefont {Jonathan C.~F.}\
  \bibnamefont {Mathews}}, \bibinfo {author} {\bibfnamefont {Anthony}\
  \bibnamefont {Laing}}, \ and\ \bibinfo {author} {\bibfnamefont {Jeremy~L.}\
  \bibnamefont {O'Brien}}} (\bibinfo {year} {2014}),\ \bibfield  {title}
  {\enquote {\bibinfo {title} {Testing foundations of quantum mechanics with
  photons},}\ }\href {http://dx.doi.org/10.1038/nphys2931} {\bibfield
  {journal} {\bibinfo  {journal} {Nature Physics}\ }\textbf {\bibinfo {volume}
  {10}},\ \bibinfo {pages} {278}}\BibitemShut {NoStop}%
\bibitem [{\citenamefont {Shalm}\ \emph {et~al.}(2015)\citenamefont {Shalm},
  \citenamefont {Meyer-Scott}, \citenamefont {Christensen}, \citenamefont
  {Bierhorst}, \citenamefont {Wayne}, \citenamefont {Stevens}, \citenamefont
  {Gerrits}, \citenamefont {Glancy}, \citenamefont {Hamel}, \citenamefont
  {Allman}, \citenamefont {Coakley}, \citenamefont {Dyer}, \citenamefont
  {Hodge}, \citenamefont {Lita}, \citenamefont {Verma}, \citenamefont
  {Lambrocco}, \citenamefont {Tortorici}, \citenamefont {Migdall},
  \citenamefont {Zhang}, \citenamefont {Kumor}, \citenamefont {Farr},
  \citenamefont {Marsili}, \citenamefont {Shaw}, \citenamefont {Stern},
  \citenamefont {Abell\'an}, \citenamefont {Amaya}, \citenamefont {Pruneri},
  \citenamefont {Jennewein}, \citenamefont {Mitchell}, \citenamefont {Kwiat},
  \citenamefont {Bienfang}, \citenamefont {Mirin}, \citenamefont {Knill},\ and\
  \citenamefont {Nam}}]{boulder}%
  \BibitemOpen
  \bibfield  {author} {\bibinfo {author} {\bibnamefont {Shalm}, \bibfnamefont
  {Lynden~K}}, \bibinfo {author} {\bibfnamefont {Evan}\ \bibnamefont
  {Meyer-Scott}}, \bibinfo {author} {\bibfnamefont {Bradley~G.}\ \bibnamefont
  {Christensen}}, \bibinfo {author} {\bibfnamefont {Peter}\ \bibnamefont
  {Bierhorst}}, \bibinfo {author} {\bibfnamefont {Michael~A.}\ \bibnamefont
  {Wayne}}, \bibinfo {author} {\bibfnamefont {Martin~J.}\ \bibnamefont
  {Stevens}}, \bibinfo {author} {\bibfnamefont {Thomas}\ \bibnamefont
  {Gerrits}}, \bibinfo {author} {\bibfnamefont {Scott}\ \bibnamefont {Glancy}},
  \bibinfo {author} {\bibfnamefont {Deny~R.}\ \bibnamefont {Hamel}}, \bibinfo
  {author} {\bibfnamefont {Michael~S.}\ \bibnamefont {Allman}}, \bibinfo
  {author} {\bibfnamefont {Kevin~J.}\ \bibnamefont {Coakley}}, \bibinfo
  {author} {\bibfnamefont {Shellee~D.}\ \bibnamefont {Dyer}}, \bibinfo {author}
  {\bibfnamefont {Carson}\ \bibnamefont {Hodge}}, \bibinfo {author}
  {\bibfnamefont {Adriana~E.}\ \bibnamefont {Lita}}, \bibinfo {author}
  {\bibfnamefont {Varun~B.}\ \bibnamefont {Verma}}, \bibinfo {author}
  {\bibfnamefont {Camilla}\ \bibnamefont {Lambrocco}}, \bibinfo {author}
  {\bibfnamefont {Edward}\ \bibnamefont {Tortorici}}, \bibinfo {author}
  {\bibfnamefont {Alan~L.}\ \bibnamefont {Migdall}}, \bibinfo {author}
  {\bibfnamefont {Yanbao}\ \bibnamefont {Zhang}}, \bibinfo {author}
  {\bibfnamefont {Daniel~R.}\ \bibnamefont {Kumor}}, \bibinfo {author}
  {\bibfnamefont {William~H.}\ \bibnamefont {Farr}}, \bibinfo {author}
  {\bibfnamefont {Francesco}\ \bibnamefont {Marsili}}, \bibinfo {author}
  {\bibfnamefont {Matthew~D.}\ \bibnamefont {Shaw}}, \bibinfo {author}
  {\bibfnamefont {Jeffrey~A.}\ \bibnamefont {Stern}}, \bibinfo {author}
  {\bibfnamefont {Carlos}\ \bibnamefont {Abell\'an}}, \bibinfo {author}
  {\bibfnamefont {Waldimar}\ \bibnamefont {Amaya}}, \bibinfo {author}
  {\bibfnamefont {Valerio}\ \bibnamefont {Pruneri}}, \bibinfo {author}
  {\bibfnamefont {Thomas}\ \bibnamefont {Jennewein}}, \bibinfo {author}
  {\bibfnamefont {Morgan~W.}\ \bibnamefont {Mitchell}}, \bibinfo {author}
  {\bibfnamefont {Paul~G.}\ \bibnamefont {Kwiat}}, \bibinfo {author}
  {\bibfnamefont {Joshua~C.}\ \bibnamefont {Bienfang}}, \bibinfo {author}
  {\bibfnamefont {Richard~P.}\ \bibnamefont {Mirin}}, \bibinfo {author}
  {\bibfnamefont {Emanuel}\ \bibnamefont {Knill}}, \ and\ \bibinfo {author}
  {\bibfnamefont {Sae~Woo}\ \bibnamefont {Nam}}} (\bibinfo {year} {2015}),\
  \bibfield  {title} {\enquote {\bibinfo {title} {Strong loophole-free test of
  local realism},}\ }\href {\doibase 10.1103/PhysRevLett.115.250402} {\bibfield
   {journal} {\bibinfo  {journal} {Physical Review Letters}\ }\textbf {\bibinfo
  {volume} {115}},\ \bibinfo {pages} {250402}}\BibitemShut {NoStop}%
\bibitem [{\citenamefont {Simon}\ and\ \citenamefont {Irvine}(2003)}]{simon}%
  \BibitemOpen
  \bibfield  {author} {\bibinfo {author} {\bibnamefont {Simon}, \bibfnamefont
  {Christoph}}, \ and\ \bibinfo {author} {\bibfnamefont {William T.~M.}\
  \bibnamefont {Irvine}}} (\bibinfo {year} {2003}),\ \bibfield  {title}
  {\enquote {\bibinfo {title} {{Robust Long-Distance Entanglement and a
  Loophole-Free Bell Test with Ions and Photons}},}\ }\href {\doibase
  10.1103/PhysRevLett.91.110405} {\bibfield  {journal} {\bibinfo  {journal}
  {Physical Review Letters}\ }\textbf {\bibinfo {volume} {91}},\ \bibinfo
  {pages} {110405}}\BibitemShut {NoStop}%
\bibitem [{\citenamefont {Smolin}(2008)}]{Smolin2008}%
  \BibitemOpen
  \bibfield  {author} {\bibinfo {author} {\bibnamefont {Smolin}, \bibfnamefont
  {Lee}}} (\bibinfo {year} {2008}),\ \href
  {https://www.basicbooks.com/titles/lee-smolin/three-roads-to-quantum-gravity/9780465013241/}
  {\emph {\bibinfo {title} {{Three Roads to Quantum Gravity}}}}\ (\bibinfo
  {publisher} {Basic books})\BibitemShut {NoStop}%
\bibitem [{\citenamefont {Stapp}(2011)}]{stapp2011mindful}%
  \BibitemOpen
  \bibfield  {author} {\bibinfo {author} {\bibnamefont {Stapp}, \bibfnamefont
  {Henry~P}}} (\bibinfo {year} {2011}),\ \href
  {http://www.springer.com/gp/book/9783642180750} {\emph {\bibinfo {title}
  {Mindful Universe: Quantum Mechanics and the Participating Observer}}}\
  (\bibinfo  {publisher} {Springer Science \& Business Media})\BibitemShut
  {NoStop}%
\bibitem [{\citenamefont {Torgerson}\ \emph {et~al.}(1995)\citenamefont
  {Torgerson}, \citenamefont {Branning}, \citenamefont {Monken},\ and\
  \citenamefont {Mandel}}]{mandel1995}%
  \BibitemOpen
  \bibfield  {author} {\bibinfo {author} {\bibnamefont {Torgerson},
  \bibfnamefont {Justin~R}}, \bibinfo {author} {\bibfnamefont {David}\
  \bibnamefont {Branning}}, \bibinfo {author} {\bibfnamefont {Carlos~H.}\
  \bibnamefont {Monken}}, \ and\ \bibinfo {author} {\bibfnamefont {Leonard}\
  \bibnamefont {Mandel}}} (\bibinfo {year} {1995}),\ \bibfield  {title}
  {\enquote {\bibinfo {title} {{Experimental demonstration of the violation of
  local realism without Bell inequalities}},}\ }\href {\doibase
  https://doi.org/10.1016/0375-9601(95)00486-M} {\bibfield  {journal} {\bibinfo
   {journal} {Physics Letters A}\ }\textbf {\bibinfo {volume} {204}}~(\bibinfo
  {number} {5}),\ \bibinfo {pages} {323}}\BibitemShut {NoStop}%
\bibitem [{\citenamefont {USAF}(2018)}]{gps}%
  \BibitemOpen
  \bibfield  {author} {\bibinfo {author} {\bibnamefont {USAF},}} (\bibinfo
  {year} {2018}),\ \href@noop {} {\enquote {\bibinfo {title} {The global
  positioning system},}\ }\bibinfo {howpublished}
  {\url{https://www.gps.gov/systems/gps/}},\ \bibinfo {note} {[Online; accessed
  15-February-2018]}\BibitemShut {NoStop}%
\bibitem [{\citenamefont {Weihs}\ \emph {et~al.}(1998)\citenamefont {Weihs},
  \citenamefont {Jennewein}, \citenamefont {Simon}, \citenamefont
  {Weinfurter},\ and\ \citenamefont {Zeilinger}}]{weihs}%
  \BibitemOpen
  \bibfield  {author} {\bibinfo {author} {\bibnamefont {Weihs}, \bibfnamefont
  {Gregor}}, \bibinfo {author} {\bibfnamefont {Thomas}\ \bibnamefont
  {Jennewein}}, \bibinfo {author} {\bibfnamefont {Christoph}\ \bibnamefont
  {Simon}}, \bibinfo {author} {\bibfnamefont {Harald}\ \bibnamefont
  {Weinfurter}}, \ and\ \bibinfo {author} {\bibfnamefont {Anton}\ \bibnamefont
  {Zeilinger}}} (\bibinfo {year} {1998}),\ \bibfield  {title} {\enquote
  {\bibinfo {title} {{Violation of Bell's Inequality under Strict Einstein
  Locality Conditions}},}\ }\href
  {https://journals.aps.org/prl/abstract/10.1103/PhysRevLett.81.5039}
  {\bibfield  {journal} {\bibinfo  {journal} {Physical Review Letters}\
  }\textbf {\bibinfo {volume} {81}}~(\bibinfo {number} {23}),\ \bibinfo {pages}
  {5039}}\BibitemShut {NoStop}%
\bibitem [{\citenamefont {Young}(1804)}]{young}%
  \BibitemOpen
  \bibfield  {author} {\bibinfo {author} {\bibnamefont {Young}, \bibfnamefont
  {Thomas}}} (\bibinfo {year} {1804}),\ \bibfield  {title} {\enquote {\bibinfo
  {title} {Experimental demonstration of the general law of the interference of
  light},}\ }\href {http://rstl.royalsocietypublishing.org/content/94/1.1}
  {\bibfield  {journal} {\bibinfo  {journal} {Philosophical Transactions of the
  Royal society of London}\ }\textbf {\bibinfo {volume} {94}}~(\bibinfo
  {number} {1804}),\ \bibinfo {pages} {1}}\BibitemShut {NoStop}%
\bibitem [{\citenamefont {Zarkeshian}\ \emph {et~al.}(2017)\citenamefont
  {Zarkeshian}, \citenamefont {Deshmukh}, \citenamefont {Sinclair},
  \citenamefont {Goyal}, \citenamefont {Aguilar}, \citenamefont {Lefebvre},
  \citenamefont {Puigibert}, \citenamefont {Verma}, \citenamefont {Marsili},
  \citenamefont {Shaw}, \citenamefont {Nam}, \citenamefont {Heshami},
  \citenamefont {Oblak}, \citenamefont {Tittel},\ and\ \citenamefont
  {Simon}}]{zarkeshian}%
  \BibitemOpen
  \bibfield  {author} {\bibinfo {author} {\bibnamefont {Zarkeshian},
  \bibfnamefont {P}}, \bibinfo {author} {\bibfnamefont {C.}~\bibnamefont
  {Deshmukh}}, \bibinfo {author} {\bibfnamefont {N.}~\bibnamefont {Sinclair}},
  \bibinfo {author} {\bibfnamefont {S.~K.}\ \bibnamefont {Goyal}}, \bibinfo
  {author} {\bibfnamefont {G.~H.}\ \bibnamefont {Aguilar}}, \bibinfo {author}
  {\bibfnamefont {P.}~\bibnamefont {Lefebvre}}, \bibinfo {author}
  {\bibfnamefont {M.~Grimau}\ \bibnamefont {Puigibert}}, \bibinfo {author}
  {\bibfnamefont {V.~B.}\ \bibnamefont {Verma}}, \bibinfo {author}
  {\bibfnamefont {F.}~\bibnamefont {Marsili}}, \bibinfo {author} {\bibfnamefont
  {M.~D.}\ \bibnamefont {Shaw}}, \bibinfo {author} {\bibfnamefont {S.~W.}\
  \bibnamefont {Nam}}, \bibinfo {author} {\bibfnamefont {K.}~\bibnamefont
  {Heshami}}, \bibinfo {author} {\bibfnamefont {D.}~\bibnamefont {Oblak}},
  \bibinfo {author} {\bibfnamefont {W.}~\bibnamefont {Tittel}}, \ and\ \bibinfo
  {author} {\bibfnamefont {C.}~\bibnamefont {Simon}}} (\bibinfo {year}
  {2017}),\ \bibfield  {title} {\enquote {\bibinfo {title} {Entanglement
  between more than two hundred macroscopic atomic ensembles in a solid},}\
  }\href {https://www.nature.com/articles/s41467-017-00897-7} {\bibfield
  {journal} {\bibinfo  {journal} {Nature communications}\ }\textbf {\bibinfo
  {volume} {8}}~(\bibinfo {number} {1}),\ \bibinfo {pages} {906}}\BibitemShut
  {NoStop}%
\bibitem [{\citenamefont {Zeilinger}(2010)}]{zeilinger2010}%
  \BibitemOpen
  \bibfield  {author} {\bibinfo {author} {\bibnamefont {Zeilinger},
  \bibfnamefont {Anton}}} (\bibinfo {year} {2010}),\ \href
  {https://us.macmillan.com/danceofthephotons/antonzeilinger/9781429963794/}
  {\emph {\bibinfo {title} {{Dance of the Photons: from Einstein to Quantum
  Teleportation}}}}\ (\bibinfo  {publisher} {Farrar, Straus and
  Giroux})\BibitemShut {NoStop}%
\bibitem [{\citenamefont {\ifmmode~\dot{Z}\else \.{Z}\fi{}ukowski}\ \emph
  {et~al.}(1993)\citenamefont {\ifmmode~\dot{Z}\else \.{Z}\fi{}ukowski},
  \citenamefont {Zeilinger}, \citenamefont {Horne},\ and\ \citenamefont
  {Ekert}}]{zukowski}%
  \BibitemOpen
  \bibfield  {author} {\bibinfo {author} {\bibnamefont {\ifmmode~\dot{Z}\else
  \.{Z}\fi{}ukowski}, \bibfnamefont {M}}, \bibinfo {author} {\bibfnamefont
  {A.}~\bibnamefont {Zeilinger}}, \bibinfo {author} {\bibfnamefont {M.~A.}\
  \bibnamefont {Horne}}, \ and\ \bibinfo {author} {\bibfnamefont {A.~K.}\
  \bibnamefont {Ekert}}} (\bibinfo {year} {1993}),\ \bibfield  {title}
  {\enquote {\bibinfo {title} {{``Event-ready-detectors'' Bell experiment via
  entanglement swapping}},}\ }\href {\doibase 10.1103/PhysRevLett.71.4287}
  {\bibfield  {journal} {\bibinfo  {journal} {Physical Review Letters}\
  }\textbf {\bibinfo {volume} {71}},\ \bibinfo {pages} {4287}}\BibitemShut
  {NoStop}%
\bibitem [{\citenamefont {Zurek}(1991)}]{decoherence}%
  \BibitemOpen
  \bibfield  {author} {\bibinfo {author} {\bibnamefont {Zurek}, \bibfnamefont
  {Wojciech~H}}} (\bibinfo {year} {1991}),\ \bibfield  {title} {\enquote
  {\bibinfo {title} {{Decoherence and the Transition from Quantum to
  Classical}},}\ }\href {https://doi.org/10.1063/1.881293} {\bibfield
  {journal} {\bibinfo  {journal} {Physics Today}\ }\textbf {\bibinfo {volume}
  {44}}~(\bibinfo {number} {10}),\ \bibinfo {pages} {36}}\BibitemShut {NoStop}%
\end{thebibliography}

\end{document}